\documentclass[a4paper]{article}
\pdfoutput=1
\usepackage{graphicx} 
\usepackage{jcappub}
\usepackage{url}
\usepackage{graphicx}
\usepackage{tikz}
\usepackage{subcaption,siunitx,booktabs}
\usepackage{amsmath,amssymb,amstext,amssymb,amsfonts,amsthm}
\usepackage{fontawesome} 
\usepackage{xspace}
\usepackage{hyperref}
\usepackage{appendix}
\usepackage{float}
\usepackage{physics}
\usepackage{soul}

\usepackage{bm}
\usepackage{comment}
\usepackage{color}
\usepackage[normalem]{ulem}
\usepackage{subcaption}
\usepackage{todonotes}
\usepackage[version=3]{mhchem}
\usepackage{aas_macros}
\usepackage[compat=1.0.0]{tikz-feynman}
\usepackage{cancel}
\usepackage{feynmp-auto}
\usepackage{stackengine} 

\usepackage{scalerel}
\usetikzlibrary{svg.path}

\definecolor{orcidlogocol}{HTML}{A6CE39}
\tikzset{
  orcidlogo/.pic={
    \fill[orcidlogocol] svg{M256,128c0,70.7-57.3,128-128,128C57.3,256,0,198.7,0,128C0,57.3,57.3,0,128,0C198.7,0,256,57.3,256,128z};
    \fill[white] svg{M86.3,186.2H70.9V79.1h15.4v48.4V186.2z}
                 svg{M108.9,79.1h41.6c39.6,0,57,28.3,57,53.6c0,27.5-21.5,53.6-56.8,53.6h-41.8V79.1z M124.3,172.4h24.5c34.9,0,42.9-26.5,42.9-39.7c0-21.5-13.7-39.7-43.7-39.7h-23.7V172.4z}
                 svg{M88.7,56.8c0,5.5-4.5,10.1-10.1,10.1c-5.6,0-10.1-4.6-10.1-10.1c0-5.6,4.5-10.1,10.1-10.1C84.2,46.7,88.7,51.3,88.7,56.8z};
  }
}

\newcommand\orcidicon[1]{\href{https://orcid.org/#1}{\mbox{\scalerel*{
\begin{tikzpicture}[yscale=-1,transform shape]
\pic{orcidlogo};
\end{tikzpicture}
}{|}}}}

\newcommand{\OO}{\mathcal{O}}
\newcommand{\HH}{\mathcal{H}}
\newcommand{\lcdm}{\Lambda {\rm CDM}}
\newcommand{\neff}{N_{\rm eff}}

\newcommand{\rhoc}{\rho_{\rm crit}}

\DeclareSIUnit{\parsec}{pc}

\emergencystretch=\maxdimen

\usepackage{listings}
\definecolor{softgreen}{rgb}{0.85,0.95,0.85}
\definecolor{softblue}{rgb}{0.99,0.99,1.0}
\definecolor{softpurple}{rgb}{0.95,0.90,1.0}
\definecolor{darkblue}{rgb}{0,0.2,0.4}
\definecolor{darkgreen}{rgb}{0.1,0.4,0.1}
\definecolor{darkpurple}{rgb}{0.4,0.1,0.4}

\lstdefinestyle{pystyle}{
    language = Python,  
    backgroundcolor=\color{softblue},
    commentstyle=\color{darkgreen},
    keywordstyle=\color{darkblue},
    stringstyle=\color{darkpurple},
    basicstyle=\ttfamily\small\selectfont,,
    breakatwhitespace=false,         
    breaklines=true,                 
    captionpos=b,                    
    keepspaces=true,                 
    numbers=none,                    
    numbersep=10pt,                  
    numberstyle=\tiny\color{darkblue},
    showspaces=false,                
    showstringspaces=false,
    showtabs=false,                  
    tabsize=2,
    morekeywords={as,len,True,False},
    columns=flexible,
    frame=single,
    framesep=2pt,
    framerule=0.4pt,
    rulecolor=\color{darkblue},
    xleftmargin=0pt,
    aboveskip=10pt,
    belowskip=10pt,
    alsoletter={_}
}

\lstdefinestyle{mathstyle}{
    language = Mathematica,  
    commentstyle=\color{codegreen},
    keywordstyle=\color[rgb]{0,0,0.75},
    keywordstyle=[1]\color[rgb]{0,0,0.75},
    keywordstyle=[2]\color[rgb]{0,0,0.75},
    keywordstyle=[3]\color[rgb]{0,0,0.75},
    keywordstyle=[4]\color[rgb]{0,0,0.75},
    commentstyle=\color[rgb]{0.133,0.545,0.133},
    stringstyle=\color{codepurple},
    basicstyle={\ttfamily \small},
    breakatwhitespace=false,         
    breaklines=true,                 
    captionpos=b,                    
    keepspaces=false,                 
    numbers=none,                    
    numbersep=5pt,                  
    showspaces=false,                
    showstringspaces=false,
    showtabs=false,                  
    tabsize=2,
    morekeywords={True, False, len},
    columns=flexible
}

\newcommand{\githubicon}{\href{https://github.com/cgiovanetti/LINX}{\faGithub}\xspace}

\definecolor{userInputColor}{HTML}{4287f5}
\definecolor{userCallColor}{HTML}{fcba03}
\definecolor{codeCallColor}{HTML}{039c33}
\definecolor{codeOutputColor}{HTML}{000000}
\definecolor{dictColor}{HTML}{9c1609}

\title{ABCMB: A Python+JAX Package for the Cosmic Microwave Background Power Spectrum}

\author[a]{Zilu Zhou \orcidicon{0000-0002-2800-958X},}
\emailAdd{zz1994@nyu.edu}
\affiliation[a]{Center for Cosmology and Particle Physics, Department of Physics, New York University, New York, NY 10003, USA}
\author[b, c]{Cara Giovanetti \orcidicon{0000-0003-1611-3379},}
\emailAdd{cgiovanetti@lbl.gov}
\affiliation[b]{Theory Group, Lawrence Berkeley National Laboratory, Berkeley, CA 94720, USA}
\affiliation[c]{Leinweber Institute for Theoretical Physics, University of California, Berkeley, CA 94720, USA}
\author[d]{and Hongwan Liu \orcidicon{0000-0003-2486-0681}}
\emailAdd{hongwan@bu.edu}
\affiliation[d]{Physics Department, Boston University, Boston, MA 02215, USA}

\abstract{We present ABCMB, a differentiable Einstein-Boltzmann solver for the cosmic microwave background (CMB).  ABCMB is an analysis-ready code package capturing important effects to linear order in $\Lambda$CDM cosmology.  It computes the CMB power spectrum and includes effects like lensing, polarization, massive neutrinos, and a state-of-the-art treatment of Big Bang Nucleosynthesis and recombination.  ABCMB has sub-percent-level agreement with CLASS and can be run on a GPU with competitive, and sometimes even faster, run times, owing to ABCMB's favorable run time scaling with maximum multipole moment. It is refactored compared to previous codes and takes advantage of object-oriented programming to improve extensibility, meaning new physics can be added to it without the need for modifying source files.  ABCMB provides accurate and stable gradients to the user, making Fisher analyses straightforward, and enabling the use of efficient gradient-based sampling methods.}

\date{\today}

\begin{document}
\maketitle

\section{Introduction}\label{sec:intro}

The Cosmic Microwave Background (CMB) is a pillar of precision cosmology, with extraordinarily detailed measurements making it a rich source of information about the early universe.  The success of \textit{Planck}~\cite{Aghanim_2018}, the Atacama Cosmology Telescope (ACT)~\cite{Louis_2025}, and the South Pole Telescope (SPT)~\cite{Bianchini_2020}/SPT-3G~\cite{camphuis2025spt3gd1cmbtemperature} in measuring CMB anisotropies to high precision has led to world-leading inference of cosmological parameters.  Near-future results from the Simons Observatory~\cite{simons} and SPT-3G can reduce uncertainties on these parameters even further.

As such, consistency with the CMB is a high barrier that any cosmological model must clear, be it the Lambda Cold Dark Matter ($\Lambda$CDM) paradigm or models of cosmology involving new physics.  Indeed, an analysis of effects on the CMB power spectrum is central to investigating many models of dark matter or other physics beyond the Standard Model (BSM) (see e.g.\ refs.~\cite{Lesgourgues:2015wza,Buen-Abad:2017gxg,DiValentino:2017oaw,Poulin:2018cxd,Boddy:2018wzy,Poulin:2018dzj,Aloni_2022,zhou2024,Herold_2025}).  
Performing such analyses often involves modifying and running public Einstein-Boltzmann solver codes, such as CLASS~\cite{lesgourgues_2011a,lesgourgues_2011b,lesgourgues_2011c,lesgourgues_2011d} or CAMB~\cite{Lewis_1999,Howlett_2012}, that compute the CMB anisotropy power spectrum in user-defined cosmologies.

Although existing Boltzmann codes are sufficiently fast for many applications, it is also easy to find examples of analyses involving complicated posteriors (e.g.\ bimodal posteriors), the comparison of a large number of different models, or high-dimensional parameter spaces, in which full parameter estimation with one of these codes becomes computationally expensive, requiring millions of calls to the Boltzmann code in traditional Markov-Chain Monte Carlo (MCMC) methods.  This is especially true for analyses involving a large number of extra model or nuisance parameters, e.g.\ those involving the CMB in combination with Big Bang Nucleosynthesis (BBN)~\cite{LINX_short}, large-scale structure~\cite{Doux_2018}, ultraviolet luminosity functions~\cite{Barron:2025dys} and others. 

Solutions to this problem fall into two primary categories.  The first is emulation, in which the cost of an individual code call is reduced by orders of magnitude using machine learning to mimic output of a more expensive Einstein-Boltzmann or likelihood calculation~\cite{Spurio_Mancini_2022,Piras_2023,El_Gammal_2023,Nygaard_2023,Bonici_2024,ole,janken2025clientnewtoolemulating,hough2026kosmulatorpythonframeworkcosmological}.  While traditional emulators are limited by the need to retrain them for analyses involving new physics, recently released packages~\cite{ole} have used online learning strategies to dramatically reduce this additional training time.  
The second strategy is differentiability, in which the Einstein-Boltzmann solver returns gradients of its output so that an investigator can use efficient, gradient-based sampling algorithms that require fewer code calls than methods that do not use gradients for sufficiently complex posteriors~\cite{li_2023,Hahn_2024,sletmoen_2025,list_2025}.  This approach maintains the reliability of first-principles computation while still offering performance improvements over non-differentiable codes.  The SPT collaboration, for example, has released a differentiable likelihood package to be used in conjunction with these differentiable codes~\cite{Balkenhol_2024}, and used it in analyses of the first SPT-3G data release~\cite{camphuis2025spt3gd1cmbtemperature}.   A differentiable version of the likelihoods used in the \textit{Planck} analysis is also available~\cite{benabed_clipy}, opening the door to differentiable analysis pipelines for \textit{Planck}.  Having a differentiable Einstein-Boltzmann solver is essential to take full advantage of these tools.  These two strategies are not mutually exclusive, as indeed many emulators are also differentiable.  

Modifying existing codes to include new particle interactions or entirely new fluid species can also be challenging.  While many public codes employ modular organization and are easy to browse, adding new parameters or functions may require declaring and defining new variables in separate source files, parsing large data structures (which may also be defined in source files separate from the ones the user wishes to modify), and generally being very familiar with the interior organization of the codes.   Einstein-Boltzmann solvers tend to have large source files, and so the need to understand their interior design can be a hindrance to including this new physics.

In this paper, we introduce Autodifferentiable Boltzmann solver for the CMB (ABCMB), a new differentiable Einstein-Boltzmann solver.  ABCMB computes the matter power spectrum and the CMB temperature and E-mode power spectra, achieving comparable accuracy to CLASS over a wide range of $\Lambda$CDM parameters, including the effects of massive neutrinos and lensing.  It has competitive, and indeed sometimes faster, run times as compared to CLASS when run on a GPU for calculations of equivalent precision.  

Primary breakthroughs of ABCMB include:
\begin{enumerate}
    \item Ease of use and extensibility.  New physics can be added to ABCMB without ever needing to open a source file, and the code is written in Python so that it can be easily learned and used by physicists. 
    \item A high precision recombination calculation required for state-of-the-art analyses.  This is achieved through the companion code HyRex, a differentiable version of HYREC-2~\cite{HYREC2}, which is not currently implemented in other differentiable Einstein-Boltzmann solvers.
    \item A fast, state-of-the-art BBN calculation.  ABCMB ships with LINX~\cite{LINX_long} and can compute BBN abundances given user input, rather than interpolating pre-tabulated results, enabling precise BBN theoretical predictions.
\end{enumerate}

ABCMB and HyRex use JAX~\cite{jax2018github,deepmind2020jax}, making them faster than pure Python and backend-agnostic so they can be run on CPU or GPU, and easily interpretable to Python users, keeping barriers to entry low.  These codes are fully differentiable, meaning functional gradients can be computed using autodifferentiation (AD), making the code particularly powerful for Fisher forecasting and when used in combination with efficient, gradient-based parameter estimation methods.   ABCMB is available publicly via PyPI\footnote{\href{https://pypi.org/project/ABCMB/}{https://pypi.org/project/ABCMB/}} and GitHub\footnote{\href{https://github.com/TonyZhou729/ABCMB}{https://github.com/TonyZhou729/ABCMB}}.  HyRex, included with ABCMB, is also available as a standalone package.\footnote{\href{https://github.com/TonyZhou729/HyRex}{https://github.com/TonyZhou729/HyRex}}

In addition to introducing ABCMB, in this paper we take advantage of the code's differentiability to explore the relationships between the CMB power spectra and various cosmological parameters.  In particular, because LINX is included with ABCMB, we are able to explore the interdependence of BBN abundances, the baryon density, the effective number of neutrinos, and the CMB power spectra.

The rest of the paper is organized as follows.  The next section is dedicated to practical use of the code; after a brief discussion of JAX and its utility for physics codes, we provide a high-level overview of the philosophy of ABCMB\@.  We proceed through a basic tutorial, followed by more detailed instructions for using the code. In section~\ref{sec:grad}, we explore the advantages of the differentiability of ABCMB\@.  We show how ABCMB can be used to better understand how the CMB and matter power spectra behave when cosmological parameters are varied, and can be combined with other differentiable tools to reveal parameter dependencies that have not previously been explored.  We validate the code against CLASS and HYREC-2~\cite{HYREC2} and benchmark the code performance in section~\ref{sec:performance}, and conclude and discuss directions for future work in section~\ref{sec:conclusion}.  

The paper also includes four appendices.  The first provides a complete index of run options and input parameters, beginning with an overview of options for BBN in appendix~\ref{app:BBN} and detailing all other options in appendix~\ref{app:options}.  Appendix~\ref{app:physics} details the physics implementation in ABCMB, including computation strategies for recombination (appendix~\ref{sec:HyRex}) and reionization (appendix~\ref{app:reion}), perturbations (appendix~\ref{sec:perturbations}), transfer and power spectra (appendix~\ref{app:spectra}), and lensing (appendix~\ref{app:lensing}).   Appendix~\ref{app:approx}  indexes approximation schemes used in the code.  Appendix~\ref{app:species} provides further detail about the equations describing each fluid species in ABCMB, and appendix~\ref{app:gradients} includes reference figures of gradients of the ABCMB output computed with respect to a number of cosmological parameters.

\section{Using ABCMB}

\subsection{JAX}\label{sec:JAX}
ABCMB uses JAX, a library that can be imported to any Python script and used to write fast and differentiable code.  The benefits of JAX for open-source physics codes are discussed in more detail in ref.~\cite{LINX_long} (section IIIA) in the context of the BBN code LINX\@.  In short, Python code written with JAX can be Just-In-Time (JIT) compiled for faster run times, is backend-agnostic (meaning code that runs on standard CPUs can also be trivially GPU-accelerated), and is automatically parallelized over all available devices with JIT automatic parallelism. The JAX \texttt{vmap} transformation can be applied to any function to vectorize over its JAX array inputs.  All of these features lead to superior performance over pure Python.  Finally, JAX code is autodifferentiable, meaning gradients of all functions are computed by applying the chain rule to operations close to compiler-level (as opposed to numerical derivatives computed with finite differences), unlocking gradient-based sampling algorithms including Hamiltonian Monte Carlo (HMC) (see e.g.\ refs.~\cite{2011hmcm.book..113N,betancourt2015hamiltonian}) for efficient parameter estimation.  

\subsection{Architecture and philosophy}

The ABCMB architecture is a significant departure from predecessor codes.  We have refactored subroutines to improve extensibility, hardcoding as little as possible about the species in the ABCMB cosmology or calculation of various thermodynamic quantities.  

The basic inputs required to perform a computation consists of a dictionary of parameters, and a collection of fluids, each defined as an independent fluid module. $\Lambda$CDM fluids---such as baryons, cold dark matter (CDM), and dark energy---already come predefined in ABCMB\@.  Each fluid module contains methods for background quantities like energy density or equation of state, as well as for computing fluid perturbations according to the fluids' Boltzmann hierarchies.  
The dictionary of parameters can be used to either set properties within these fluids, e.g.\ the CDM energy density $\omega_{\textrm{cdm}}$ and neutrino mass $m_\nu$, or to set other global values required for the computation, such as the scalar amplitude $A_s$, the spectral index $n_s$, and the reionization optical depth $\tau_{\rm reion}$. 
New fluids can be defined by creating a new fluid module, which contains required methods for the fluid's background and perturbation quantities, and passing that module into ABCMB at initialization. 
Parameters of the new fluid can be included in the parameter dictionary as necessary; all new fluid parameters must be assigned a value in this dictionary for a successful computation. 

While each fluid module is independent, methods within each module are written to accept both the parameter dictionary and a list of all perturbations, so that fluids can be easily coupled to one another. 
The computation of background quantities, perturbations, and spectra are each performed within their own centralized modules, which are aware of the full collection of fluids defined in the calculation. 
These modules can loop through all of the fluids and call methods within them, allowing them to construct e.g.\ the full set of coupled ordinary differential equations needed to solve for the evolution of perturbations, or integrate the Hubble parameter as a function of time.  As an example, the module that computes perturbations constructs a single vector including density, velocity, shear, etc.\ perturbations for every fluid in a cosmology.  When this module calls methods related to a given fluid, it passes a list of all fluids, and this entire vector of perturbations $y$, to the method.  A method need only reference the indices of $y$ related to another fluid to access the current values of the latter's Boltzmann hierarchy, and therefore couple to it.  Fluids contain attributes that make finding those indices trivial, attributes which can be accessed through the list of fluids that is also passed to all methods.

We stress that these centralized background, perturbation and spectrum modules \textit{are designed to minimize the need for users to modify them}: users should only need to modify the methods of individual fluids within each fluid module, or define new fluids as necessary.  
This philosophy makes extensions to ABCMB easy: the user does not need to modify any source files to include new fluids!  Instead, the user need only define a new fluid class, which contains all of the background and perturbation quantities relevant to itself and pass in the new fluid at initialization.  The background, perturbations, and spectrum modules will ask the new fluid for e.g.\ its energy density, or its perturbation equations, or for its attribute declaring whether it behaves as matter at late times, as they are needed.  This behavior will occur without ABCMB being explicitly instructed to do so, so long as a new fluid module is defined with the attributes and methods the code expects.  ABCMB provides template base classes, discussed in more detail below, to make the process of defining new fluids as streamlined as possible.

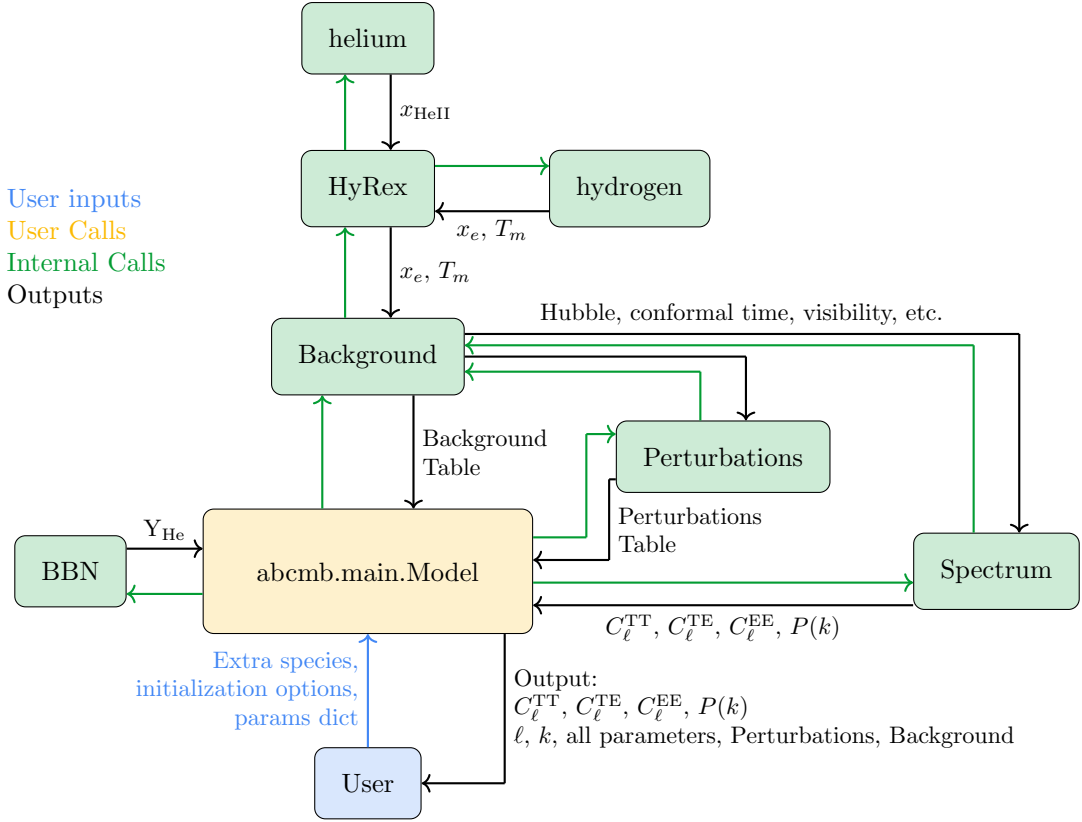
\begin{figure}
    \centering
    \begin{tikzpicture}
        \coordinate (user) at (-10,-10);
        \node[draw, fill=userInputColor!20, rectangle, rounded corners, inner sep=10pt,anchor=west] (userBlock) at (user) {User};

        \coordinate (extraSpeciesArrowStart) at ($(userBlock.north)$);

        \draw[->,thick,userInputColor] (extraSpeciesArrowStart) -- ++(0,1.5) node[midway, left, align=right,scale=0.9]{Extra species,\\initialization options,\\params dict} coordinate (extraSpeciesArrow); 

        \coordinate (Model) at ($(userBlock.north)+(0,1.5)$);
        \node[draw, fill=userCallColor!20, rectangle, rounded corners, inner sep=20pt,anchor=south] (ModelBlock) at (Model.north) {abcmb.main.Model};

        \coordinate (speciesArrowStart) at ($(ModelBlock.north)-(0.6,0)$);
        \coordinate (ModelParamsDictArrowStart) at (ModelBlock.north)+(0.3,0);

        \draw[->,thick,codeCallColor] (speciesArrowStart) -- ++(0,1.5) coordinate (speciesArrow); 

        \coordinate (BBNParamsDictArrowStart) at ($(ModelBlock.west)-(0,0.3)$);

        \draw[->,thick,codeCallColor] (BBNParamsDictArrowStart) -- ++(-1,0)  coordinate (BBNParamsDictArrow);

        \node[draw, fill=codeCallColor!20, rectangle, rounded corners, inner sep=10pt,anchor=east,align=center] (BBNBlock) at ($(BBNParamsDictArrow.west)+(0,0.3)$) {BBN};

        \draw[->,thick,codeOutputColor] ($(BBNBlock.east)+(0,0.3)$)-- ++(1,0) node[midway, above, align=center,scale=0.9]{Y$_{\rm{He}}$} coordinate (b);

        \coordinate (Background) at ($(ModelBlock.north)+(0,1.5)$);
        \node[draw, fill=codeCallColor!20, rectangle, rounded corners, inner sep=10pt,anchor=south] (backgroundBlock) at (Background.north) {Background};

        \coordinate (hyrexArrowStart) at ($(backgroundBlock.north)-(0.3,0)$);

        \draw[->,thick,codeCallColor] (hyrexArrowStart) -- ++(0,1.2)  coordinate (hyrexArrow);

        \node[draw, fill=codeCallColor!20, rectangle, rounded corners, inner sep=10pt,anchor=south] (hyrexBlock) at ($(hyrexArrow.north)+(0.3,0)$) {HyRex};

        \draw[->,thick,codeCallColor] ($(hyrexBlock.north)-(0.3,0)$) -- ++(0,1)  coordinate (heliumArrow);

        \node[draw, fill=codeCallColor!20, rectangle, rounded corners, inner sep=10pt,anchor=south] (heliumBlock) at ($(heliumArrow.north)+(0.3,0)$) {helium};

        \draw[->,thick,codeOutputColor] ($(heliumBlock.south)+(0.3,0)$) -- ++(0,-1) node[midway, right, align=left,scale=0.9]{$x_{\rm{HeII}}$} coordinate (heliumArrowBack);

        \draw[->,thick,codeCallColor] ($(hyrexBlock.east)+(0,0.3)$) -- ++(1.5,0)  coordinate (hydrogenArrow);

        \node[draw, fill=codeCallColor!20, rectangle, rounded corners, inner sep=10pt,anchor=west] (hydrogenBlock) at ($(hydrogenArrow.east)-(0,0.3)$) {hydrogen};

        \draw[->,thick,codeOutputColor] ($(hydrogenBlock.west)-(0,0.3)$) -- ++(-1.5,0) node[midway, below, align=center,scale=0.9]{$x_e,\,T_m$} coordinate (hydrogenArrowBack);

        \draw[->,thick,codeOutputColor] ($(hyrexBlock.south)+(0.3,0)$) -- ++(0,-1.2) node[midway, right, align=left,scale=0.9]{$x_e,\,T_m$} coordinate (hyrexArrowBack);

        \draw[->,thick,codeOutputColor] ($(backgroundBlock.south)+(0.6,0)$) -- ++(0,-1.5) node[midway, right, align=left,scale=0.9]{Background\\Table} coordinate (hyrexArrowBack);

        \coordinate (perturbations) at ($(ModelBlock.east)+(2.5,2)$);
        \node[draw, fill=codeCallColor!20, rectangle, rounded corners, inner sep=10pt,anchor=north] (perturbationsBlock) at (perturbations.east) {Perturbations};
        
        \coordinate (pertArrowStart) at ($(ModelBlock.east)+(0,0.45)$);
        \coordinate (pertArrowTurn) at ($(ModelBlock.east)+(.7,0.45)$);
        \draw[-,thick,codeCallColor] (pertArrowStart) -- (pertArrowTurn);

        \coordinate (pertArrowEnd) at ($(perturbationsBlock.west)+(0,0.3)$);

        \draw[-,thick,codeCallColor] (pertArrowTurn) -- ($(pertArrowTurn |- pertArrowEnd)$) coordinate (secondPertArrowTurn);

        \draw[->,thick,codeCallColor] (secondPertArrowTurn) -- (pertArrowEnd);

        \coordinate (backgroundToPerturbations) at ($(backgroundBlock.east)+(0,0)$);
        \coordinate(backgroundToPerturbationsBend) at (perturbationsBlock.north |- backgroundToPerturbations);

        \draw[-,thick,codeOutputColor] (backgroundToPerturbations) -- ($(backgroundToPerturbationsBend)+(0.3,0)$);
        \draw[->,thick,codeOutputColor] ($(backgroundToPerturbationsBend)+(0.3,0)$) -- ($(perturbationsBlock.north)+(0.3,0)$);

        \draw[-,thick,codeCallColor] ($(perturbationsBlock.north)-(0.3,0)$) -- ($(backgroundToPerturbationsBend)-(0.3,.2)$);
        \draw[->,thick,codeCallColor] ($(backgroundToPerturbationsBend)-(0.3,.2)$) -- ($(backgroundBlock.east)-(0,.2)$);

        \coordinate (pertOutArrowStart) at ($(perturbationsBlock.west)+(0,-0.3)$);
        \coordinate (pertOutArrowTurn) at ($(secondPertArrowTurn)+(.3,0)$);
        
        \draw[-,thick,codeOutputColor] (pertOutArrowStart) -- ( pertOutArrowTurn |- pertOutArrowStart);

        \coordinate (yheightModel2) at ($(ModelBlock.east)+(0,0.15)$);
        \coordinate (secondPertOutArrowTurn) at (pertOutArrowTurn |-  yheightModel2);

        \draw[-,thick,codeOutputColor] ( pertOutArrowTurn |- pertOutArrowStart) -- ( secondPertOutArrowTurn) node[midway, right, align=left,scale=0.9,yshift=-0.15cm]{Perturbations\\Table};
        \draw[->,thick,codeOutputColor]  (secondPertOutArrowTurn) -- (yheightModel2) ;

        \coordinate (spectrum) at ($(ModelBlock.east)+(5,0)$);
        \node[draw, fill=codeCallColor!20, rectangle, rounded corners, inner sep=10pt,anchor=west] (spectrumBlock) at (spectrum.east) {Spectrum};

        \coordinate (modelToSpectrumStart) at  ($(ModelBlock.east) +(0,-0.15)$);
        \coordinate (modelToSpectrumEnd) at (spectrumBlock.west |- modelToSpectrumStart);
        \draw[->,thick,codeCallColor](modelToSpectrumStart) -- (modelToSpectrumEnd);

        \coordinate (backgroundToSpectrum) at ($(backgroundBlock.east)+(0,0.3)$);
        \coordinate(backgroundToSpectrumBend) at (spectrumBlock.north |- backgroundToSpectrum);

        \draw[-,thick,codeOutputColor] (backgroundToSpectrum) -- ($(backgroundToSpectrumBend)+(0.3,0)$) node[midway, above, align=center,scale=0.9]{Hubble, conformal time, visibility, etc.};
        \draw[->,thick,codeOutputColor] ($(backgroundToSpectrumBend)+(0.3,0)$) -- ($(spectrumBlock.north)+(0.3,0)$);

        \draw[-,thick,codeCallColor] ($(spectrumBlock.north)-(0.3,0)$) -- ($(backgroundToSpectrumBend)-(0.3,0.15)$);
        \draw[->,thick,codeCallColor] ($(backgroundToSpectrumBend)-(0.3,0.15)$) -- ($(backgroundBlock.east)+(0,0.15)$);

        \coordinate(spectrumArrowHome) at ($(ModelBlock.east)-(0,.45)$);
        \coordinate (spectrumArrowOut) at (spectrumBlock.west |-spectrumArrowHome);
        \draw[->,thick,codeOutputColor] (spectrumArrowOut) -- (spectrumArrowHome) node[midway, below, align=center,scale=0.9]{$C_{\ell}^{\rm{TT}}$, $C_{\ell}^{\rm{TE}}$, $C_{\ell}^{\rm{EE}}$, $P(k)$};

        \coordinate (outputArrowBegin) at ($(ModelBlock.south)+(1.8,0)$);
        \coordinate (outputArrowTurn) at (outputArrowBegin |- userBlock.east);
        \draw[-,thick,codeOutputColor] (outputArrowBegin) -- (outputArrowTurn) node[midway, right, align=left,scale=0.9]{Output:\\$C_{\ell}^{\rm{TT}}$, $C_{\ell}^{\rm{TE}}$, $C_{\ell}^{\rm{EE}}$, $P(k)$\\$\ell$, $k$, all parameters, Perturbations, Background};
        \draw[->, thick, codeOutputColor] (outputArrowTurn) -- (userBlock.east);

        \node[anchor=north,align=left] (legendUserInputs) at (-13,-2) {\textcolor{userInputColor}{User inputs}\\
        \textcolor{userCallColor}{User Calls}\\
        \textcolor{codeCallColor}{Internal Calls}\\
        \textcolor{codeOutputColor}{Outputs}};

    \end{tikzpicture}
    \caption{A code block diagram for ABCMB\@.  Control is managed by \texttt{abcmb.main.Model}, which dispatches a BBN calculation, the computation of background quantities (including recombination), the evolution of perturbations, and the integration of those perturbations to compute a CMB power spectrum.  The user will rarely have cause to call these submodules directly, apart from perhaps the HyRex module which can be run standalone.  Instead these submodules return desired outputs through \texttt{abcmb.main.Model} to the user.}
    \label{fig:code_block}
\end{figure}

Fig.~\ref{fig:code_block} shows the code block diagram for an ABCMB computation, organized by the computation modules (with methods of the fluid modules called by each of them as necessary).  Control is managed by \texttt{abcmb.main.Model}, with its \texttt{\_\_call\_\_} the primary function used for computation.

\subsubsection{Capabilities}
ABCMB is intended for use as a state-of-the-art Einstein-Boltzmann solver suitable for cosmological analyses, and currently includes a wide range of features needed to perform many of the typical analyses of existing CMB data with no loss of precision compared to other standard Boltzmann codes.  In section~\ref{sec:performance}, we will verify that our treatment of lensing, E-mode polarization, and massive neutrinos, in addition to our base Einstein-Boltzmann solver in $\Lambda$CDM, agrees with state-of-the-art codes at a level of roughly $2-4$ permille.  

We include detailed treatment of recombination on par with HYREC-2, and push beyond the capabilities of state-of-the-art codes with a more detailed treatment of BBN\@.  The inclusion of a BBN calculation is especially relevant in light of new measurements of the primordial helium-4 abundance~\cite{aver2026_IV}; BBN currently provides tighter constraints on $N_{\rm{eff}}$ than CMB experiments, even in combination~\cite{yeh2026_V}, making it crucial to include BBN information in any analysis involving this parameter.

All of this is notwithstanding the technical capabilities of ABCMB---it is fully differentiable and has competitive run times as compared to CLASS\@. Indeed, since ABCMB is optimized to run on GPU, it scales more favorably with the number of modes $\ell$ considered than codes like CLASS do.  When considering modes up to $\ell=4000$, as is required to analyze data from ACT DR6~\cite{Louis_2025}, ABCMB can even outperform the CLASS run time (see section~\ref{sec:performance}).  

ABCMB is easy to modify, and was written with new physics in mind.  BSM approaches to the Hubble tension~\cite{DiValentino:2017oaw,Aloni_2022,zhou2024}, neutrino mass hierarchies~\cite{Herold_2025}, interacting dark matter~\cite{Lesgourgues:2015wza,Buen-Abad:2017gxg}, and more can be added to ABCMB without needing to hack source files.  Indeed, a few examples of modifications like these are demonstrated on the \href{https://github.com/TonyZhou729/ABCMB/blob/main/example_notebooks/ABCMB_Fluids.ipynb}{ABCMB GitHub}.

Nevertheless, there are many desirable features that are not as yet included in ABCMB\@.  To name a few examples, it currently does not allow the user to explore non-flat cosmologies (e.g.\ the curvature parameter is fixed to 0).  In addition, ABCMB only allows for computation in the synchronous gauge, though this does not hinder the accuracy or capabilities of the code.  Not all approximation schemes that improve the performance of other CMB codes are included in ABCMB (see section~\ref{sec:performance} and appendix~\ref{app:approx}). ABCMB only computes E-mode polarization, and not B-modes. These enhancements and more are left to future work.

\subsection{Basic tutorial}
To compute the CMB power spectrum with ABCMB, the user should begin by initializing an instance of \texttt{abcmb.main.Model}.  This wrapper class includes a number of options that can be specified at initialization, which will be discussed below.  After the model is initialized, cosmological parameters can be specified in a dictionary and passed into the initialized \texttt{Model}.  To run with default parameters and options (see appendix~\ref{app:options}), we can skip inputs for both initialization and call:

\begin{lstlisting}[language=Python]
from abcmb.main import Model

# initialize the Model with default initialization options
model = Model()

# pass nothing into model to use default parameters
output = model()

# extract Cl, Pk, ells, k, and params as JAX arrays
ClTT = output.ClTT
ClTE = output.ClTE
ClEE = output.ClEE
Pk   = output.Pk

l          = output.l
k          = output.k
out_params = output.params
\end{lstlisting}
On call, \texttt{model()} outputs an object \texttt{output}. This object contains the CMB spectra $C_\ell^{\rm TT},\ C_\ell^{\rm TE},\ C_\ell^{\rm EE}$,\footnote{Not to be confused with $D_{\ell} \equiv (\ell(\ell+1)/2\pi) C_{\ell}$; ABCMB outputs $C_{\ell}$.} as well as the linear matter power spectrum $P(k)$. Also included are a vector of multipoles $\ell$ for the CMB spectra, a vector of wavenumbers $k$ for the matter power spectrum, as well as a full list of parameters used for this particular computation. Additional outputs related to the background and perturbation computation are discussed below.

To specify non-default cosmological parameters, use a dictionary containing the keys ABCMB expects (see appendix~\ref{app:options} for a complete list):

\begin{lstlisting}[language=Python]
from abcmb.main import Model
# we will need some functions from JAX's numpy; we cannot use regular numpy and still
# have fast code
import jax.numpy as jnp 

# default initialization options
model = Model()

# pass new values for LCDM parameters, and Neff
params = {"omega_b"   : 0.0224, # Omega_b h^2
          "omega_cdm" : 0.13, # Omega_cdm h^2
          "h"         : 0.68,
          "A_s"       : 2.1e-9,
          "n_s"       : 0.98,
          "tau_reion" : 0.056,
          "Neff"      : 3.1}

# all other params use default values
output = model(params)

\end{lstlisting}
New $C_{\ell}$'s extracted from this \texttt{output} will reflect the new parameter values selected.  

As this line is run, \texttt{model} is just-in-time (JIT) compiled---future calls of this function that use inputs of the same sizes and datatypes as above will use a cached compiled version of the code and will evaluate quickly (see section~\ref{sec:performance}).

ABCMB organizes the background quantities (including recombination) and fluid perturbations into separate \texttt{Background} and \texttt{Perturbations} modules.  Relevant quantities stored in these modules include values of fluid perturbations, the free electron fraction during recombination, the value of the Hubble parameter at all times, the redshift and sound horizon at baryon decoupling, etc.\:

\begin{lstlisting}[language=Python]
from abcmb.main import Model
import jax.numpy as jnp 

# initialize and call as before
model = Model()
output = model()

# After computation, we can access perturbations and background objects
PT = output.PT
BG = output.BG

# with BG exposed, access the redshift and sound horizon at decoupling in Mpc
z_d  = BG.z_d()
rs_d = BG.rs_d()

# with PT exposed, we can access the log scale factor during computation
perturbations_scale_factor = PT.lna

# get one value of wavenumber k at a corresponding index
k_index = 100 # can be anything from 0 to len(PT.k) (600 by default)
k_val = PT.k[k_index]

# get the photon density perturbation for all scale factors at given k index.  The values of
# the scale factor are already exposed in perturbations_scale_factor
photon_perturbations = PT.species_perturbations["Photon"]
delta_g = photon_perturbations["delta"][:, k_index]

\end{lstlisting}
See \href{https://abcmb.readthedocs.io/en/latest/}{ABCMB documentation} for a complete list of methods and attributes of these objects.

\subsubsection{Adding new species}

Finally, it is possible to input new fluids to \texttt{abcmb.main.Model} to create user-defined cosmologies.  Massive neutrinos are not included in the default ABCMB cosmology but are predefined and can be added easily.  For example, CMB analyses with massive neutrinos frequently set massive neutrino parameters in the following way:
\begin{lstlisting}
from abcmb.main import Model
from abcmb import species # we will need the species module

# define a tuple of species to add.  MassiveNeutrino is already defined in species.py
user_species = (
    species.MassiveNeutrino,
)

# add the new species to ABCMB via input to Model at initialization
LCDM_mnu = Model(user_species = user_species)

# MassiveNeutrinos have some extra parameters you can specify
params_LCDM_mnu = {
    "N_nu_massless" : 2.,      # Scaling of massless neutrino energy density  
                                  # (literal number of species if there are no massive
                                  # neutrinos)
    "N_nu_massive"  : 1.,      # Number of species of massive neutrinos
    "T_nu_massive"  : 0.71611, # Massive neutrino effective temperature ratio.
    "m_nu_massive"  : 0.06,    # Neutrino mass, in eV
}

output_mnu = LCDM_mnu(params_LCDM_mnu)

\end{lstlisting}

ABCMB also ships with a variety of base classes whose attributes and methods can be inherited to quickly define new fluids, as will be discussed in more detail in section~\ref{sec:structure}.  As an example to illustrate the relative ease with which user-defined fluids can be added to ABCMB, below we sketch how to add a new perturbed fluid to the code (more concrete examples are explored in depth on the \href{https://github.com/TonyZhou729/ABCMB/blob/main/example_notebooks/ABCMB_Fluids.ipynb}{ABCMB GitHub}):

\begin{lstlisting}[language=Python]
from abcmb.main import Model
from abcmb.species import StandardFluid
import abcmb.constants as cnst

import jax.numpy as jnp

# Define a new class that describes the relevant background and perturbation functions 
# for this new species
class myFluid(StandardFluid):
    """
    My fluid.
    Required input parameters: params[`myParam'].
    """
    
    # Number of moments in the Boltzmann Hierarchy
    num_equations = 2 # Pick a non free-streaming species, which requires density 
                      # and velocity perturbations

    name = "myFluid"

    def __init__(self, first_idx, specs):
        super().__init__(first_idx, specs)
    
    def rho(self, lna, params):
        """
        Energy density at log scale factor lna.
        Should be in units of eV/cm^3.
        """
        ...

    def P(self, lna, params):
        """
        Pressure at log scale factor lna.
        """
        ...

    def y_ini(self, k, tau_ini, args):
        """
        Adiabatic superhorizon initial conditions for the new fluid.
        """
        params = args
        myNewParam = params["myParam"] # we will give this a value in the params dict later
        ...

    def y_prime(self, k, lna, metric_h_prime, metric_eta_prime, y, args):
        """
        Derivatives of the fluid's perturbations w.r.t. lna.
        """
        ...

    def output_perturbations(self, lna, modes, args):
        """
        Instructions for which perturbation equations to output
        """
        ...
        
\end{lstlisting}
Note that all calculations in ABCMB take place in the synchronous gauge when defining methods related to perturbations.

Once these methods are defined, we pass the fluid into \texttt{Model} at initialization through the \texttt{user\_species} argument.  Then, when specifying the input parameters, the user need only provide a value for the key they used to define their model parameter---this new parameter does not need to be declared elsewhere.
\begin{lstlisting}[language=Python]
user_species = (
    myFluid,
)

myModel = Model(user_species = user_species)

# the methods we defined above are expecting an entry in the params dict for "myParam", which 
# has no default value and therefore must be specified here:
params = {
    "myParam" : .3,
}

output_myFluid = myModel(params)
\end{lstlisting}
This will compute the CMB power spectrum in a cosmology with all $\Lambda$CDM fluids and this new fluid.  We emphasize that \textit{ABCMB allows the user to add this new fluid without ever opening a single source file}, avoiding the need to hack multiple source files in languages other than Python to add new species, a procedure that is typical of other Boltzmann solvers. 

Below, we provide more details about advanced functionality, other initialization options, and other abstract base classes that can be used to define new fluids.  Additional examples are also available on GitHub \githubicon.

\subsection{Detailed functionality}\label{sec:structure}

\subsubsection{Initialization and fluid species}
The user primarily interfaces with \texttt{abcmb.main.Model} in order to run the entire code.  \texttt{Model} includes a number of initialization options which the user can specify as keyword arguments; see appendix~\ref{app:options} for a complete list.

\texttt{Model} also contains a list of fluid species. By default, an ABCMB cosmology includes CDM, dark energy, baryons, photons, and massless neutrinos.  The user can choose to add additional fluids to \texttt{Model}, which are appended to the end of the list.  These additional fluids can be other species that ship with ABCMB (e.g.\ massive neutrinos), or user-defined by inheriting from the ABCMB \texttt{Fluid} object or its child classes.

We utilize the equinox~\cite{kidger2021equinox} package to manage object oriented programming, subclassing and inheritance. A parent object from the ABCMB \texttt{species} module specifies all attributes and methods a fluid species must contain. We have included several parent fluids that the user may choose to define a custom fluid. 

The most fundamental of these parent classes is the \texttt{Fluid} class.  To specify a \texttt{Fluid}, the user must specify all of the following methods: 
\begin{itemize}
    \item \texttt{rho}: The energy density $\rho$ of a fluid with respect to log scale factor $\ln a$, in $\SI{}{eV \,cm^{-3}}$.
    \item \texttt{P}: The pressure $p$ of a fluid with respect to log scale factor, in $\SI{}{eV \,cm^{-3}}$.
    \item \texttt{y\_ini}: A vector of initial conditions for all perturbation modes of the fluid, as a function of wavenumber, and initial conformal time for the fluid.
    
    The evolution of perturbed fluids in ABCMB is described by a differential equation for the Boltzmann hierarchy $F$ of the fluid in the synchronous gauge.  For each moment $F_{l}$, the user should specify $\frac{dF_{l}}{d\ln a}$ for the fluid, which may be a function of other moments of the fluid, moments of other fluids, or other parameters.  The values for the perturbations $F_{l}$ for all fluids are stored in a vector \texttt{y}.

    \texttt{y\_ini} is the initial condition for all of a fluid's $F_{l}$ at an initial conformal time $\tau_0$.  It should have a length equal to the fluid's number of $l$ moments (described below).

    \item \texttt{y\_prime}: A vector of derivatives with respect to $\ln a$ for the perturbed moments of the fluid (e.g.\ the functions $\frac{dF_{l}}{d\ln a}$ for all $l$ moments), given wavenumber, log scale factor, metric perturbations, and current fluid perturbation values.  This vector should also have a length equal to the fluid's number of $l$ moments.
    \item \texttt{rho\_delta}: To define this method and those that follow, we use the formalism of ref.~\cite{Ma:1995ey}, in the synchronous gauge.  See appendix~\ref{app:physics} for more details.  The perturbed line element in the Friedmann-Robertson-Walker metric has the form
    \begin{equation*}
    ds^2 = a^2(\tau)\left(-d\tau^2 + (\delta_{ij}+h_{ij})dx_i dx_j\right).
    \end{equation*}
    For scalar perturbations $h_{ij}$ can be written in Fourier space as
    \begin{equation}
    h_{ij}(\vec x,\tau) = \int\,d^3 \vec{k} \, e^{i\vec k\cdot \vec x} \left[ \hat{k}_i\hat{k}_j h_m(\vec k,\tau)+ \left(\hat{k}_i\hat{k}_j -\frac{1}{3}\delta_{ij}\right)6\eta(\vec k,\tau) \right],
    \end{equation}
    where, throughout this paper, we write the first metric perturbation as $h_m$ so that it may be disambiguated from the dimensionless Hubble constant $h$.  One can solve for the metric perturbation evolution and find
    \begin{align}
    h_m' &= \frac{2k^2 \eta}{\HH^2} + \frac{8\pi Ga^2}{\HH^2}\sum_i \bar{\rho}_i\delta_i \label{eq:hprime_mainbody}\\
    \eta' &= \frac{4\pi Ga^2}{\HH k^2}\sum_i \left(\bar{\rho}_i+\bar{p}_i\right)\theta_i \label{eq:etaprime_mainbody}\, ,
    \end{align}
    where primes denote derivatives with respect to $\ln a$.  The density perturbation for a fluid $i$, $\delta_i$, is defined as $(\rho_i-\bar\rho_i
)/\bar\rho_i$, where $\bar\rho_i$ the homogeneous background density and $\rho_i$ the actual energy density, including perturbations.  Similarly the velocity perturbation $\theta_i\equiv\nabla\cdot \vec{v}_i$, or in Fourier space $ik^jv_j$.  $\HH=aH$ is the conformal Hubble parameter, $G$ is Newton's constant, and $\bar{\rho}_i$ and $\bar{p}_i$ denote the background energy density and pressure.  The sums in eqs.~\eqref{eq:hprime_mainbody} and~\eqref{eq:etaprime_mainbody} run over all fluid species in a cosmology.

    \texttt{rho\_delta}, therefore, is a method that computes the fluid's contribution to this sum.  It returns $\bar\rho_i\delta_i$ for the given species, to be included in the sum in eq.~\eqref{eq:hprime_mainbody} when computing $h'_m$.  This method takes a log scale factor and a vector of magnitudes of perturbations of all fluids \texttt{y}, and returns the fluid's  $\bar\rho_i\delta_i$ in units $\SI{}{eV\,cm^{-3}}$.
    
    \item \texttt{rho\_plus\_P\_theta}: A method to compute a fluid's $\left(\bar\rho_i+\bar p_i\right)\theta_i$, to be included in the sum in eq.~\eqref{eq:etaprime_mainbody} when computing $\eta'$.  This method takes a log scale factor and a vector of magnitudes of perturbations of all fluids \texttt{y}, and returns the fluid's $\left(\bar\rho_i+\bar p_i\right)\theta_i$ in units $\SI{}{eV\,cm^{-3}\,Mpc^{-1}}$.  Note the difference in units from \texttt{rho\_delta} and \texttt{rho\_plus\_P\_sigma} (below), as the velocity perturbation $\theta_i$ carries units of $\SI{}{Mpc^{-1}}$. 
    \item \texttt{rho\_plus\_P\_sigma}: The CMB source functions (see eqs.~\eqref{eq:source_functions}) depend on higher derivatives of the metric perturbations, and involve a sum over $\left(\bar\rho_i+\bar p_i\right)\sigma_i$ for each fluid, where $\sigma_i$ is the fluid's shear perturbation.  \texttt{rho\_plus\_P\_sigma} is the fluid's contribution to this sum, in units $\SI{}{eV\,cm^{-3}}$, given the log scale factor and perturbation magnitudes.
    \item \texttt{output\_perturbations}: A book-keeping method to instruct which perturbation equations of the fluid should be saved in a perturbation dictionary in \texttt{PerturbationTable}. This is an opportunity for the user to view the perturbation evolution of their new species. For instance, \texttt{Baryon.output\_perturbations} saves both the density and velocity perturbations so that $\delta_b$ and $\theta_b$ are both visible after evaluation of a model.
\end{itemize}
\texttt{Fluid} also includes a method \texttt{w} (the equation of state), which is automatically defined to be the fluid's \texttt{P}/\texttt{rho} and does not need to be implemented by the user.

In addition, a \texttt{Fluid} includes the following fields (static attributes):
\begin{itemize}
    \item \texttt{first\_idx}: an integer index that tracks the position of the \texttt{Fluid}'s density perturbation in ABCMB's internal array of perturbations \texttt{y}.  This parameter defaults to 0 but is automatically set internally when \texttt{Model} is initialized, and so it does not need to be set by the user.
    \item \texttt{num\_equations}: the number of moments in the \texttt{Fluid}'s Boltzmann hierarchy, which defaults to 0.  In the array of perturbations \texttt{y}, all perturbations for fluid are stored in \texttt{y[first\_idx:first\_idx + num\_equations]}  For all fluids, the density perturbation $\delta_i$ is stored at position \texttt{first\_idx}, $\theta_i$ at \texttt{first\_idx+1}, $\sigma_i$ at \texttt{first\_idx+2}, and so on, until the highest mode in the hierarchy is stored at index \texttt{first\_idx+num\_equations-1}.
    \item \texttt{name}: a string naming the fluid, which defaults to an empty string.
    \item \texttt{is\_matter}: a boolean that indicates whether the fluid contributes to the matter overdensity today.  Defaults to \texttt{False}.
\end{itemize}
These four fields should be treated as static---once a \texttt{Fluid} is initialized, the user should not attempt to update these attributes. 

While any new fluid can be defined as a child class of \texttt{Fluid}, ABCMB also includes classes that inherit from \texttt{Fluid} and have already defined some of these methods to expedite the process of defining new species in some broad, general cases.  

The first is \texttt{BackgroundFluid}, which is used for fluids that do not have perturbations.  A fluid that inherits from \texttt{BackgroundFluid} has its perturbations methods set to empty arrays or 0.  The user need only define a \texttt{BackgroundFluid}'s \texttt{rho} and \texttt{P}, and optionally \texttt{name} and \texttt{is\_matter} (though we expect the default \texttt{is\_matter=False} to be appropriate in most cases).  \texttt{num\_equations} is fixed to 0 in this parent class.  For example, one can define a dynamical dark energy scenario by inheriting from \texttt{BackgroundFluid}:
\begin{lstlisting}
from abcmb.species import BackgroundFluid 
import jax.numpy as jnp

class DynamicalDarkEnergy(BackgroundFluid): # inherits from BackgroundFluid!
    """
    Dynamical dark energy implementation using the CPL parameterization.

    Required input parameters (passed in at runtime): params[`w0DE'], params[`waDE']

    Required derived parameters (computed given other parameters): params[`omega_Lambda']
    """

    name = "DynamicalDarkEnergy" 
    
    def __init__(self, first_idx, specs):
        super().__init__(first_idx, specs) # super init as usual in Python

    def rho(self, lna, params):
        """
        Energy density at log scale factor lna,
        in units of eV/cm^3.
        """
        rho_i = params["omega_Lambda"] * (3.*cnst.H0_over_h**2/8./jnp.pi/cnst.G)
        a = jnp.exp(lna)

        # Below is obtained by solving continuity equation \dot{\rho} = -3H\rho*(1+w)
        return rho_i * a**(-3.*(1.+params["w0DE"]+params["waDE"])) * \
             jnp.exp(3*params["waDE"]*(a-1))

    def P(self, lna, params):
        """
        Pressure at log scale factor lna.
        """
        a = jnp.exp(lna)
        w = params["w0DE"] + (1-a)*params["waDE"]
        return w*self.rho(lna, params)
\end{lstlisting}
This new fluid can be passed to the \texttt{user\_species} argument of \texttt{Model} at initialization and will be treated appropriately so long as values of \texttt{w0DE} and \texttt{waDE} are specified in the \texttt{params} dict; we will demonstrate how to do this in the next subsection.  Note that we specify which parameters we need to define in the fluid's doc string; this is done as a convenience for the user in all fluids that ship with ABCMB\@.

Before specifying values of these parameters, though, we must initialize a model with this new fluid.  This fluid can even replace the default dark energy fluid in ABCMB by instructing ABCMB not to use the default $\Lambda$CDM species:

\begin{lstlisting}
from abcmb.main import Model

# Since we're not using the default species set, we need to add all the other 
# fluids too.  They are already defined for us in species.
user_species = {
    species.Baryon,
    species.Photon,
    species.ColdDarkMatter,
    species.MasslessNeutrino,
    DynamicalDarkEnergy # our new species
}

# be sure to instruct Model not to populate the LCDM species, since we are replacing one
DDEModel = Model(user_species=user_species, use_LCDM_species=False)
\end{lstlisting}

The above is appropriate for fluids without perturbations.  Our other parent class, \texttt{StandardFluid}, is used for fluids with perturbations.  A \texttt{StandardFluid} still requires the user to specify \texttt{rho}, \texttt{P}, \texttt{y\_ini}, and \texttt{y\_prime}, as well as all of the attributes of a \texttt{Fluid}.  However, the required \texttt{rho\_delta}, \texttt{rho\_plus\_P\_theta}, and \texttt{rho\_plus\_P\_sigma} are automatically defined for a \texttt{StandardFluid} based on the fluid's \texttt{self.rho} and \texttt{self.P}:
\begin{itemize}
    \item \texttt{rho\_delta}: \texttt{y[self.first\_idx]*self.rho}, where \texttt{y[self.first\_idx]} is the full vector of perturbations \texttt{y} sliced at the index of the density perturbation for this fluid \texttt{self.first\_idx}.
    \item \texttt{rho\_plus\_P\_theta}: \texttt{y[self.first\_idx+1](self.rho+self.P)}, where \texttt{y[self.first\_idx+1]} is the velocity perturbation for this fluid.
    \item \texttt{rho\_plus\_P\_sigma}: \texttt{y[self.first\_idx+2](self.rho+self.P)}, where \texttt{y[self.first\_idx+2]} is the shear perturbation for this fluid.
\end{itemize}
This streamlines the process of defining new perturbed fluids; such a fluid can inherit from this parent class to avoid needing to specify these three methods.

The user is free to specify additional methods in a new species.  This is done routinely in ABCMB for \texttt{StandardFluid}s; ABCMB's \texttt{species.Baryon} is an example of a child class of \texttt{StandardFluid} that defines these necessary methods as well as auxiliary methods unique to this fluid.  \texttt{species.Baryon} also illustrates the ease with which fluids can be coupled to one another, without needing to hardcode couplings into source files:
\begin{lstlisting}
from abcmb.species import StandardFluid
import abcmb.constants as cnst
import jax.numpy as jnp

class Baryon(StandardFluid):
    """
    Baryon fluid species implementation (abbreviated)
    """
    
    name = "Baryon"
    num_equations = 2
    is_matter = True

    def __init__(self, first_idx, specs):
        super().__init__(first_idx, specs)

    def rho(self, lna, args):
        ...

    def P(self, lna, args):
        ...

    # An example of an auxiliary method not inherited from Standard Fluid.
    # See also how we can easily access aspects of the photon fluid!
    def cs2(self, lna, args):
        """
        Baryon sound speed squared.
        """
        BG, params, species_list, species_dict = args
        # Get photon class from list
        i = species_dict["Photon"]
        photon = species_list[i]

        Tm = BG.Tm(lna, params) # Baryon temp
        Tg = BG.TCMB(lna, params) # Photon temp
        # we'll need another function, mean_mass, to compute this sound speed
        mu = self.mean_mass(lna, (BG,params))
        R = 4.*photon.rho(lna, params)/3./self.rho(lna, params)

        return Tm/mu * (5./3. - 2./3.*mu*R/cnst.me/BG.aH(lna, params)/BG.tau_c(lna, params) \
               * (Tg/Tm - 1.))

    # Another auxiliary function
    def mean_mass(self, lna, args):
        ...

    def y_ini(self, k, tau_ini, args):
        ...

    def y_prime(self, k, lna, metric_h_prime, metric_eta_prime, y, args):
        # we include y_prime to highlight how fluids are coupled to one another--nothing 
        # is hardcoded here!
        
        BG, params, species_list, species_dict = args
        # species dict tells us at what index the photon is stored
        i = species_dict["Photon"]

        # now we access the actual photon object with species_list
        photon = species_list[i]

        aH = BG.aH(lna, params)
        cs2 = self.cs2(lna, args) # use our custom method
        R = 4.*photon.rho(lna, params)/3./self.rho(lna, params)
        tau_c = BG.tau_c(lna, params)

        delta = y[self.first_idx]
        theta = y[self.first_idx+1]

        # get the photon's velocity perturbation with photon.first_idx.  This index is the 
        # location of the first mode of the Boltzmann hierarchy, so add one to get to the 
        # velocity perturbation
        theta_g = y[photon.first_idx+1]
        delta_prime = -theta/aH-metric_h_prime/2.

        # compute baryon theta prime including photon slip
        theta_prime = -theta + cs2*k**2*delta/aH + R/tau_c/aH*(theta_g-theta)
        
        return jnp.array([delta_prime, theta_prime])

    def output_perturbations(self, lna, modes, args):
        return {
            "delta": modes[self.first_idx],
            "theta": modes[self.first_idx + 1],
        }
\end{lstlisting}
While we do not demonstrate it here, even the child classes defined in ABCMB can be used as parent classes for other fluids.  The \href{https://github.com/TonyZhou729/ABCMB/blob/main/example_notebooks/ABCMB_Fluids.ipynb}{ABCMB GitHub} includes an example of a child class that inherits from ABCMB's \texttt{species.ColdDarkMatter}.

Keeping a consistent structure across all fluids allows the rest of the program to operate while remaining agnostic to the specifics of any individual fluid. All new physics the user may wish to introduce to our program can be specified in one place, while leaving other modules unmodified.  

Note that the user does not (and should not) specify the cosmological parameters during initialization. In other words, the user may specify that there is a new fluid in their model, and can indicate where a new parameter will be used with a new key in the parameter dictionary, but no value for that parameter should be specified at this stage.  Only the fields---the immutable attributes of an object---are specified during this step.  Attempting to change a field mid-way through an analysis will trigger recompilation, and so model parameters whose values may change as the model is called repeatedly should not be assigned values until the next step.

\subsubsection{Parameters and computation}
After an instance of \texttt{Model} has been initialized with the desired initialization options and fluids, the user can specify cosmological parameters through dictionary inputs to the initialized \texttt{Model}'s \texttt{params} parameter.  

ABCMB natively allows a number of different parameters to define a cosmology---see appendix~\ref{app:options} for a complete list and default values.  If in defining new fluids the user needs additional parameters, these parameters need only be passed in with the expected key.  For example, the \texttt{DDEModel} that was defined in the previous subsection required two new parameters, \texttt{params[`w0DE']} and \texttt{params[`waDE']}:
\begin{lstlisting}[language=Python]
# continuing from dynamical dark energy example above, where we initialized DDEModel

# omega_b and Neff are ABCMB parameters, which can always be specified.  w0DE and waDE
# were used to define DynamicalDarkEnergy methods, and now we need to give them values:
params = {
    "omega_b" : 0.0224,
    "Neff"    : 3.1,
    "w0DE"    : -0.8,
    "waDE"    : -0.6,
}

# all other params use default values; call the code and store the output
output_DDE = DDEModel(params)
\end{lstlisting}

These values in the \texttt{params} dictionary, as well as default values for all other parameters, are automatically passed to all methods of all fluids.  Since \texttt{DDEModel} was already initialized with the \texttt{DynamicalDarkEnergy} fluid, it will automatically be able to access these values for \texttt{params[`w0DE']} and \texttt{params[`waDE']}.

When called, ABCMB will first compute derived parameters from the user's input parameters, such as total radiation energy density (which is fully specified given other fluid parameters) or the Hubble constant $H_0$ (which is specified by the user's input dimensionless Hubble constant $h$).  A full list of derived parameters is included in appendix~\ref{app:options}.  

Depending on the selected initialization options (see appendix~\ref{app:all_options} for more information), the calculation of derived parameters may also include a BBN calculation.  ABCMB can dispatch LINX to compute the helium-4 mass fraction Y$_{\rm{He}}$ given other cosmological parameters.  LINX allows the user to modify BBN nuisance parameters not typically included in interpolation tables, such as the neutron lifetime or uncertainties on nuclear reaction rates, in order to demonstrate their affect on the CMB power spectrum, .  Alternatively, ABCMB can interpolate over a table of pre-tabulated values of Y$_{\rm{He}}$, but this will not include the effects of BBN nuisance parameters.  The user may also input Y$_{\rm{He}}$ directly.  We note that LINX does not (currently) share the same extensibility strategy as ABCMB, and may need to be modified separately for new physics. For generic new physics scenarios (other than nonstandard $N_{\rm eff}$), there are no BBN interpolation tables available for Boltzmann codes, and so if BBN is significantly impacted by new physics LINX will need to be modified in these scenarios.  See appendix~\ref{app:BBN} for more information about ABCMB and BBN\@.

After all parameter values are set, \texttt{Model} moves on to dispatch the various modules for computation of the CMB and matter power spectra.

The \texttt{Background} module is called next, which is responsible for computing various background quantities.  These include the total energy density, Hubble, conformal time, and other quantities of interest such as the redshift of baryon decoupling and the sound horizon at decoupling.  \texttt{Background} is also responsible for managing recombination, and the companion code HyRex has been integrated into this module for this purpose (see appendix~\ref{sec:HyRex}).  \texttt{Background} stores critical background functions which appear frequently in the calculation of recombination, perturbations, and power spectra, including Hubble, the visibility function, and conformal time.  These functions are also directly accessible to the user through the \texttt{BG} attribute of the object returned by \texttt{Model}. 

Once the background cosmology is computed, ABCMB uses the output from \texttt{Background} to compute cosmological perturbations in the \texttt{Perturbations} module.  This is the most computationally expensive aspect of the calculation, requiring integration of Boltzmann hierarchies for all species in the user-specified cosmology, tracking the evolution of all of these moments.  The physics and equations implemented in this module are described in appendix~\ref{sec:perturbations}; these quantities, including density, velocity, and shear perturbations (or higher moments) are also accessible to the user through the \texttt{PT} attribute of the object returned by \texttt{Model}, with the option to append the perturbations of any user defined species as well. 

With the output from the perturbations module, \texttt{Model} finally dispatches the computation of the power spectra in the \texttt{Spectrum} module.  This module integrates transfer functions against tabulated Bessel functions to compute the matter power spectrum and $C_\ell$'s, optionally including effects from lensing.  This calculation is described in more detail in appendices~\ref{app:spectra} and~\ref{app:lensing}.
\\

\noindent
$\mathbf{N_{\rm{eff}}}$.  $N_{\rm{eff}}$ is an optional input in the \texttt{params} dictionary in ABCMB (it can also be treated as a derived parameter, described below). If the user chooses to specify \texttt{params[`Neff']}, ABCMB will automatically adjust the massless neutrino energy density to provide a cosmology with the desired $N_{\rm{eff}}$, with this parameter defined as
\begin{equation}
    N_{\rm{eff}}=\left(\frac{\rho_R - \rho_\gamma}{\rho_\nu^{\rm{inst}}}\right)_0.\label{eq:Neffdef}
\end{equation}
$\rho_R$ is the total energy density in radiation,\footnote{ABCMB actually tallies the energy density in all species $\rho_{\rm{tot}}$ to compute $N_{\rm{eff}}$, but since it performs this calculation deep in radiation domination, $\rho_{\rm{tot}}\approx\rho_R$ to precision far beyond what can be measured today.} $\rho_\gamma$ is the energy density in photons and $\rho_\nu^{\rm{inst}}$ is the energy density in one neutrino species assuming that species decoupled instantaneously (so its temperature is $T_\nu^{\rm{inst}}=\left(4/11\right)^{1/3}T_\gamma$).  The subscript ``0'' indicates that this ratio should be computed well after electron/positron annihilation (or any other transitions in a user-defined cosmology that would change the number of total number of effectively massless degrees of freedom in the early universe).  We note that in Boltzmann codes, $N_{\rm{eff}}$ is typically treated as a constant anyway, since modes around the time of electron-positron annihilation freeze-out lie well beyond experimental reach.  Nevertheless it can be useful to insist on this ``post-annihilation'' definition of $N_{\rm{eff}}$ in case ABCMB is used in conjunction with LINX for BBN, since these energy densities do evolve during BBN\@.

This prescription decouples user input radiative species from the calculation of $N_{\rm{eff}}$.  In other words, the user is free to specify a new fluid whose energy density contributes to $\rho_R$, and yet can still input $N_{\rm{eff}}$ independently of the parameters of the fluid.  For example, consider a cosmology with user-defined dark radiation, whose energy density $\rho_d$ contributes meaningfully to $\rho_R$.  If the user inputs $N_{\rm{eff}}=4.5$ and yet $\left(\rho_d/\rho_\nu^{\rm{inst}}\right)_0=0.5$ given other input fluid parameters, ABCMB will still realize a cosmology with $N_{\rm{eff}}=4.5$ by adjusting the energy density in massless neutrinos.  More specifically, ABCMB adjusts the number of massless neutrino species; this procedure is discussed in more detail in the next section.

This prescription works for cosmologies with massless neutrinos, cosmologies with massive neutrinos, and cosmologies with user-defined radiative fluids, so long as the user's requested $N_{\rm{eff}}$ is not so small that it would require a negative number of massless neutrinos to realize (in which case ABCMB will throw an error and exit).  This is also demonstrated in more detail in the next section.

ABCMB also allows the user to treat \texttt{params[`Neff']} as a derived parameter, meaning the user can specify all other physical parameters contributing to $N_{\rm{eff}}$, and allow ABCMB to compute and store the resulting $N_{\rm{eff}}$.  In this case, the user is required to specify a number of massless neutrino species.  All other parameters that contribute to $N_{\rm{eff}}$---the massless neutrino temperature, the number and temperature of massive neutrino species, parameters controlling the energy densities of user-defined fluids---must also be specified, or left at their default values in the case of other neutrino parameters.  This option may be more desirable if the user wishes to explore cosmologies where changes to $N_{\rm{eff}}$ occur due to the inclusion of a specific new fluid, or if the scenario of interest is particularly sensitive to the neutrino temperature.

The user switches between these options based on the keys they choose to specify in the \texttt{params} dictionary.  If the user specifies \texttt{params[`Neff']}, the first approach will be used and the number of massless neutrinos will be adjusted to account for all other input parameters.  If instead the user specifies the number of massless neutrino species \texttt{params[`N\_nu\_massless']}, ABCMB will compute $N_{\rm{eff}}$ given this and all other inputs.  If the user attempts to specify both \texttt{params[`Neff']} and \texttt{params[`N\_nu\_massless']}, the code will print a warning and exit without running any computations.  If neither parameter is specified, the code defaults to $N_{\nu, {\rm massless}}=3$ and adjusts $\neff$ accordingly (properly accounting for any user-defined fluid contributions to $N_{\rm{eff}}$). Default values are used for the massless neutrino temperature, the number of massive neutrinos, and the massive neutrino temperature in all scenarios, unless the user provides other values for these parameters (see appendix~\ref{app:options}).

\subsubsection{Neutrino parameters}\label{sec:neutrinos}
We include detailed descriptions of the implementations of all fluids in appendices~\ref{app:physics} and~\ref{app:species}.  However, since the details of the neutrino implementation are especially relevant for specifying $N_{\rm{eff}}$ and understanding user input parameters like the massive neutrino temperature, we define parameters related to these fluids here.
\\

\noindent
\textbf{Massless neutrinos.}  The massless neutrino energy density is parametrized according to 
\begin{equation}
    \rho_{\nu,\rm{massless}}(a) = N_{\nu,\rm{massless}}\frac{7\pi^2}{120}T^4_{\nu,\rm{massless}}a^{-4},\label{eq:rho_massless_def}
\end{equation}
where $T_{\nu,\rm{massless}}$ is the temperature of massless neutrinos today and $N_{\nu,\rm{massless}}$ is the number of massless neutrinos.  $a$ is the scale factor at the given temperature.  In analyses where all neutrinos are massless, $N_{\nu,\rm{massless}}$ is 3 by default and $T_{\nu,\rm{massless}}$ is $0.71636856 \times T_{\gamma}$, $T_{\gamma}$ the photon temperature, to recover a default $N_{\rm{eff}}$ of 3.044~\cite{Akita:2020szl,Froustey:2020mcq,Bennett:2020zkv}.  These parameters can be adjusted by specifying \texttt{params[`N\_nu\_massless']} ($N_{\nu,\rm{massless}}$) and \texttt{params[`T\_nu\_massless']} ($T_{\nu,\rm{massless}}/T_{\gamma}$ after decoupling).

Above, we discussed that the number of massless neutrino species $N_{\nu,\rm{massless}}$ may adjust in response to a user input $N_{\rm{eff}}$.  In particular, after the user inputs a value for $N_{\rm{eff}}$, ABCMB loops through all fluids present in the user-specified cosmology---with the exception of massless neutrinos---and computes each fluid's contribution to $N_{\rm{eff}}$.  If the calculated $N_{\rm{eff}}$ does not match the user specified $N_{\rm{eff}}$, ABCMB will make up the difference by adjusting $N_{\nu,\rm{massless}}$.  In other words, ABCMB solves eq.~\eqref{eq:Neffdef} for $N_{\nu,\rm{massless}}$ given all fluids and $N_{\rm{eff}}$, and fixes this parameter to the value required to provide the requested $N_{\rm{eff}}$ in a user's specified cosmology.
\\

\noindent
\textbf{Massive neutrinos.} We follow the CLASS prescription for treating massive neutrinos. Massive neutrinos have three new parameters: $N_{\nu,\rm{massive}}$, the number of massive neutrinos, $T_{\nu,\rm{massive}}$, a temperature associated with massive neutrinos, and $m_\nu$, the neutrino masses, which are assumed to be degenerate.  These parameters are derived based on the results obtained in ref.~\cite{Mangano_2005}, which carefully studies the phase-space density of each neutrino flavor including distortions to the Fermi-Dirac distribution due to non-instantaneous decoupling. 

Let $f_{\rm exact}(p, a)$ be the exact phase space density of one massive neutrino (as computed in ref.~\cite{Mangano_2005}) as a function of momentum $p$ and scale factor $a$.\footnote{In fact, we assume that all massive neutrinos have the same phase distribution by taking average of the phase space of each neutrino flavor.} 
Then $T_{\nu, {\rm massive}}$ is set by ensuring that the number density $n_{\nu,{\rm massive}}(a = 1)$---and likewise also $\rho_{\nu,{\rm massive}}(a = 1) = m_\nu n_{\nu,{\rm massive}}(a = 1)$---is equal to the number density of a free-streaming fermion that decoupled relativistically and instantaneously, with present-day temperature $T_{\nu, {\rm massive}}$, i.e. 
\begin{alignat}{1}
\int \frac{d^3 \vec{p}}{(2\pi)^3} f_{\rm exact} (p,a = 1)  = \int \frac{d^3\vec{p}}{(2\pi)^3} \frac{1}{e^{p/T_{\nu,{\rm massive}}} + 1} \,,
\end{alignat}
where the left-hand side can be evaluated using the results of ref.~\cite{Mangano_2005}, leading to the $\Lambda$CDM value for three degenerate neutrinos of\footnote{We note that this result is commonly extrapolated to cases where only one neutrino is massive, or where $N_{\rm{eff}}$ takes on a different value. We caution that extrapolation of the results in ref.~\cite{Mangano_2005} to this regime has not been validated, to our knowledge.}
\begin{alignat}{1}
\left( \frac{T_{\nu,{\rm massive}}}{T_\gamma} \right)_0 \approx 0.71611 \,.
\end{alignat}
At all $a < 1$, we make the further assumption that the phase space density can be approximated as
\begin{alignat}{1}
f(p, a) \approx \left( e^{pa/T_{\nu,\rm{massive}}} + 1 \right)^{-1} \,,
\label{eq:approx_massive_nu_f}
\end{alignat}
i.e.\ it can be treated as free-streaming, with $T = a^{-1} T_{\nu,{\rm massive}}$. 
The massive neutrino energy density is then computed at all redshifts as 
\begin{alignat}{1}
\rho_{\nu,{\rm massive}}(a) = 2N_{\nu,{\rm massive}} \int \frac{d^3 \vec{p}}{(2\pi)^3} f(p,a) \sqrt{p^2 + m_\nu^2} \,,
\end{alignat}
with the approximate $f(p,a)$ defined in eq.~\eqref{eq:approx_massive_nu_f}. 
This simplification, however, results in a $\rho_{\nu,{\rm massive}}$ that has fractional error from the true value $\rho_{\nu,{\rm massive, exact}}$ of
\begin{alignat}{1}
\frac{\rho_{\nu, {\rm massive, exact}}(a)}{\rho_{\nu,{\rm massive}}(a)} - 1 = \frac{\int \frac{d^3 \vec{p}}{(2\pi)^3} \delta f(p) \sqrt{p^2 + m_\nu^2 a^2}}{\int \frac{d^3 \vec{p}}{(2\pi)^3} [e^{p/T_{\nu,{\rm massive}}} + 1]^{-1} \sqrt{p^2 + m_\nu^2 a^2}} \stackrel{a \to 0}{\to} \frac{\int \frac{d^3 \vec{p}}{(2\pi)^3} p \, \delta f(p) }{\int \frac{d^3 \vec{p}}{(2\pi)^3} p [e^{p/T_{\nu,{\rm massive}}} + 1]^{-1}}\,,
\end{alignat}
where
\begin{alignat}{1}
\delta f(p) \equiv f(p, a = 1) - \frac{1}{e^{p/T_{\nu,\rm massive}} + 1} \,,
\end{alignat}
i.e.\ the spectral distortion relative to the assumed distribution in eq.~\eqref{eq:approx_massive_nu_f} at $a = 1$. 
At sufficiently early times when $T_{\nu,{\rm massive}} \gg m_\nu a$, the ratio goes to a constant that is independent of $m_\nu$ and $a$. 

To compensate for this error, as well as the error associated with assuming that all massive neutrinos have the same phase space, we simply set $N_{\nu,{\rm massless}}$ such that the total $N_{\rm eff}$ is equal to the user-specified value, which in $\Lambda$CDM is 3.044. Specifically, ABCMB performs the same procedure as in the massless case: it loops through all species, except massless neutrinos, and computes the contribution of each to $N_{\rm{eff}}$. When it comes time to add the massive neutrino contribution, it computes the approximate energy density $\rho_{\nu,\rm{massive}}$ with temperature $T_{\nu,\rm{massive}}$, knowing this is not the correct energy density in massive neutrinos at early times. Instead, ABCMB adjusts the energy density in massless neutrinos by adjusting $N_{\nu,\rm{massless}}$---exactly as it did in scenarios that do not involve massive neutrinos. This means that $N_{\nu,\rm{massless}}$ is not truly the ``number’' of massless neutrino species anymore in massive neutrino cosmologies. Instead, $N_{\nu,\rm{massless}}$ absorbs a correction from the massive neutrino energy density being computed from $T_{\nu,\rm{massive}}$.\footnote{This is similar to the treatment in CLASS, where even in cosmologies where all three neutrinos are massive, the user still must include some small nonzero ``\texttt{N\_ur}‘’ to achieve an $N_{\rm{eff}}$ of 3.044.}

With the temperature set as described above, ABCMB computes the energy density and momentum of massive neutrinos by integrating their phase space with Gaussian quadrature---see appendix~\ref{app:massive_neutrinos} for more detail.

\section{Gradients of ABCMB outputs}\label{sec:grad}
A major advantage of code written in JAX is the effortless calculation of gradients with autodifferentiation (AD).  We emphasize that gradients computed with JAX, and therefore ABCMB, are not a result of numerical differentiation, but instead use AD (i.e. decomposing all functions into low-level operations like addition, multiplication, etc.\ whose functional derivatives are known, and applying the chain rule).  This makes the resulting gradients more stable than traditional numerical derivatives and more reliable for techniques like Fisher forecasting (as recently explored in ref.~\cite{Hahn_2024}), and is more efficient than traditional methods like finite differences.  As discussed in section~\ref{sec:JAX}, this feature can also be used with more sophisticated sampling techniques than traditional Markov Chain Monte Carlo (MCMC), leading to fewer function calls and more efficient parameter estimation for high-dimensional or complex posteriors.

ABCMB uses forward AD (or forward accumulation) by default.  For a function
\begin{equation}
    y(x)=f(g(h(x))),
\end{equation}
forward AD computes $\partial{y} / \partial x$ by first computing $\partial h / \partial x$, followed by $\partial g / \partial h$, and finally followed by $\partial f / \partial g$, computing partial derivatives from inside to outside.  By contrast, reverse AD computes derivatives in reverse order, beginning with $\partial f / \partial g$ and ending with $\partial h / \partial x$.  In either case, the value of the partial derivative is propagated either forward or backward through these chains as subsequent derivatives are computed. 

While these two operations are mathematically equivalent, their performance can vary depending on the scenario.  If $y :\mathbb{R}^n\rightarrow\mathbb{R}^m$, forward AD is more efficient than reverse AD for $m>n$, while reverse AD is more efficient for $n>m$.  This is because forward AD requires a new pass through the chain for each independent input variable (but only one pass to propagate each input variable through \textit{all} outputs), while reverse AD instead needs a separate pass through the chain for all output variables (but only one pass to propagate to \textit{all} inputs).  

ABCMB has a large number of outputs---it is expected that several thousand $C_\ell$ and $P(k)$ values are computed in a single evaluation, and so forward AD is more appropriate.  The same is true for HyRex.  We include illustrations of gradients with respect to a number of cosmological parameters in appendix~\ref{app:gradients}.  Each of these is computed using forward AD with the built-in JAX function \texttt{jax.jacfwd}.\footnote{For applications involving likelihoods, where the ABCMB output is collapsed to a single float, one might expect reverse AD to be more efficient.  While ABCMB supports reverse AD, these efficiency gains do not materialize due to high memory requirements.  Further optimization along this path is left to future work.}

\begin{figure}[t]
\centering
\includegraphics[width=.9\textwidth]{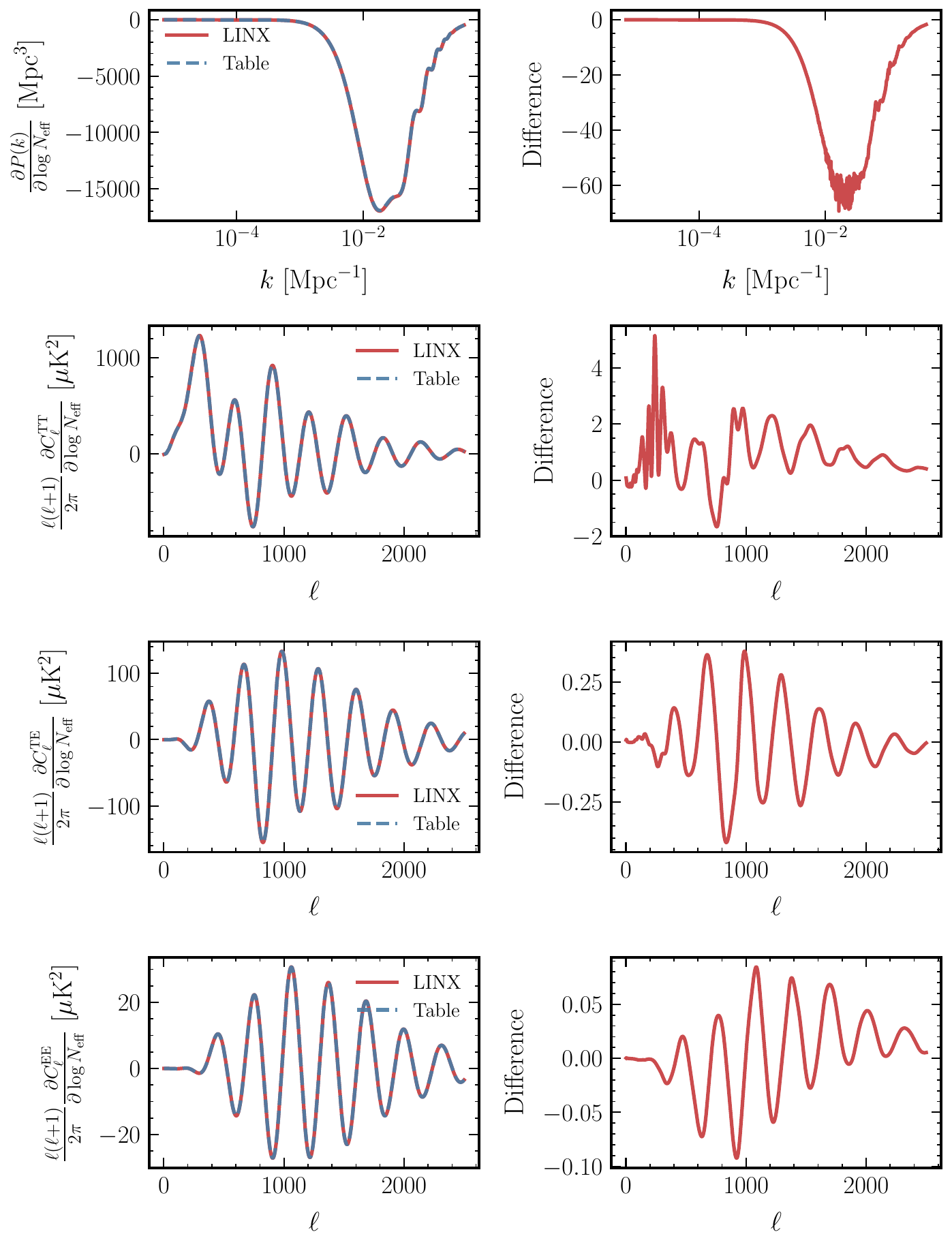}
\caption{Gradients of matter and CMB power spectra with respect to $N_{\rm eff}$. The blue dashed curves were computed with helium abundance interpolated from the CLASS default BBN table, and the red solid curves used the LINX prediction for the abundance given $N_{\rm eff}$ with a different reaction network. The right panels of each spectrum gradient shows the absolute difference between the two approaches, with identical units and normalization as the left panels where applicable.}
\label{figure:linx_gradients_neff}
\end{figure}

\begin{figure}[t]
\centering
\includegraphics[width=.9\textwidth]{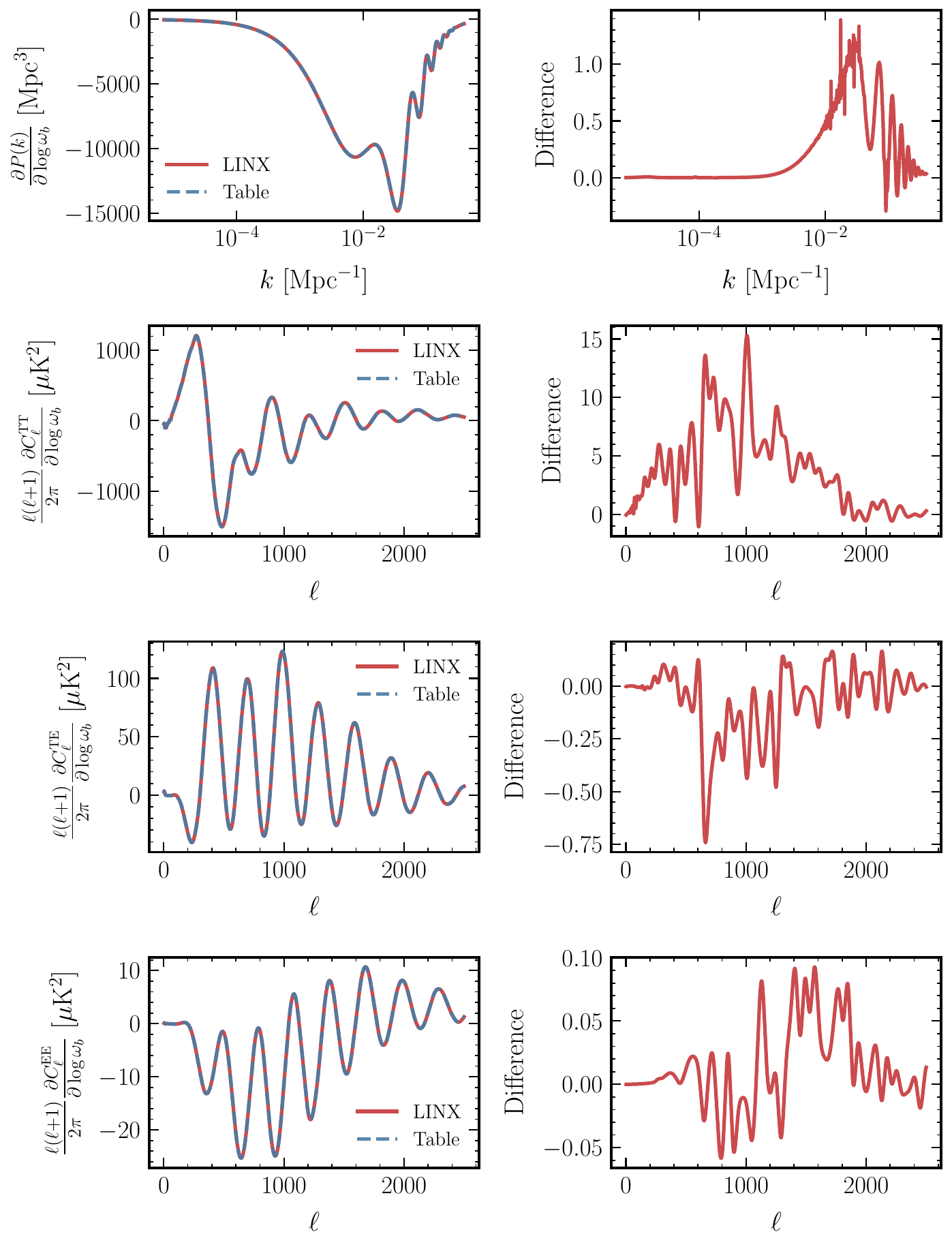}
\caption{Gradients of matter and CMB power spectra with respect to the baryon density $\omega_{\rm b}$. The blue dashed curves were computed with helium abundance interpolated from the CLASS default BBN table, and the red solid curves used the LINX prediction for the abundance given $\omega_{\rm b}$ with a different reaction network. As in fig.~\ref{figure:linx_gradients_neff}, the right panels of each spectrum gradient shows the absolute difference between the two approaches, with identical units and normalization as the left panels where applicable.}
\label{figure:linx_gradients}
\end{figure}

When ABCMB runs with LINX, the entire pipeline from BBN through recombination and the CMB power spectrum is fully differentiable.  As such we are able to take derivatives of these quantities with respect to both the baryon density and $N_{\rm{eff}}$ through all of these epochs.  LINX also allows the user to choose different reaction networks---different sets of experimentally-determined rates for the reactions active during BBN---and so here we compare the results from differentiating through ABCMB only, through its tabulated results using the reaction network of ref.~\cite{Consiglio_2018}, versus differentiating through both ABCMB and LINX, with the LINX default reaction network from ref.~\cite{Pitrou_2018}.   We show these results in figures~\ref{figure:linx_gradients_neff} and~\ref{figure:linx_gradients}.  The effects of using a different reaction network with LINX when computing gradients with respect to these parameters is as large as percent-level in some spectra.  

In new physics analyses, the ability to take gradients across epochs becomes especially important, in particular if scenarios of interest have model parameters that have a strong impact on BBN, or additional sensitivity to BBN parameters like the neutron lifetime or individual nuclear reaction rates.

\section{Performance and validation}\label{sec:performance}
We validate the output of HyRex and ABCMB against that of HYREC-2 and CLASS\@.  We separately validate each of these components, and we further separate the matter power spectrum, the CMB power spectrum, lensing, massive neutrinos, and polarization when validating our Einstein-Boltzmann solver.  We also report the performance of the code on different hardware.  We note that there is some variation of these reported run times with the version of JAX used, with version 0.4.28 and earlier running faster on CPU and versions 0.6.1 and later running faster on GPU\@.  We report run times below only for versions $\geq$ 0.6.1, as there is no support for earlier versions of JAX on the newest versions of Python.

To ensure our code has good agreement with the benchmark codes HYREC-2 and CLASS across a wide range parameter values, we randomly generate 50 samples of the $\lcdm$ parameters.  These are $\{h, \omega_b, \omega_{\rm cdm}\}$ (the Hubble parameter in units of \SI{100}{\kilo\meter\per\second\per\mega\parsec}, the present-day ratio of the baryon energy density to the critical energy density times $h^2$, and similarly for CDM) for recombination and $\{h, \omega_b, \omega_{\rm cdm}, A_s, n_s, \tau_{\rm reion}\}$ (additionally including the scale of the primordial power spectrum $A_s$, the scalar tilt $n_s$, and the optical depth to reionization $\tau_{\rm{reion}}$) for the CMB power spectra. These parameters are uniformly sampled over the $\pm5\sigma$ range reported by Planck 2018, with the exception of the Hubble parameter, which is sampled over the wider interval $h\in [0.6, 0.8]$ in light of the Hubble tension.  We summarize these parameter ranges in table~\ref{tab:test_params} along with other fixed cosmological parameters.

\begin{table}[h]
\centering
\begin{tabular}{l|c}
\hline
\textbf{Varied}            & \textbf{Range}                  \\ \hline
$h$                           & {[}0.6, 0.8{]}         \\ 
$\omega_b$                    & {[}0.02162, 0.02312{]} \\
$\omega_{\rm cdm}$            & {[}0.114, 0.126{]}   \\
$10^9 A_s$                    & {[}1.931, 2.256{]} \\
$n_s$                         & {[}0.9439, 0.9859{]}   \\
$\tau_{\rm reion}$               & {[}0.0179, 0.0909{]}            \\
\hline
\textbf{Fixed}            & \textbf{Value}                  \\ \hline
$N_{\rm eff}$                 & 3.044                  \\
$m_\nu$                       & 0.06 eV                \\
$T_{\nu, {\rm massive}}$      & 0.71611                \\
$N_{\nu, {\rm massive}}$      & 1                      \\
$\Delta z_{\rm reion}$        & 0.5                    \\
$z_{\rm{HeII\,reion}}$         & 3.5                    \\
$\Delta z_{\rm{HeII\,reion}}$ & 0.5                    \\
$\beta_{\rm{reion}}$          & 1.5                    \\ \hline
\end{tabular}
\caption{Parameters used to perform $x_e$ comparison against HYREC-2, and $C_\ell$, $P(k)$ comparisons against CLASS\@. The six $\lcdm$ parameters are uniformly sampled 50 times in the indicated range. These ranges correspond to the $5\sigma$ range reported by Planck 2018, and for $h$ it is further widened in light of the Hubble tension. $N_{\rm eff}$ is fixed in these tests, and Y$_{\rm He}$ is inferred using the CLASS BBN table. In tests including one massive neutrino, we use the massive neutrino related parameters listed in the table, which are otherwise excluded. The four reionization parameters $\Delta z_{\rm reion}$, $z_{\rm{HeII\,reion}}$, $\Delta z_{\rm{HeII\,reion}}$, and $\beta_{\rm{reion}}$ are further discussed in appendix~\ref{app:reion}.} 
\label{tab:test_params}
\end{table}

For each random set of parameters, we compare the output of our code to the benchmark, and at each redshift we record the maximum error incurred across all 50 evaluations.  The residuals are defined throughout as $|\textrm{benchmark} - \textrm{ABCMB}|/\textrm{benchmark}$.

\subsection{HyRex and HYREC-2}
\begin{figure}[h]
\centering
\includegraphics[width=\textwidth]{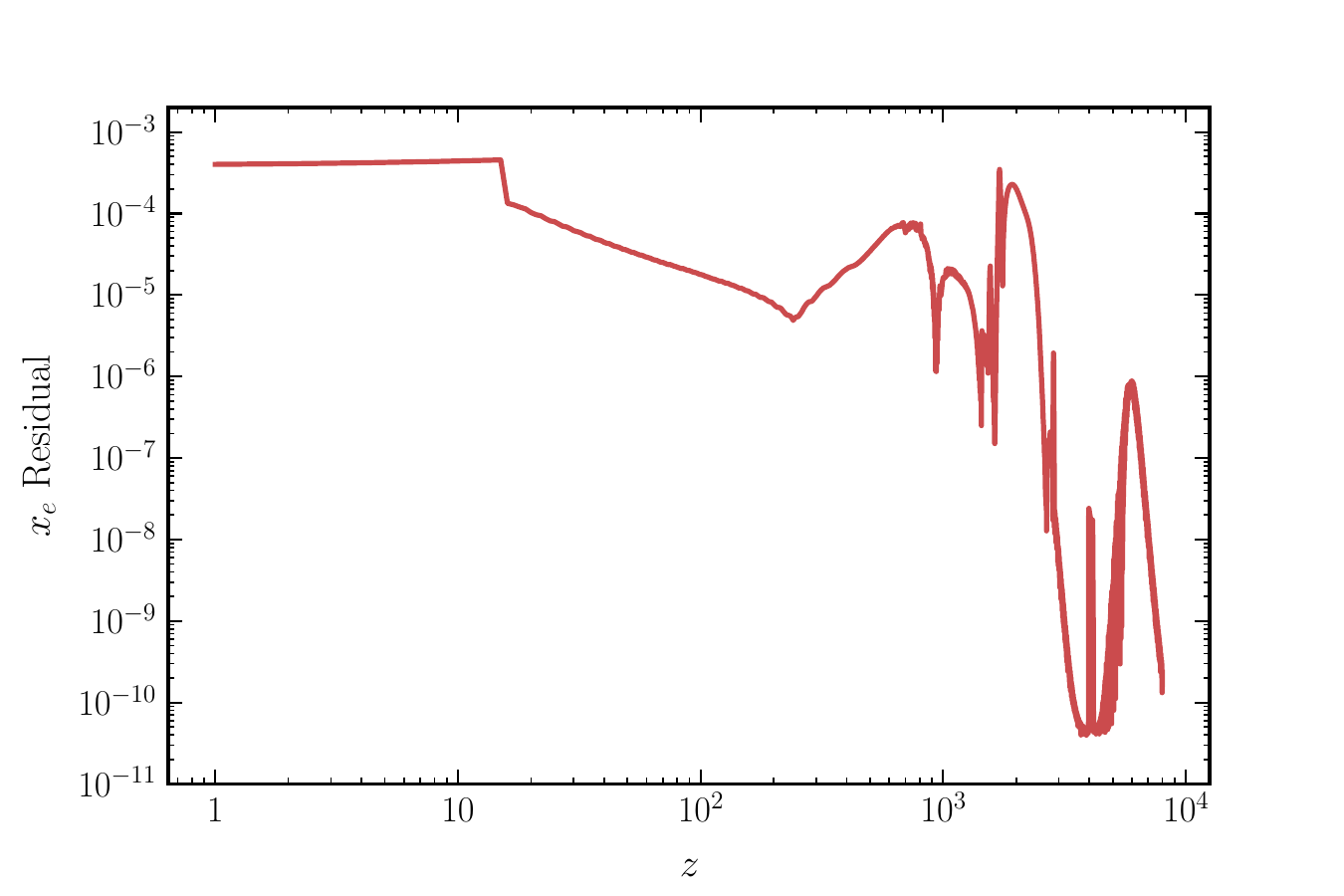}
\caption{Maximum residual comparison between HyRex and HYREC-2 for $x_e$, the free electron fraction across redshift. We make 50 comparisons using randomly generated values for the $\lcdm$ parameters $\{h, \omega_b, \omega_{\rm cdm}\}$ in the range specified in table~\ref{tab:test_params}, and report the maximum error occurred at each redshift across all runs. The two codes agree to $<0.1\%$, with better agreement at early times more relevant for the CMB\@.} 
\label{figure:xe_comparison}
\end{figure}

HYREC-2 is a state-of-the-art recombination code whose output HyRex must match in order to deliver the precision demanded by e.g.\ \textit{Planck}.  
We therefore follow the procedure described above, comparing the output of HyRex for each sample to that of HYREC-2.  In particular we compare the $x_e$ output of the two codes in the redshift range $1\leq z\leq 8000$.  We show the maximum error as a function of redshift in figure~\ref{figure:xe_comparison}. 

The agreement is excellent across the entire solution region.  The outputs never differ by more than 0.1\%, and the largest errors occur at late times, where recombination is only approximated with the three-level atom (TLA) and reionization is more relevant.  This is despite the use of different ODE solvers by the two codes.

\subsection{The ABCMB Einstein-Boltzmann solver and CLASS}
CLASS is widely used to compute the CMB power spectrum, and so we compare the output of the ABCMB Einstein-Boltzmann solver to that of CLASS\@.  We check the linear matter power spectrum and the CMB power spectra (temperature autocorrelation and E-mode polarization autocorrelation), in $\Lambda$CDM (with massless neutrinos), including massive neutrinos, and with and without lensing. For the matter spectrum we compute to $k_{\rm max}=0.5\ \SI{}{Mpc^{-1}}$, a small scale where non-linear corrections are known to be significant. We test the CMB spectra in the interval $2\leq\ell\leq4000$, covering the full measurement extent of both Planck 2018~\cite{Aghanim_2018} and ACT DR6~\cite{Louis_2025}. We follow the procedure described above, using CLASS as the benchmark, and record the largest residual difference compared to CLASS across all 50 runs at each $\ell$ and $k$.  For the CLASS precision settings, we use \texttt{`accurate\_lensing'=1}, which is important as the alternative approximation introduces percent level errors for $\ell > 3000$ (see appendix~\ref{app:lensing} for further discussion). Other precision settings, such as the perturbation Boltzmann hierarchy cutoffs, and settings related to the fluid approximations, are set as the CLASS default values. We use the ABCMB default settings in these comparisons (see appendix~\ref{app:options}).

\begin{figure}[h]
\centering
\includegraphics[width=\textwidth]{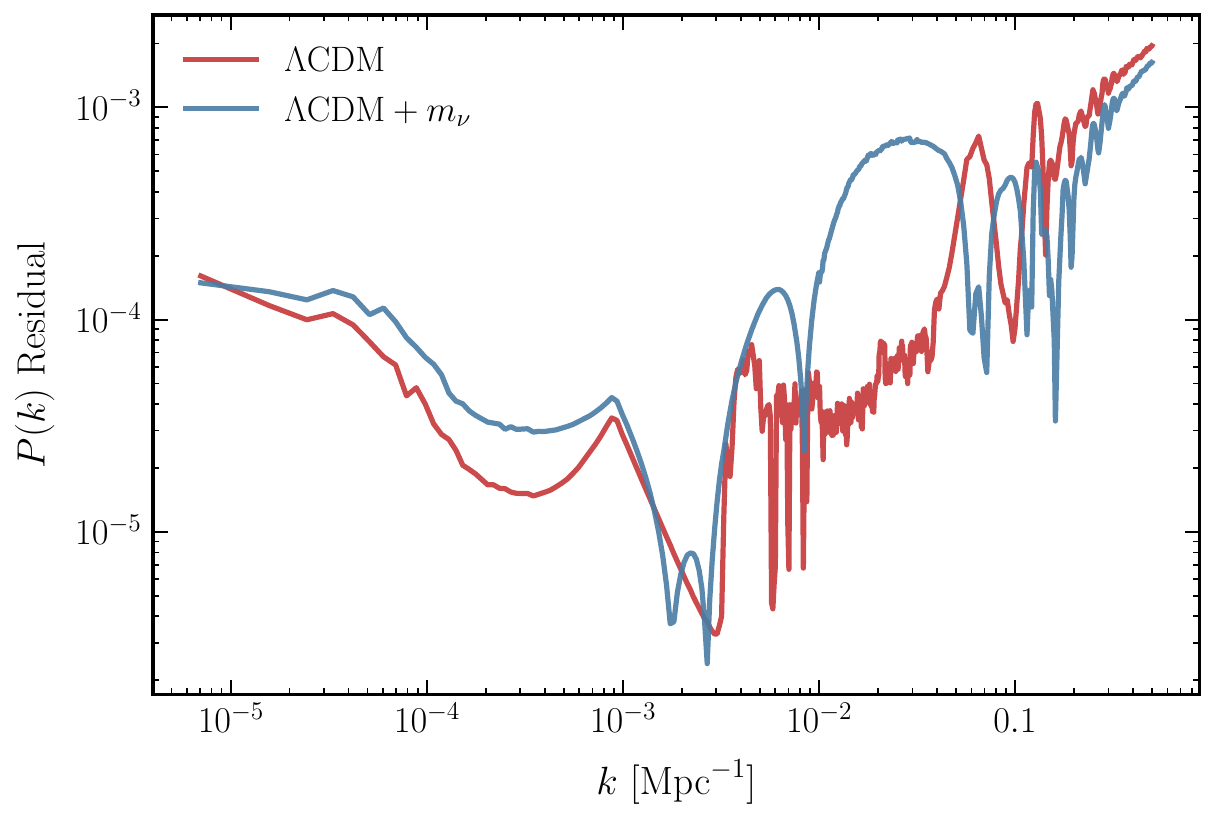}
\caption{Residuals in the linear matter power spectrum computed as $|{\rm CLASS}-{\rm ABCMB}|/{\rm CLASS}$. We make 50 comparisons using randomly generated values for the 6 $\lcdm$ parameters, and report the maximum error at each wavenumber $k$ across all runs. The range of the parameters sampled can be found in table~\ref{tab:test_params}. The red curve denotes $\lcdm$ with massless neutrinos only, while the blue curve contains one massive neutrino. The errors are comparable in the two models, with accuracy at the 2 permille level at $k=0.5\ {\rm Mpc}^{-1}$, where non-linear corrections are already important. The level of accuracy on these small scales indicate that the percent level suppression by neutrino mass is well captured in ABCMB\@.}
\label{figure:pk_residual}
\end{figure}

We show a comparison of the linear matter power spectrum in figure~\ref{figure:pk_residual}, both in $\Lambda$CDM and with massive neutrinos. In both cases we find good agreement with CLASS, to better than $2\times10^{-3}$ on all scales. This indicates ABCMB accurately captures the percent-level suppression induced by massive neutrinos at large $k$.

Next we turn to the CMB power spectra. Here we only compare the TT and EE spectra to CLASS and do not compare the TE spectrum, since the cross correlation frequently crosses zero and makes the residual comparison difficult to interpret. Since the TE spectrum uses the same transfer functions as the TT and EE spectra, their accuracy should be representative of the cross correlation as well. 

\begin{figure}[t]
\centering
\includegraphics[width=\textwidth]{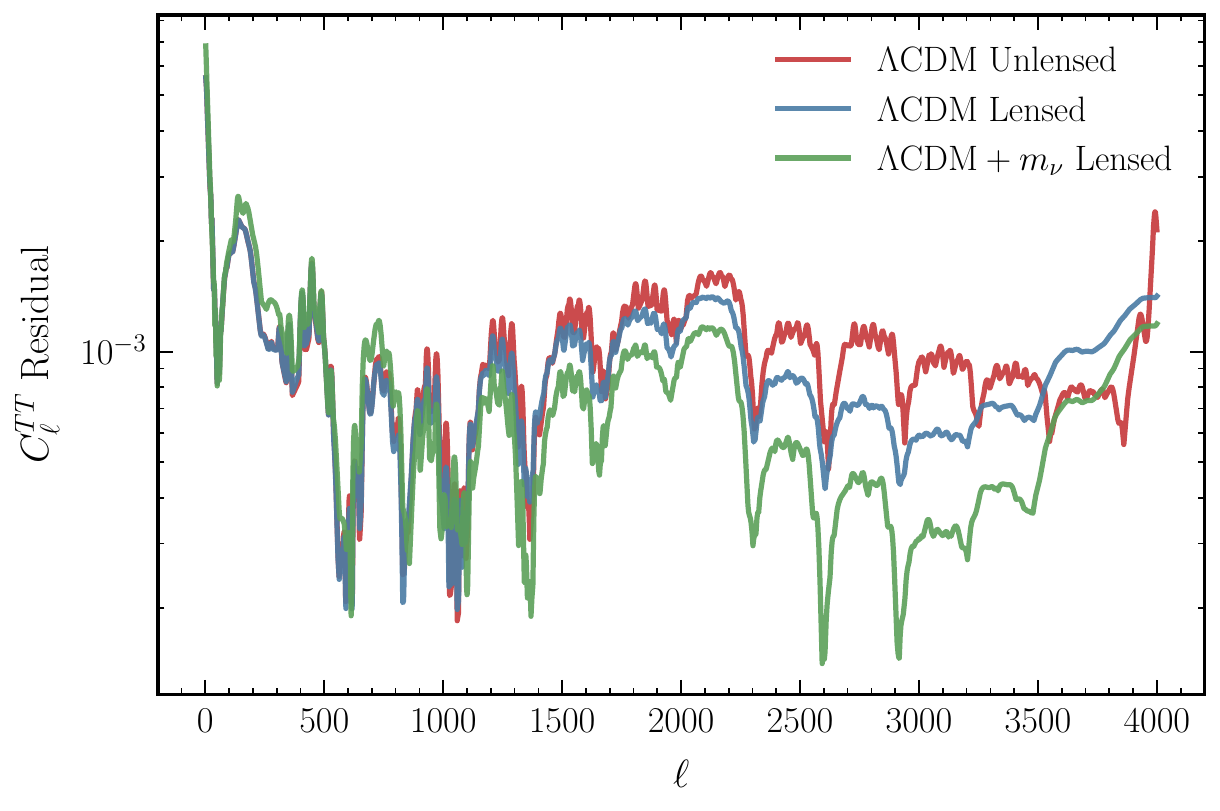}
\caption{Residuals in the CMB TT power spectrum computed as $|{\rm CLASS}-{\rm ABCMB}|/{\rm CLASS}$. We make 50 comparisons using randomly generated values for the 6 $\lcdm$ parameters, and report the maximum error at each angular scale $\ell$ across all runs. The range of the parameters sampled can be found in table~\ref{tab:test_params}. The red , blue and green curves correspond to the unlensed $\lcdm$, lensed $\lcdm$, and lensed $\lcdm$ with 1 massive neutrino, respectively. The agreement is better than 7 permille across all $\ell$ (and typically closer to 1 permille), and the presence of lensing and massive neutrino do not induce additional error above that of the unlensed spectrum. } 
\label{figure:tt_residual}
\end{figure}

\begin{figure}[t]
\centering
\includegraphics[width=\textwidth]{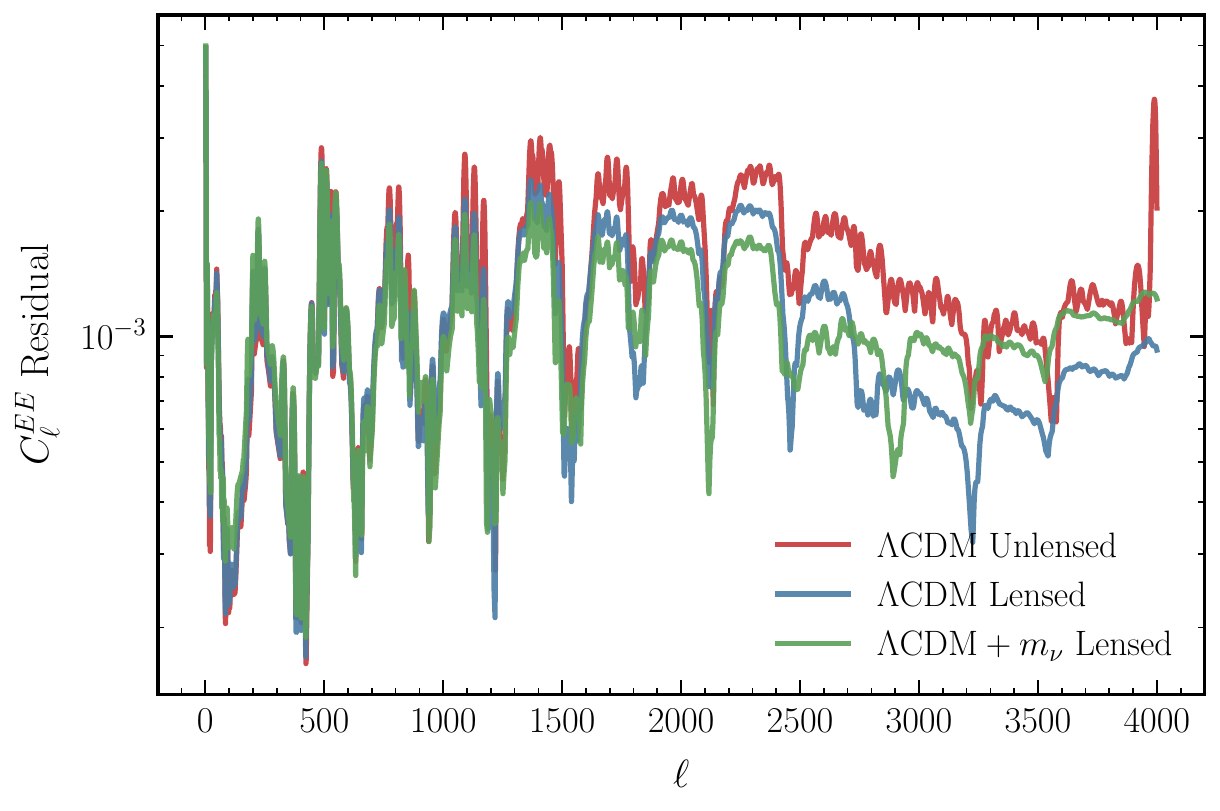}
\caption{Residuals in the CMB EE power spectrum computed as $|{\rm CLASS}-{\rm ABCMB}|/{\rm CLASS}$. We make 50 comparisons using randomly generated values for the 6 $\lcdm$ parameters, and report the maximum error at each angular scale $\ell$ across all runs. The range of the parameters sampled can be found in table~\ref{tab:test_params}. The red, blue and green curves correspond to the unlensed $\lcdm$, lensed $\lcdm$, and lensed $\lcdm$ with 1 massive neutrino, respectively. The agreement is comparable to the levels seen in the TT spectra, and again we find that lensing and massive neutrino do not induce more error.} 
\label{figure:ee_residual}
\end{figure}

We show the TT spectra residual in figure~\ref{figure:tt_residual} and the EE spectra residual in figure~\ref{figure:ee_residual}. In each figure we check the accuracy of the unlensed $\lcdm$, lensed $\lcdm$, and lensed $\lcdm+m_\nu$ combinations. We find sub-percent accuracy in all cases, and in most cases we find the error hovers around 2 permille, with exceptions only at very low $\ell$ where CMB measurements are cosmic variance dominated.  

\subsection{Performance}
As ABCMB is written in JAX, it is backend-agnostic and returns equivalent output on both CPU and GPU\@.

\paragraph{HyRex.}On an M1 Macbook Pro with 8 cores, HyRex computes $x_e$, $T_m$ and $\ln a$ during recombination and reionization in about $\SI{50}{ms}$, after a $\sim\SI{30}{s}$ compile time.  On a single Intel Xeon Platinum 8592+ 64C CPU core, the equivalent computation takes $\SI{37}{ms}$, after a $\sim\SI{28}{s}$ compile time.  Increasing the number of CPU cores available reduces compile time, but does not improve run time.  

On an NVIDIA L40S GPU run time is $\SI{1.5}{s}$, and on an H200 NVIDIA GPU it is $\SI{1.7}{s}$, with compile times around $\SI{30}{s}$ on both devices.  This slowdown is likely from use of \texttt{lax.while\_loop} and implicit use of the same function in the differential equation solvers, which relies on (slow) information exchange between the host CPU and the GPU\@.

\paragraph{ABCMB\@.}  These benchmarks include the HyRex run time, when HyRex is restricted to run on CPU (its faster backend).  We conduct the tests using the CPU and GPU devices on the NYU Torch High Performance Computing cluster. All CPU tests are run on the Intel Xeon Platinum 8592+ 64C processors, while the GPU runs were tested on the NVIDIA L40S, A100, H100, and H200 cards. We compute the lensed $\lcdm$ and $\lcdm+m_\nu$ CMB power spectra (including TT, TE and EE) up to $\ell_{\rm max}=4000$, and the matter power spectrum up to $k_{\rm max}=0.5\ {\rm Mpc}^{-1}$.  The run times are shown in tables~\ref{table:masslesstime} ($\lcdm$) and~\ref{table:massivetime} ($\lcdm+m_\nu$).  All runs are performed with 16GB memory. After an initial evaluation that compiles the code (compile time reported in fourth column), we report the average run time and standard deviation across 10 calls. We find that ABCMB performs most optimally on GPUs, outperforming its single CPU core runtime by about 10 times. We also run a test using 10 CPUs with JAX's built-in autoparallelization. We see that the compile time is considerably shorter, and the run time also sees a modest reduction. 
\begin{table}[h]
\centering
\begin{tabular}{l|c|c|c}
\hline
                       Device & Mean (s) & Variance (s) & Compile Time (s) \\ \hline
CPU Single Core      & 110.360            & 0.382                & 202         \\
CPU 10 Cores            & 60.001            & 1.429                 & 108  \\
NVIDIA L40S & 13.737            & 0.032                 & 125          \\
NVIDIA A100    & 8.964            & 0.031                 & 165          \\
NVIDIA H100    & 6.705            & 0.018                 & 120         \\ 
NVIDIA H200    & 6.402             & 0.017                 & 112         
\\\hline
\end{tabular}

\caption{ABCMB run times for $\lcdm$ with massless neutrinos only. All runs compute lensed TT, TE, and EE spectra to $\ell_{\rm max}=4000$ as well as $P(k)$ to $k_{\rm max}=0.5\ \SI{}{Mpc^{-1}}$. The rows for correspond to the devices used to conduct the tests. All run times are averaged over 10 identical runs with fluctuations characterized by the variance. We also include the compile time incurred on the initial call.}
\label{table:masslesstime}
\end{table}

\begin{table}[h]
\centering
\begin{tabular}{l|c|c|c}
\hline
                        Device & Mean (s) & Variance (s) & Compile Time (s) \\ \hline
CPU Single Core      & 256.475            & 4.940                & 346         \\
CPU 10 Cores            & 213.445           & 4.263                 & 270  \\
NVIDIA L40S & 26.980           & 0.092                 & 181          \\
NVIDIA A100    & 17.931            & 0.079                 & 211          \\
NVIDIA H100    & 12.813            & 0.059                 & 141          \\ 
NVIDIA H200    & 11.699             & 0.047                 & 129         
\\\hline
\end{tabular}
\caption{ABCMB run times for $\lcdm$ with one massive neutrino. The runs use the same precision parameter and compute identical outputs as table \ref{table:masslesstime}.}
\label{table:massivetime}
\end{table}

The ABCMB run time can be compared against existing state-of-the-art Boltzmann solvers such as CLASS.  However, given that ABCMB is GPU-optimized, while codes like CLASS and CAMB can are CPU-optimized, an apples-to-apples comparison between ABCMB and existing packages is difficult. Nevertheless, a comparison between GPU ABCMB and CPU CLASS in a realistic scenario is still informative.

An analysis using Planck, ACT and SPT for CMB, in combination with BAO, requires computing the lensed temperature and polarization spectra out to $\ell_{\rm max}\sim 4000$, and the matter power spectrum to non-linear scales; we therefore compare at the same $\ell_{\rm{max}}=4000$ and $k_{\rm{max}}=\SI{0.5}{Mpc^{-1}}$ used in the standalone ABCMB tests above. We additionally require the accurate lensing setting for CLASS which is relevant at such high $\ell$'s, and slightly impacts performance (see Appendix \ref{app:lensing}). Modern analyses typically include one massive neutrino, which also noticeably impacts performance.

ABCMB run times under these conditions are reported in table~\ref{table:massivetime}, with the fastest time around 11.7s on the NVIDIA H200 graphics card.  In contrast, on an Intel Xeon Platinum 8592+ 64C CPU, the analogous CLASS computation runs in about 13 seconds on a single core, or about 2 seconds threaded across 10 cores.

The CLASS performance is enabled by several perturbation approximation schemes that are not presently implemented in ABCMB. These include the tight-coupling approximation (TCA), the ultra-relativistic fluid approximation (UFA), and the radiation streaming approximation (RSA). TCA is used to alleviate the stiffness of the baryon-photon coupling equations at early times, and to remove the redundant higher moments (i.e. $l\geq 3$) in the photon Boltzmann hierarchy. UFA and RSA are used to drastically simplify the subhorizon modes of relativistic species, and a similar fluid approximation can be used in place of the massive neutrino Boltzmann hierarchy at late times to reduce complexity.  Without the three fluid approximations, the CLASS run time can increase by a factor of about 5.


Nevertheless, there is not (as of yet) a backend on which ABCMB runs faster than multithreaded CLASS with fluid approximations\@.  However, we note that since there are applications in which a solver may be restricted to a single device or core (e.g.\ nested sampling, or even MCMC with many chains), the single-core comparison is still informative.  Since ABCMB can also use gradient-based sampling methods it is difficult to draw further comparisons between total analysis times of these two codes for applications like parameter estimation---the fact that the ABCMB runtime is across the board comparable to that of CLASS should be taken as an encouraging sign.

\begin{figure}
    \centering
    \includegraphics[width=\linewidth]{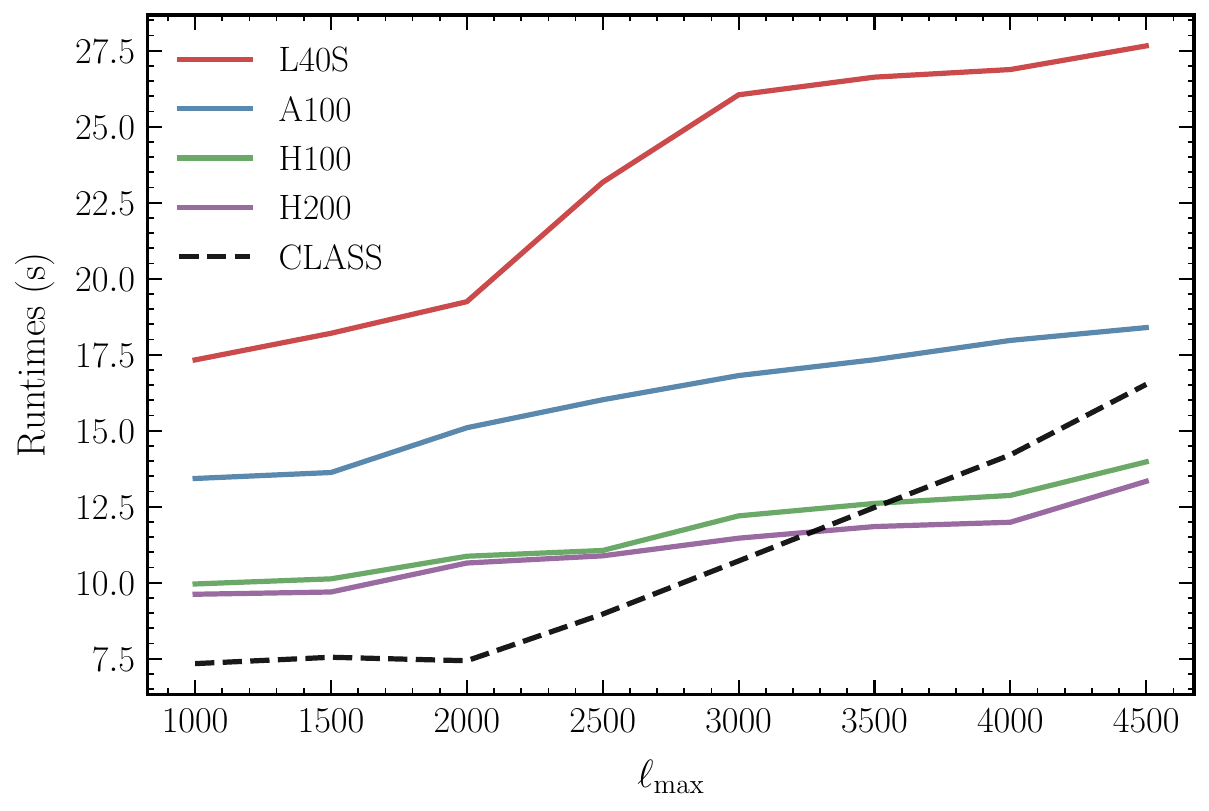}
    \caption{Single evaluation run times, after compilation, of ABCMB, on different hardware and as a function of $\ell_{\rm{max}}$, using default precision parameters and including one massive neutrino.  The scaling is nearly flat, providing a performance advantage as $\ell_{\rm{max}}$ grows.  We also show the CLASS run time on a single Intel Xeon Platinum 8592+ 64C processor for comparison; the scaling is closer to linear and therefore the run time can become longer than the ABCMB run time for large $\ell_{\rm{max}}$.}
    \label{fig:ellscale}
\end{figure}

The ABCMB run time scales favorably with $\ell_{\rm{max}}$.  This is because the code is GPU-optimized, and so evaluation can occur in parallel on a single device, until available memory is saturated.  This means that in general one may expect the ABCMB run time to become even faster than the single-core CLASS run time as $\ell_{\rm{max}}$ increases, and we illustrate these scalings in the case of a massive neutrino in figure~\ref{fig:ellscale}.  

The ABCMB default parameters are selected to produce $\sim1$ permille agreement with CLASS\@.  However, this level of precision may not always be necessary, especially in more exploratory applications.  In these scenarios, the user may enjoy $\sim25\%$ faster run times across devices by setting the option \texttt{rtol\_large\_k\_PE=1e-3}  when initializing \texttt{Model}  (see appendix~\ref{app:options}).  This parameter controls the tolerances of the differential equation solver used for perturbations at large $k$, and the default $10^{-4}$ is only needed if $\sim1$ permille agreement is required at large $\ell$.  We find that a larger tolerance of $10^{-3}$ is still sufficient to produce agreement at the level of $\sim8$ permille with CLASS for both the TT and EE spectra.  Adjusting this parameter has little effect on the agreement of $P(k)$ with CLASS, and so we certainly recommend increasing this tolerance if the user only requires the matter power spectrum for their analysis.

\paragraph{Gradients.} We repeat the timing test discussed above for computing the gradients of ABCMB output with respect to cosmological parameters, using JAX auto-differentiation. We apply forward mode auto-differentiation to the function over the 6 $\lcdm$ parameters to obtain a gradient vector $\vec{\nabla}_\theta C_{\ell}$ for each of the TT, TE, and EE spectra, as well as $\vec{\nabla}_\theta P(k)$.  We report the performance in table~\ref{table:massless_gradient} for $\lcdm$ and in table~\ref{table:massive_gradient} with one massive neutrino. CPU run times are not reported as they are too slow to be practical.

\begin{table}[h]
\centering
\begin{tabular}{l|c|c|c}
\hline
                        Device & Mean (s) & Variance (s) & Compile Time (s) \\ \hline
NVIDIA L40S & 46.924           & 0.245                 & 337          \\
NVIDIA A100    & 25.051            & 0.311                & 370          \\
NVIDIA H100    & 14.920           & 0.075                 & 242          \\ 
NVIDIA H200    & 13.336             & 0.064                 & 260         
\\\hline
\end{tabular}
\caption{ABCMB run times for forward mode auto-differentiation on $\lcdm$ outputs (with massless neutrinos). As in the timing tests for the outputs themselves, each device test consists of one initial compiling run, and 10 subsequent runs with varying cosmological parameters. }
\label{table:massless_gradient}
\end{table}

\begin{table}[h]
\centering
\begin{tabular}{l|c|c|c}
\hline
                        Device & Mean (s) & Variance (s) & Compile Time (s) \\ \hline
NVIDIA L40S & 102.697           & 0.510                 & 387          \\
NVIDIA A100    & 50.778            & 0.204                & 380          \\
NVIDIA H100    & 32.345           & 0.151                 & 334          \\ 
NVIDIA H200    & 28.229             & 0.114                 & 281       
\\\hline
\end{tabular}
\caption{ABCMB run times for forward mode auto-differentiation on outputs from $\lcdm$ with one massive neutrino. }
\label{table:massive_gradient}
\end{table}

\section{Discussion}\label{sec:conclusion}
ABCMB is an accurate, differentiable, backend-agnostic Einstein-Boltzmann solver for the CMB, offering many precision features and written with extensibility in mind.  We have validated the code  against HYREC-2 and CLASS to deliver the same level of precision in the matter power and CMB power spectra, including effects from massive neutrinos, lensing, and polarization.  This tool is useful for new physics analyses, as it provides a simple, intuitive, robust code architecture that can be modified to accommodate a number of scenarios without extensive knowledge of the inner workings of the code.  

Both $\Lambda$CDM and new physics analyses stand to benefit from the differentiability of ABCMB, as  efficient sampling algorithms capitalizing on gradient information become available for complex posteriors.  The availability of stable functional gradients means ABCMB can also be used for Fisher forecasting.  ABCMB's speed comes from JAX, not emulation, and so it can also be used in frequentist analyses that rely on dependable likelihood estimates at high resolution.

While ABCMB includes a number of desirable features for high-precision computation of the CMB power spectrum, there are many additional features that we leave to future work, some of which have already been mentioned.  All calculations in ABCMB are currently performed in flat spacetime geometries, and in the synchronous gauge.  No tensor perturbations or CMB B-modes are evaluated.  The HyRex module enforces a strict assumption that the recombination of helium and hydrogen do not overlap---while this is a good assumption in most cosmologies, future work may include relaxing this assumption.  HYREC-2 also natively supports time-varying electron mass and fine structure constant, which are eschewed in ABCMB for the time being.  Finally, ABCMB is complete only to linear order---there is currently no implementation of HALOFIT~\cite{halofit} or other nonlinear corrections to the matter power spectrum.  Curved spacetimes, the conformal Newtonian gauge, tensor perturbations, B-modes, extreme recombination histories, time-varying fundamental constants, and nonlinear corrections are left for future development.  As it stands, however, we believe that ABCMB is already capable of performing many interesting cosmological analyses. 

Future work will also include enhancements to the ABCMB performance.  Better single-GPU run time can potentially be achieved by implementing batched solvers like those used in DISCO-DJ~\cite{Hahn_2024}. There are also a number of approximations used in CLASS and CAMB in the linear perturbations equations, such as the various fluid approximations for massless and massive neutrinos, that could help make ABCMB faster. In particular, the UFA and RSA discussed in section~\ref{sec:performance} are even known to reduce error in relativistic fluid perturbations, by decoupling the photon and neutrino Boltzmann hierarchies to avoid error incurred by introducing a cutoff \cite{lesgourgues_2011b}. For these reasons, we seek to implement UFA and RSA for improvements to both speed and accuracy.  

ABCMB is already combined with LINX, so that it is trivial to perform fully differentiable joint CMB+BBN analyses including all BBN and Planck nuisance parameters, in both $\Lambda$CDM and new physics scenarios.  Particularly exciting is the prospect of further combining ABCMB with DISCO-DJ II~\cite{list_2025}; in this setup, the linear and nonlinear matter power spectra can be computed using the DISCO-DJ II $N$-body simulations, which can then be used to compute the CMB power spectrum using a ABCMB in a fully JAX, fully differentiable pipeline.  ABCMB is written with new physics in mind, and can be used to perform self-consistent, state-of-the-art analyses of BSM particle physics scenarios, potentially combining with these other probes that have been developed in JAX\@. The existence of online learning emulator frameworks such as that in Ref.~\cite{ole} also opens the possibility for fast analyses with ABCMB using ordinary, non-differentiable MCMC. These and other analyses are left to future work. 

\acknowledgments
We would like to thank Yacine Ali-Ha\"imoud, I-Kai Chen, Colin Hill, Nanoom Lee, Florian List, Siddharth Mishra-Sharma, Clark Miyamoto, Caio Nascimento, Roman Scoccimarro, Neal Weiner, Zachary Weiner, and Ian Williams for helpful discussions.  We thank Skye Wanderman-Milne and Dan Foreman-Mackey for their assistance with JAX, especially with patching together diffrax solutions in different regimes, as well as other participants of the October 2024 JAXtronomy meeting.  We thank Yacine Ali-Ha\"imoud, Julien Lesgourgues, Florian List, Marlon Josue Rivera Valladares, and Zachary Weiner for helpful feedback on a draft version of this manuscript and on our code.  We thank Katelin Schutz and Marlon Josue Rivera Valladares for beta testing our code, and we thank Catherine Welch for verifying our installation instructions.  We benefited tremendously from Hans A. Winther's cosmology lecture notes.\footnote{\url{https://cmb.wintherscoming.no/index.php}}  We use code packages numpy~\cite{harris2020array}, scipy~\cite{2020SciPy-NMeth}, diffrax~\cite{kidger2021on}, equinox~\cite{kidger2021equinox}, JAX~\cite{jax2018github,deepmind2020jax}, matplotlib~\cite{Hunter:2007}, interpax~\cite{conlin_2025}, and sphinx~\cite{sphinx}.  We use a custom interpolator developed by Yunfan Zhang.\footnote{\url{https://github.com/jax-ml/jax/issues/16182}} The authors acknowledge the use of Claude Sonnet (Anthropic) to generate most docstrings in both ABCMB and HyRex.  

C.G. is supported by the Office of High Energy Physics of the U.S. Department of Energy under contract DE-AC02-05CH11231. H.L. is supported by the U.S. Department of Energy under grant DE-SC0026297. In addition, H.L. is supported by the Cecile K. Dalton Career Development Professorship, endowed by Boston University trustee Nathaniel Dalton and Amy Gottleib Dalton.  This work was supported in part through the NYU IT High Performance Computing resources, services, and staff expertise.  This research used resources of the National Energy Research Scientific Computing Center (NERSC), a Department of Energy User Facility (project m3166).  This research used the Lawrencium computational cluster resource provided by the IT Division at the Lawrence Berkeley National Laboratory (Supported by the Director, Office of Science, Office of Basic Energy Sciences, of the U.S. Department of Energy under Contract No. DE-AC02-05CH11231).

\appendix
\section{Other options}\label{app:all_options}

\subsection{BBN}\label{app:BBN}
ABCMB natively includes three options for determining the helium-4 mass fraction,\footnote{This is as opposed to the ``helium-4 mass fraction'' sometimes used in BBN analyses, defined as Y$_{\rm{P}}=\frac{4n_{\rm{He}}}{n_{\rm{B}}}$, $n_{\rm{B}}$ the total number of baryons.}
\begin{equation}
    \rm{Y}_{\rm{He}}=\frac{\rho_{\rm{He}}}{\rho_{\rm{H}} + \rho_{\rm{He}}},
\end{equation}
from user input.  The first and simplest option is for the user to input their own float via \texttt{params[`YHe']} at runtime.

The second option is analogous to the CLASS \texttt{`BBN'} option, where user input \texttt{params[`omega\_b']} and \texttt{params[`Neff']} are used to look up a precomputed value of the helium-4 mass fraction from the public BBN code PArthENoPE~\cite{Pisanti:2007hk,Consiglio_2018, Gariazzo:2021iiu}.  We use an identical table for this option, reproducing the CLASS functionality.  This option is triggered by the user setting \texttt{bbn\_type=`Table'} (any capitalization) when initializing an instance of \texttt{abcmb.main.Model}.

Finally, ABCMB also ships with LINX, a fast and differentiable JAX BBN code~\cite{LINX_long}. Combination with LINX allows the user to sample BBN nuisance parameters such as the neutron lifetime or individual reaction rates as a part of a joint CMB+BBN analysis.  To run LINX to obtain the helium-4 abundance including effects from varying these nuisance parameters alongside cosmological parameters, the user must do the following:
\begin{enumerate}
    \item {Initialize an instance of \texttt{abcmb.main.Model} with \texttt{bbn\_type=`LINX'} (any capitalization).  Optionally include a choice of nuclear reaction network by also initializing with \texttt{linx\_reaction\_net}, which can be set to any of the (case-sensitive) LINX reaction network options (\texttt{`np\_only'}, \texttt{`key\_PRIMAT\_2023'} (default), \texttt{`full\_PRIMAT\_2023'}, \texttt{`key\_PRIMAT\_2018'}, \texttt{`key\_YOF'}, or \texttt{`key\_PArthENoPE'}).  See ref.~\cite{LINX_long} for more information about each choice of reaction network.}
    \item {Specify \texttt{params[`Delta\_Neff\_init']}, and do not specify any \texttt{params[`Neff']}, \texttt{params[`N\_nu\_massless']}, or \texttt{params[`T\_nu\_massless']}.  $N_{\rm{eff}}$ in LINX is computed by including an inert relativistic species with energy density $\rho_{\rm{ext}}$.  The user specifies the parameter
    \begin{equation}
        \Delta N_{\textrm{eff},i}\equiv\frac87 \left(\frac{11}{4}\right)^{4/3}\frac{\rho_{\textrm{ext},i}}{\rho_{\gamma,i}},
    \end{equation}
    where the subscript $i$ indicates initial quantities in the BBN calculation. 
    Since LINX starts its calculation well before electron/positron annihilation, this input parameter is \textit{not} the same as $\Delta N_{\rm{eff}}=\frac87 \left(\frac{11}{4}\right)^{4/3}\left(\frac{\rho_{\textrm{ext}}}{\rho_{\gamma}}\right)_0$, which instead is now a derived parameter.}
    \item{Optionally specify BBN nuisance parameters. These can be accessed via \texttt{params[`tau\_n\_fac']} (scaling of the input neutron lifetime from $\SI{879.4}{s}$) and \texttt{params[`nuclear\_rates\_q']} (a vector of scalings for each of the rates in the nuclear network).}
\end{enumerate}
See ref.~\cite{LINX_long} for more details about using LINX\@.  These latter two BBN options will override any user input \texttt{params[`YHe']}, and we do not recommend specifying a value for it if these options are used.

\subsection{Parameters and run options}\label{app:options}

Below is a complete list of initialization options, input parameters, and derived parameters that ship with ABCMB as of \texttt{v0.3.1}.  The can be specified individually, or packed into key/value pairs in a dictionary and unpacked during initialization of \texttt{Model}.

\subsubsection{Initialization options}
These options should be specified once as keyword arguments to \texttt{abcmb.main.Model} at initialization and used consistently for an entire analysis.  Changing these options will trigger recompilation. 
\\

\noindent
The following keyword arguments control input options:
\begin{itemize}
\item \texttt{use\_LCDM\_species}.  Boolean; whether ABCMB should use the default baryon, photon, massless neutrinos, CDM, and dark energy fluids in computation.  Setting this parameter to \texttt{True} instructs ABCMB to use all of these species, while setting it to \texttt{False} uses none of them.  Defaults to \texttt{True}. This option should be used if the user wishes to replace any of the standard species with a modified species, e.g. interacting dark matter over CDM. Note that ABCMB expects each cosmology to contain at least some version of the five $\lcdm$ fluids, and will fail if any are missing.
\item \texttt{input\_tau\_reion}.  Boolean; whether the optical depth to reionization $\tau_{\rm reion}$ is an input parameter, or if it is computed given the redshift to reionization $z_{\rm{reion}}$.  If set to \texttt{True}, the initialized \texttt{Model} should always be called with $\tau_{\rm reion}$ rather than $z_{\rm reion}$, or else a default $\tau_{\rm reion}$ value is used over the user input. Defaults to \texttt{True}.
\end{itemize}

\noindent
The following keyword arguments control output options:
\begin{itemize}
\item{\texttt{l\_min}.  The minimum $\ell$ at which to output CMB power spectra.  Defaults to 2.}
\item{\texttt{l\_max}. The maximum $\ell$ at which to output CMB power spectra.  Defaults to 2500.}
\item \texttt{k\_max}.  The maximum $k$ at which to output the matter power spectrum.  Defaults to $\SI{0.5}{Mpc^{-1}}$.
\item{\texttt{lensing}. Boolean; whether to compute corrections from lensing.  Defaults to \texttt{False}.}
\end{itemize}

\noindent
These keyword arguments control how ABCMB computes (or does not compute) Y$_{\rm{He}}$:
\begin{itemize}
\item{\texttt{bbn\_type}.  Specifies how to compute the primordial helium mass fraction.  Options are (any capitalization) \texttt{`Table'} to use precomputed yields, \texttt{`LINX'} to compute abundances given other input parameters using LINX, or \texttt{None} to use user input Y$_{\rm{He}}$.  Defaults to \texttt{None}. }
\item{\texttt{linx\_reaction\_net}.  If \texttt{bbn\_type} is set to use LINX, specify a BBN reaction network.  Options are \texttt{`np\_only'}, \texttt{`key\_PRIMAT\_2023'}, \texttt{`full\_PRIMAT\_2023'}, \texttt{`key\_PRIMAT\_2018'}, \texttt{`key\_YOF'}, or \texttt{`key\_PArthENoPE'}.  See ref.~\cite{LINX_long} for more information about each choice of reaction network.  Defaults to \texttt{`key\_PRIMAT\_2023'}.} 
\end{itemize}

\noindent
These keyword arguments control cutoffs for Boltzmann hierarchies:
\begin{itemize}
    \item \texttt{l\_max\_g}.  The number of moments $l$ to use in the Boltzmann hierarchy for photons for temperature perturbations.  Defaults to 12.
    \item \texttt{l\_max\_pol\_g}. The number of moments $l$ to use in the Boltzmann hierarchy for photons for polarization.  Defaults to 10.
    \item \texttt{l\_max\_massless\_nu}.  The number of moments $l$ to use in the Boltzmann hierarchy for massless neutrinos.  Defaults to 17.
    \item \texttt{l\_max\_massive\_nu}. The number of moments $l$ to use in the Boltzmann hierarchy for massive neutrinos.  Defaults to 17.
\end{itemize}

\noindent
The wavenumbers $k$ used by ABCMB to compute perturbations are unevenly spaced, in order to provide adequate resolution at the most critical regimes without becoming too memory intensive. We employ the same scheme as implemented in CLASS, reproducing similar formulae here so that we can indicate which parameters are controlled by user input. The $k$-grid is computed according to
\begin{align}
    k_{n+1} &= k_n + \Delta k(k_n)\cdot s(k_n) \nonumber\\
    \Delta k(k_n) &= \left(k_{\rm{step}}^{\rm{super}} + \frac12\tanh\left(\frac{k_n-k_{\rm{rec}}^{\rm{fid}}}{k_{\rm{rec}}^{\rm{fid}}k_{\rm{step}}^{\rm{transition}}}+1\right)\left(k_{\rm{step}}^{\rm{sub}}-k_{\rm{step}}^{\rm{super}}\right)\right)k_{\rm{rec}}^{\rm{fid}}\label{eq:deltak}\\
    s(k_n) &= \frac{\frac{k_n^2}{\left(H_0^{\rm{fid}}\right)^2}+1}{\left(\frac{k_n^2}{\left(H_0^{\rm{fid}}\right)^2}+\frac{1}{k_{\rm{step}}^{\rm{super,reduction}}}\right)}\, ,
    \label{eq:step_scale}
\end{align}
where 
\begin{equation*}
    k_{\rm{rec}}^{\rm{fid}} = \frac{2\pi}{r_s^{\rm{rec,fid}}}
\end{equation*}
and $k_{\rm{step}}^{\rm{sub}}$, $k_{\rm{step}}^{\rm{super}}$, $k_{\rm{step}}^{\rm{transition}}$, $k_{\rm{step}}^{\rm{super,reduction}}$, $r_s^{\rm{rec,fid}}$, and $H_0^{\rm{fid}}$ are user input.  In order to prevent recompilation each time the user provides new input parameters, the same $k$ grid must be used for each call of an initialized model.  Therefore the grid specification requires fiducial values for the wavenumber corresponding to the scale of the sound horizon at recombination $k_{\rm{rec}}^{\rm{fid}}$, the sound horizon itself at recombination $r_s^{\rm{rec,fid}}$, and the Hubble constant $H_0^{\rm{fid}}$, rather than updating the grid each time the user inputs a new value of e.g.\ $H_0$.  

The user also defines minimum and maximum $k$ via
\begin{align}
    k_{\rm{min}} &= \frac{k_{\rm{min}}' \tau_0}{\tau_0^{\rm{fid}}}\label{eq:kmin_grid}\\
    k_{\rm{max}} &=  \frac{k_{\rm{max}}'\tau_0}{\ell_{\rm{max}}\tau_0^{\rm{fid}} }\label{eq:kmax_grid}
\end{align}
where the combinations $k_{\rm{min}}'{\tau_0}$ and $\frac{k_{\rm{max}}'\tau_0}{\ell_{\rm{max}}}$ are input by the user.  The fiducial conformal time $\tau_0^{\rm{fid}}$ again takes on some fiducial value to prevent recompilation, which can be specified by the user.  Input arguments are defined as:
\begin{itemize}
    \item \texttt{k\_step\_sub}. $k_{\rm{step}}^{\rm{sub}}$ in eq.~\eqref{eq:deltak}.  Dimensionless parameter describing the number of periods of acoustic oscillation at decoupling (given by $\frac{2\pi}{r_s^{\rm{rec}}}$) by which to step in $k$ for modes inside the horizon at decoupling.  Defaults to $5\times10^{-2}$.
    \item \texttt{k\_step\_super}. $k_{\rm{step}}^{\rm{super}}$ in eq.~\eqref{eq:deltak}.  Dimensionless parameter describing the number of periods of acoustic oscillation at decoupling by which to step in $k$ for modes outside the horizon at decoupling. Defaults to $2\times10^{-3}$.
    \item \texttt{k\_step\_transition}. $k_{\rm{step}}^{\rm{transition}}$ in eq.~\eqref{eq:deltak}.  Modulates a smooth transition between \texttt{k\_step\_sub} and \texttt{k\_step\_super}.  Defaults to $2\times10^{-1}$.
    \item \texttt{k\_step\_super\_reduction}. $k_{\rm{step}}^{\rm{super,reduction}}$ in eq.~\eqref{eq:step_scale}.  For the very largest scales, \texttt{k\_step\_super} is reduced by this amount (as can be seen by taking a limit of eq.~\eqref{eq:step_scale}).  Defaults to $1\times10^{-1}$.
    \item \texttt{k\_min\_tau0}. $k_{\rm{min}}' *{\tau_0}$ in eq.~\eqref{eq:kmin_grid}; minimum $k$ scaled by an initial conformal time parameter $\tau_0$.  Defaults to $1\times10^{-1}$.  This sets the internal parameter \texttt{k\_min} to the right-hand side of eq.~\eqref{eq:kmin_grid}.
    \item \texttt{k\_max\_tau0\_over\_l\_max}. $\frac{k_{\rm{max}}'\tau_0}{\ell_{\rm{max}}}$ in eq.~\eqref{eq:kmax_grid}; maximum $k$ scaled by an initial conformal time parameter $\tau_0$ per maximum $\ell$.  Defaults to 1.8.  Sets the internal parameter \texttt{k\_max\_cmb} to the right-hand side of eq.~\eqref{eq:kmax_grid}.
    \item \texttt{H0\_fid}.  A fiducial value for the Hubble constant used to compute the $k$ grid, as shown in eq.~\eqref{eq:step_scale}.  Defaults to $2.255560\times10^{-4}\si{Mpc^{-1}}$.
    \item \texttt{rs\_rec\_fid}.  A fiducial value for the sound horizon at recombination, used to specify $k_{\rm{rec}}^{\rm{fid}}$.  Defaults to $\SI{144.6279}{Mpc}$.
    \item \texttt{tau0\_fid}.  A fiducial value for conformal time today, as in eqs.~\eqref{eq:kmin_grid} and~\eqref{eq:kmax_grid}.  Defaults to $1.418668\times 10^4 \si{Mpc}$.
\end{itemize}

\noindent
A separate $k$-grid, indicated below by $\tilde{k}$, is used for integrating transfer functions, in particular 
\begin{equation}
    \tilde{k}_{n+1} = \tilde{k}_n + \frac{\tilde{k}_{\rm{period}}\tilde{k}_{\rm{transfer}}^{\rm{linstep}}\tilde{k}_n}{\tilde{k}_n + \frac{\tilde{k}_{\rm{transfer}}^{\rm{linstep}}}{\tilde{k}_{\rm{transfer}}^{\rm{logstep}}}\left(\frac{1}{\si{Mpc}}\right)}\label{eq:k_transfer_spacing},
\end{equation}
where
\begin{equation}
    \tilde{k}_{\rm{period}}= \frac{2\pi}{\tau_0^{\rm{fid}} - \tau_{\rm{rec}}^{\rm{fid}}}.\label{eq:defkperiod}
\end{equation}
$\tau_0^{\rm{fid}} - \tau_{\rm{rec}}^{\rm{fid}}$ is the lookback time to recombination.  $\tau_0^{\rm{fid}}$ is the same parameter that appears in eqs.~\eqref{eq:kmin_grid} and~\eqref{eq:kmax_grid}.  $\tau_{\rm{rec}}^{\rm{fid}}$, $\tilde{k}_{\rm{transfer}}^{\rm{linstep}}$, and $\tilde{k}_{\rm{transfer}}^{\rm{logstep}}$ are user-defined.  This parametrization ensures logarithmic growth for small $k$ and linear growth for large $k$.  The grid is computed between \texttt{k\_min} and \texttt{k\_max\_cmb}.  The input arguments are
\begin{itemize}
    \item \texttt{k\_transfer\_linstep}. $\tilde{k}_{\rm{transfer}}^{\rm{linstep}}$ in eq.~\eqref{eq:k_transfer_spacing}; the size of the step in $\tilde{k}$ deep in the linear regime, for large $\tilde{k}$.  Defaults to $4.5\times10^{-1}$.
    \item \texttt{k\_transfer\_logstep}. $\tilde{k}_{\rm{transfer}}^{\rm{logstep}}$ in eq.~\eqref{eq:k_transfer_spacing}; $\log\left(\tilde{k}_{\rm{transfer}}^{\rm{logstep}}\right)$ is the size of the step in $\log\tilde{k}$ deep in the logarithmic regime, for small $\tilde{k}$.  Defaults to 170.
    \item \texttt{tau\_rec\_fid}.  $\tau_{\rm{rec}}^{\rm{fid}}$ in eq.~\eqref{eq:defkperiod}.  Defaults to \SI{281.040565}{Mpc}.
\end{itemize}
As discussed above, our construction of these grids differs from the approach of CLASS by the inclusion of fiducial cosmological parameters.  This is to ensure fixed array sizes at compile time and prevent recompilation.  These static fiducial parameters are sufficient to reproduce the CLASS result within $\pm5\sigma$ of the \textit{Planck} best-fit cosmological parameters (see section~\ref{sec:performance}).  If the user is interested in parameter values that differ from these fiducial values by $\mathcal{O}(1)$, then new $r_s^{\rm{rec,fid}}$, $H_0^{\rm{fid}}$, $\tau_0^{\rm{fid}}$, or $\tau_{\rm{rec}}^{\rm{fid}}$ may need to be computed and input by the user at initialization.

Otherwise, the keyword arguments listed above that control the perturbations and transfer $k$ grids have no impact on the underlying physics. They are merely defined such that various integrals over $k$ can be done efficiently while still retaining as much precision as possible. The scheme in eqs.~\eqref{eq:deltak}-\eqref{eq:defkperiod} that we adapt from CLASS is optimized for precisely this purpose, and we encourage the user not to modify them without a clear reason. 
\\

\noindent
The following keyword argument sets the characteristics of the primordial power spectrum:
\begin{itemize}
    \item \texttt{k\_pivot}. The pivot scale; see eq.~\eqref{eq:PK_pivot}.  Defaults to \SI{0.05}{Mpc^{-1}}.
\end{itemize}

\noindent
We adopt the CLASS strategy for determining the initial scale factor $a_i(k)$ at which to begin integrating perturbations with wavenumber $k$.  $a_i(k)$ has to satisfy two conditions. First, it must be early enough such that the baryons and photons are tightly coupled, i.e. $a_i$ has to be sufficiently small such that the ratio
\begin{equation}
    R_{\rm tc} = \HH(a_i) \tau_c(a_i)\label{eq:k_start_small} \ll 1,
\end{equation}
where $\tau_c = (a n_e \sigma_T)^{-1}$ is the Compton scattering time, with $\sigma_T$ the Thomson cross section. This relationship is $k$-independent, and so sets a maximum value for $a_i$ for all modes.

Second, for perturbations on large scales, we need to ensure that integration begins while the mode is superhorizon, i.e.\ for each $k$, we must choose $a_i$ such that the ratio
\begin{equation}
    R_{\rm large} =\frac{k}{\HH(a_i)} \ll 1\label{eq:k_start_large},
\end{equation}
which sets a value of $a_i$ for each $k$-mode. 

Both $R_{\rm tc}$ and $R_{\rm large}$ are constants much smaller than 1 that can be specified as user inputs in order to guarantee an appropriate choice of $a_i$.  Given these two variables, ABCMB solves for the appropriate $a_i$ for each $k$, i.e.\ $a_i$ such that $R_{\rm tc} = \mathcal{H}(a_i) \tau_c(a_i)$ and $R_{\rm large} = k / \mathcal{H}(a_i)$, and chooses the smaller value of $a_i$ as the initial scale factor to start the integration at.


In the code, these initial conditions are set by the following two keyword arguments:
\begin{itemize}
    \item \texttt{R\_tc}. $R_{\rm{tc}}$ in eq.~\eqref{eq:k_start_small}.  For small $k$ modes this condition will be satisfied first and will provide the initial condition in $\ln a$ for integration.  Defaults to $1.5\times10^{-3}$. 
    \item \texttt{R\_large}. $R_{\rm{large}}$ in eq.\eqref{eq:k_start_large}. For large $k$ modes this condition will be satisfied first and will provide the initial condition in $\ln a$ for integration. Defaults to $7\times10^{-2}$.
\end{itemize}

\noindent
These inputs determine settings for the differential equation solver used for perturbations:
\begin{itemize}
    \item \texttt{max\_steps\_PE}.  Maximum number of steps used to solve the differential equation describing perturbations for a single $k$ mode.  Defaults to 2048.
    \item \texttt{k\_split\_PE}.  Value of $k$ at which to shift tolerances in solving the differential equation  differential equation describing perturbations for a single $k$ mode (see below).  Defaults to $1\times10^{-2}\ {\rm Mpc}^{-1}$.
    \item \texttt{rtol\_small\_k\_PE}.  The relative tolerance used by the perturbations differential equation solver below \texttt{k\_split\_PE}.  Defaults to $1\times10^{-5}$.
    \item \texttt{rtol\_large\_k\_PE}.  The relative tolerance used by the perturbations differential equation solver above \texttt{k\_split\_PE}.  Defaults to $1\times10^{-4}$.
    \item \texttt{atol\_small\_k\_PE}.  The absolute tolerance used by the perturbations differential equation solver below \texttt{k\_split\_PE}.  Defaults to $1\times10^{-10}$.
    \item \texttt{atol\_large\_k\_PE}.  The absolute tolerance used by the perturbations differential equation solver above \texttt{k\_split\_PE}.  Defaults to $1\times10^{-6}$.
    \item \texttt{pcoeff\_PE}.  The \texttt{pcoeff} value used in the adaptive stepsize controller for the differential equation solved for perturbations; see diffrax~\cite{kidger2021on} documentation for more information.  Defaults to 0.25.
    \item \texttt{icoeff\_PE}.  The \texttt{icoeff} value used in the adaptive stepsize controller for the differential equation solved for perturbations; see diffrax documentation for more information.  Defaults to 0.8.
    \item \texttt{dcoeff\_PE}.  The \texttt{dcoeff} value used in the adaptive stepsize controller for the differential equation solved for perturbations; see diffrax documentation for more information.  Defaults to 0.
\end{itemize}

\noindent
These input parameters control the overall scaling of the different physical contributions to the CMB temperature transfer function; they can be adjusted to highlight the various contributions for pedagogical purposes:
\begin{itemize}
    \item \texttt{scale\_sw}.  Scaling of the Sachs-Wolfe contribution to the CMB temperature transfer function; 0 shuts this contribution off.  Defaults to 1.
    \item \texttt{scale\_isw}.  Scaling of the integrated Sachs-Wolfe contribution to the CMB temperature transfer function; 0 shuts this contribution off.  Defaults to 1.
    \item \texttt{scale\_dop}.  Scaling of the Doppler contribution to the CMB temperature transfer function; 0 shuts this contribution off.  Defaults to 1.
    \item \texttt{scale\_pol}.  Scaling of the polarization contribution to the CMB temperature transfer function; 0 shuts this contribution off.  Defaults to 1.
\end{itemize}

\noindent
Finally, this input argument controls the method used for autodifferentiation, should the user request a gradient:
\begin{itemize}
    \item \texttt{adjoint}.  The method to use for computing adjoints through \texttt{diffrax} differential equations.  Defaults to \texttt{diffrax.ForwardMode}, and \texttt{diffrax.RecursiveCheckpointAdjoint} is recommended for reverse AD.
\end{itemize}

\subsubsection{Cosmological parameters}
The following input parameters can be specified by the user after initialization, at runtime, through a dictionary passed to the initialized \texttt{Model}.  If unspecified, default values will be used.
\begin{itemize}
    \item{\texttt{params[`omega\_b']}.  Total energy density in baryons, compared to the critical matter density and multiplied by $h^2$.  Defaults to 0.02237.}
    \item{\texttt{params[`omega\_cdm']}.  Total energy density in CDM, compared to the critical matter density and multiplied by $h^2$.  Defaults to 0.120.}
    \item{\texttt{params[`Neff']}.  As defined in eq.~\eqref{eq:Neffdef}.  This parameter cannot be specified simultaneously with \texttt{params[`N\_nu\_massless']}; see section~\ref{sec:neutrinos}.  This parameter cannot be specified if LINX options are used for BBN\@.  Defaults to 3.044.}
    \item{\texttt{params[`h']}.  Hubble parameter.  Defaults to 0.6736.}
    \item{\texttt{params[`A\_s']}. Amplitude of fluctuations.  Defaults to $2.1\times10^{-9}$.}
    \item{\texttt{params[`n\_s']}. Scalar spectrum power law index.  Defaults to 0.9649.}
    \item{\texttt{params[`YHe']}. Primordial helium mass fraction, $\frac{\rho_{\rm{He}}}{\rho_{\rm{H}} + \rho_{\rm{He}}}$. Defaults to 0.245 if \texttt{Model} is not initialized with \texttt{bbn\_type}.}
    \item{\texttt{params[`TCMB0']}. CMB temperature today, in eV\@.  Defaults to $2.34865418\times10^{-4}$.}
    \item{\texttt{params[`N\_nu\_massless']}.  Massless neutrino energy density as given by eq.~\eqref{eq:rho_massless_def}; \texttt{params[`N\_nu\_massless']} is $N_{\nu,\rm{massless}}$.  This parameter cannot be specified simultaneously with \texttt{params[`Neff']}, and is adjusted internally according to section~\ref{sec:neutrinos} depending on input value of \texttt{params[`Neff']}, if \texttt{params[`Neff']} is specified.}
    \item{\texttt{params[`T\_nu\_massless']}.  The \textit{ratio} of the massless neutrino temperature ($T_{\nu,\rm{massless}}$ in eq.~\eqref{eq:rho_massless_def}) to the photon temperature, at late times.  Defaults to $0.71636856$ to capture effects from non-instantaneous neutrino decoupling.}
    \item{\texttt{params[`N\_nu\_massive']}.  Number of massive neutrino species.  Defaults to 0.}
    \item{\texttt{params[`m\_nu\_massive']}.  Mass of massive neutrino species, in eV\@.  Defaults to 0.06 (but not used if \texttt{params[`N\_nu\_massive']} is 0).  All massive neutrinos have the same mass.}
    \item{\texttt{params[`T\_nu\_massive']}. The temperature $\textit{ratio}$ of massive neutrinos to the CMB temperature, at late times. Defaults to $0.71611$, as per ref.~\cite{Mangano_2005}(see section~\ref{sec:neutrinos}).}
    
    \item {\texttt{params[`tau\_reion']}}. Optical depth to reionization, defined as
    \begin{align}
        \tau_{\rm reion} &= \int d\ln{a} \frac{a \sigma_T n_Hx_e^{\rm reion}}{\HH} \, ,
        \label{eq:tau_reion}
    \end{align}
    where $x_e^{\rm reion}$ denotes the free electron fraction coming purely from reionization\footnote{The tiny amount contributed by the free electrons after recombination is neglected in this integral. This is a good assumption in $\lcdm$, but new physics such as decaying dark matter can add an early reionization component. We leave the generalization to this class of models to future work.}, as given by equations~\ref{eq:h_reion} and \ref{eq:he_reion}. Defaults to 0.05430842. This integrated quantity has a one-to-one correspondence with the parameter $z_{\rm reion}$, the redshift of hydrogen and HeI reionization, once the remaining four reionization parameters are fixed. If $\tau_{\rm reion}$ is passed in, the appropriate value for $z_{\rm reion}$ is found with eq.~\eqref{eq:tau_reion} using a root finder. For this reason in ABCMB only one of $\tau_{\rm reion}$ or $z_{\rm reion}$ can be specified, and to switch between these options one should set the keyword argument \texttt{input\_tau\_reion} to \texttt{True} or \texttt{False} appropriately. 
    \item{\texttt{params[`z\_reion']}. Redshift of hydrogen and HeI reionization (see eq.~\eqref{eq:h_reion}).  Defaults to 7.67.  Set to 0 to turn off this reionization. If specified, ABCMB will compute the integral~\eqref{eq:tau_reion} to find the optical depth to reionization automatically. As discussed above, only one of $\tau_{\rm reion}$ or $z_{\rm reion}$ can be specified, and to switch between these options one should set the keyword argument \texttt{input\_tau\_reion} to \texttt{True} or \texttt{False} appropriately. }
    \item{\texttt{params[`Delta\_z\_reion']}. Width of the $\tanh$ step for hydrogen and HeI reionization (see eq.~\eqref{eq:h_reion}).  Defaults to 0.5.  Set to 0 to turn off this reionization.}
    \item{\texttt{params[`z\_reion\_He']}. Redshift of HeII reionization (see eq.~\eqref{eq:he_reion}).  Defaults to 3.5.  Set to 0 to turn off this reionization.}
    \item{\texttt{params[`Delta\_z\_reion\_He']}. Width of the $\tanh$ step for HeII reionization (see eq.~\eqref{eq:he_reion}).  Defaults to 0.5. Set to 0 to turn off this reionization.}
    \item{\texttt{params[`exp\_reion']}. Exponent $\beta_{\rm{reion}}$ that appears in the $\tanh$ parametrization for reionization (see eq.~\eqref{eq:y_reio}).  Defaults to 1.5.}
    \item{\texttt{params[`nuclear\_rates\_q']}.  Optional; used only if \texttt{bbn\_type=`LINX'}.  An array of scalings of nuclear rates in the LINX BBN reaction network.  This array should include one value $q_i$ for each reaction in the network (12 reactions if a ``key'' network is used).  Each $q_i$ which scales the median values for its corresponding reaction rates $\overline{r}_i$ according to $\log r_i (T) = \log \overline{r}_i (T) + q_i \sigma_i (T)$, where $r_i (T)$ is the rate that LINX will use for the specified reaction.  $q_i$ is a unit Gaussian random variable.}
    \item{\texttt{params[`tau\_n\_fac']}. Optional; used only if \texttt{bbn\_type=`LINX'}.  This input parameter is a multiplicative scaling of the fiducial neutron lifetime $\SI{879.4}{s}$.}
\end{itemize}

\noindent
Finally, given the set of user input parameters, the following parameters are derived at runtime.  User inputs will be overwritten, and therefore should not be input.
\begin{itemize}
    \item{\texttt{params[`omega\_m']}.  The total matter density compared to the critical matter density and multiplied by $h^2$.  This is computed via summing over all fluid energy densities whose flag \texttt{is\_matter=True}.}
    \item{\texttt{params[`R\_b']}.  The baryon fraction.  Given by \texttt{params[`omega\_b']} / \texttt{params[`omega\_m']}.}
    \item{\texttt{params[`H0']}.  Value of the Hubble constant today in \SI{}{km/s/Mpc}.  Given by \texttt{params[`h']}$\times \SI{100}{km/s/Mpc}$.}
    \item{\texttt{params[`omega\_r']}.  Total energy density in radiation, compared to the critical matter density and multiplied by $h^2$.  This is calculated as $\frac{8\pi G\rho_R a_i^4 h^2}{3  H_0^2}$, where the total radiation energy density $\rho_R$ is computed by ABCMB given all input fluids and their energy densities at some very small scale factor $a_i$, and $G$ is Newton's constant. This assumes radiation domination at $a_i$.}
    \item{\texttt{params[`R\_nu']}. The fraction of radiation in neutrinos.  Given by $\rho_\nu/\rho_R$, where $\rho_R$ is computed as described above and $\rho_\nu$ is similarly computed at some very small scale factor $a_i$.  This parameter determines adiabatic initial conditions for perturbations; while modes inside the horizon during $e^+e^-$ annihilation should in principle be described by a different \texttt{R\_nu}, the effects of neglecting this correction do not affect modes we will measure in the foreseeable future.}
    \item{\texttt{params[`om']}}. Defined as $\omega\equiv \frac{\Omega_m}{\sqrt{\Omega_r}}H_0$, important in determining the conformal time and the scale factor at early times, and used in setting adiabatic initial conditions for various fluids. See CAMB notes\footnote{\href{https://cosmologist.info/notes/CAMB.pdf}{https://cosmologist.info/notes/CAMB.pdf}} for more details.
    \item{\texttt{params[`omega\_Lambda']}. Total energy density in dark energy, compared to the critical matter density and multiplied by $h^2$.  Inferred from other densities such that it is given by $\texttt{params[`h']}^2 - \texttt{params[`omega\_r']} - \texttt{params[`omega\_m']}$.}   
\end{itemize}

\section{Physics implementation in ABCMB}\label{app:physics}
In this appendix, we detail the physics implemented in the various ABCMB modules.  We begin with a detailed overview of the companion code HyRex, used for computing the free electron fraction and matter temperature during recombination.  We then move on to the calculation of cosmological perturbations and spectra.

\subsection{Recombination}\label{sec:HyRex}
When the user calls ABCMB, the code first computes the free electron fraction $x_e$ and the matter temperature $T_m$ as a function of scale factor $a$ in the HyRex module.  These are used as background quantities in the evolution of the perturbation equations used to compute the CMB power spectrum.

We follow the implementation in HYREC-2, whose accuracy exceeds that of RECFAST~\cite{RECFAST} and rivals that of the earlier code HYREC~\cite{HYREC}, while running in a fraction of HYREC's run time~\cite{HYREC2}.   HYREC-2 reduces computational overhead by revising the Effective Multi-Level Atom (EMLA)~\cite{Ali_Haimoud_2010} down to an effective 4-level atom, where corrections to the Lyman-$\alpha$ escape rate are precomputed from HYREC\@.  These corrections have predictable scalings with cosmological parameters and therefore do not need to be recomputed for new sets of parameters.  We follow this procedure as well, rewriting HYREC-2 in a differentiable and user-friendly JAX implementation, and also including a simple reionization model.  This calculation separates helium recombination from hydrogen recombination (as is done in HYREC-2), which are further separated into various thermodynamic phases.

Throughout this section, fractional abundances $x_i$ are always the number density of the subscripted state over the number density of all hydrogen states, $x_i\equiv n_i/n_H$.  Calculations in HyRex use units \si{eV} for energies and temperatures, \si{eV/cm^3} for energy densities, and \si{s} for time. Hubble is computed in \si{s^{-1}}.  

\subsubsection{Helium recombination}
We follow the treatment of helium recombination in HYREC \cite{HYREC}, which occurs in two phases: first, doubly-ionized helium HeIII recombines to singly-ionized helium HeII, and then, at a later time, that singly-ionized helium recombines to neutral helium HeI\@.

HeII recombination is fast compared to the expansion rate, so we begin by tracking the fractions of doubly- and singly-ionized helium in Saha equilibrium, $x_{\rm{HeIII}}$ and $x_{\rm{HeII}}$, respectively: 
\begin{align*}
\frac{x_{\rm{HeIII}}\left(1+f_{\rm{He}}+x_{\rm{HeIII}}\right)}{f_{\rm{He}}-x_{\rm{HeIII}}}&=\frac{(2\pi \mu_{e/\textrm{HeIII}} T)^{3/2}}{h_P^3 n_{\rm{H}}} e^{-E_{I_2}/T},\\
x_{\rm{HeII}}&=f_{\rm{He}}-x_{\rm{HeIII}}\\
x_e&=1+f_{\rm{He}}+x_{\rm{HeIII}}
\end{align*}
where $E_{I_2}$ is the second ionization energy of helium, $T$ is the radiation temperature, $\mu_{e/\textrm{HeIII}}$ is the reduced mass of the electron-HeIII system, $h_P$ is Planck's constant (to be disambiguated from the dimensionless equivalent of the Hubble constant $h$), and $n_{\rm{H}}$ is the hydrogen number density.  $f_{\rm{He}}$ is the number abundance of helium nuclei relative to hydrogen.  Solving the first of these equations for $x_{\rm{HeIII}}$ provides all of the information needed to track $x_e$ in this epoch, and so each of these fractions is tracked from when integration starts ($z=8000$ by default) until most HeIII has recombined to HeII ($x_{\rm{HeIII}} < 10^{-9}$ by default).

After crossing this threshold, we switch to a post-Saha equilibrium expansion to describe fast recombination to HeI, given by
\begin{align}
\frac{x_{\rm{HeII}}(1+x_{\rm{HeII}})}{f_{\rm{He}}-x_{\rm{HeII}}}& = 4 \frac{(2\pi \mu_{e/\textrm{HeII}} T)^{3/2}}{h_P^3 n_{\rm{H}}} e^{-E_{I_1}/T}\nonumber\\
x_e^{\rm{Saha}}&=1+x_{\rm{HeII}}\nonumber\\
x_e &= x_e^{\rm{Saha}} + \Delta x_e\label{eq:PostSahaHe},
\end{align}
where $E_{I_1}$ is the first ionization energy of helium and $\mu_{e/\textrm{HeII}}$ is the reduced mass of the electron-HeII system.  $x_e^{\rm{Saha}}$ is the Saha equilibrium value for $x_e$, which differs from the true value of $x_e$ by a small amount $\Delta x_e = \frac{d x_e^{\rm{Saha}}}{dt}/\frac{\partial{\dot{x}_e}}{\partial x_e}|_{x_e^{\rm{Saha}}}$.  

A large $\Delta x_e$ ($10^{-5}$ by default, which is more precise than the default HYREC-2 threshold $5\times10^{-4}$) indicates a significant departure from Saha equilibrium.  This suggests slower recombination of helium and requires solving for recombination of $x_{\rm{HeII}}$ to neutral $x_{\rm{HeI}}$ in full detail, including effects from radiative transfer to hydrogen.  This involves the construction and solution of an ODE, which is provided in ref.~\cite{HYREC}.  We solve this ODE with diffrax~\cite{kidger2021on}, terminating when $x_{\rm{HeII}}$ drops below $10^{-4}$.

This calculation describes the evolution of $x_e$ down to a redshift of about 1700 (with the exact value depending on the user's specified cosmology).  This history is returned to the main HyRex module, and its stopping redshift is used as an initial condition for hydrogen recombination.  We assume these epochs do not overlap---this is a reasonable assumption, given that hydrogen recombination tends not to begin in earnest until redshift $z\sim1600$, as verified in ref.~\cite{HYREC}.

\begin{figure}[t]
    \centering
    \includegraphics[width=0.7\linewidth]{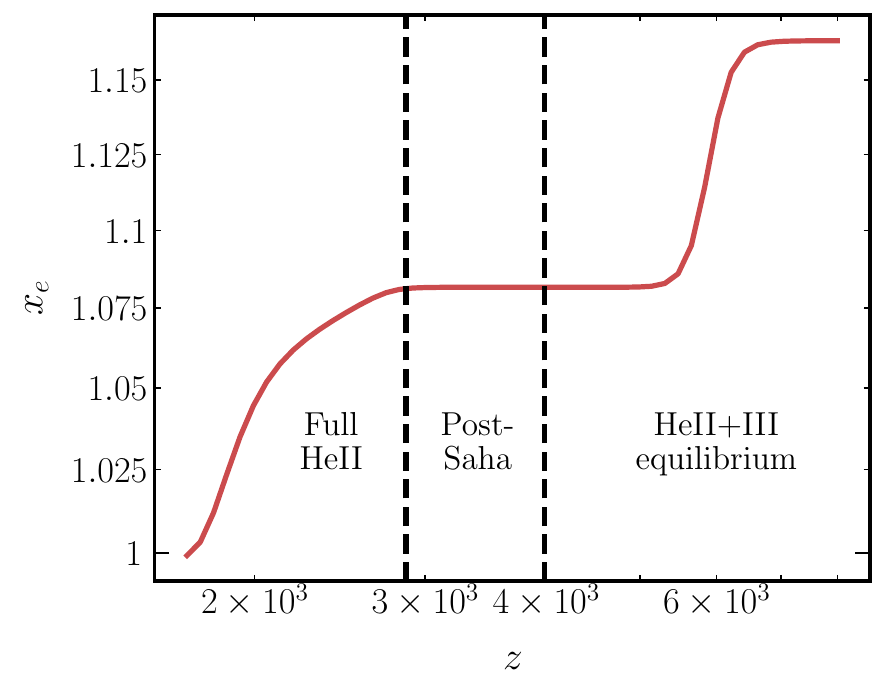}
    \caption{The free electron fraction $x_e$ as a function of redshift $z$ during helium recombination only, as computed by HyRex. The relevant cosmological parameters chosen are $\textrm{Y}_\textrm{He}=0.245$, $N_{\rm{eff}}=3.044$, $h=0.7$, $\Omega_bh^2=0.0224$, and $\Omega_{\rm{CDM}}h^2=0.12$, and the dashed lines separate different solving regimes as described in the main text.}
    \label{fig:helium_recomb}
\end{figure}

This computation is summarized in figure~\ref{fig:helium_recomb}, where we delineate the three regimes described above.

\subsubsection{Hydrogen recombination}
The calculation of hydrogen recombination requires a good deal more detail and care to obtain the precision required by the CMB, as the CMB probes hydrogen recombination directly (helium recombination primarily affects the Silk damping scale~\cite{Hu_1995}).
Hydrogen that recombines directly to the ground state emits high-energy photons that quickly lead to another ionization event, leading to no net change in ionization. Therefore, recombination to excited states dominates (``case B'' recombination).  The escape rate of Lyman emissions from hydrogen recombining to excited states and de-exciting therefore must be modeled carefully, and transition rates from other excited states must also be taken into account.  Meanwhile, the hydrogen $2s$ state is metastable, requiring emission of two photons to decay to the ground state, further complicating the set of equations that must be solved.  Detailed discussions of these phenomena are available in e.g.\ Refs.~\cite{Peebles_1968, Zeldovich_1968, Hirata_2009, Ali-Haimoud:2010puu,Grin_2010}. 

As in HyRec \cite{HYREC}, the hydrogen recombination is broken into four parts: a post-Saha equilibrium phase, an EMLA + two-photon processes phase, an EMLA phase where two-photon $2s\rightarrow1s$ decays are no longer active, and a phase described by the simplified three-level atom (TLA) model.

At early times, during helium recombination, hydrogen remains in Saha equilibrium.  Though hydrogen departs from Saha equilibrium while helium is still recombining, the Saha-equilibrium free electron fraction still makes for a reasonable initial condition to begin integrating the post-Saha equilibrium phase, incurring errors only as large as about one part in $10^{4}$, as verified by ref.~\cite{HYREC}.  Our initial condition is therefore
\begin{equation}
x_e^{\rm{Saha}}=\frac{\sqrt{s^2+4s}-s}{2}\label{eq:SahaH},
\end{equation}
where
\begin{equation}
s=\frac{(2\pi \mu_{e/\textrm{H}} T)^{3/2}}{h_P^3 n_{\rm{H}}} e^{-E_{R}/T}\nonumber,
\end{equation}
$E_{R}$ is the Rydberg constant, and  $\mu_{e/\textrm{H}}$ is the reduced mass for the hydrogen/electron system.  We then compute $x_e$ as given by eq.~\eqref{eq:PostSahaHe} (with  $x_e^{\rm{Saha}}$ given by eq.~\eqref{eq:SahaH}).  Note two-photon processes are already relevant at this stage, and so the derivatives defining $\Delta x_e$ now depend on the effective recombination coefficients, with $\dot{x}_e$ given by \cite{HYREC2}
\begin{align}
    \dot{x}_e &= -\sum_{\ell=s,p} C_{2\ell}\left(n_{\rm H} x_e^2 \mathcal{A}_{2\ell} - g_{2\ell}x_{1s} e^{-E_{21}/T_\gamma}\mathcal{B}_{2\ell}\right)\, ,
    \label{eq:xHIIdot}
\end{align}
where $\mathcal{A}_{2\ell}$, $\mathcal{B}_{2\ell}$ are the effective recombination and effective photo-ionization coefficients, respectively, $C_{2\ell}$ is the generalized Peebles-C factor, and $g_{2\ell}$ counts the degrees of freedom of the $2s$ and $2p$ hydrogen states, which are respectively 1 and 3. In HYREC-2, effects of two-photon processes are fully captured in $C_{2\ell}$.

\begin{figure}[t]
    \centering
    \includegraphics[width=0.9\linewidth]{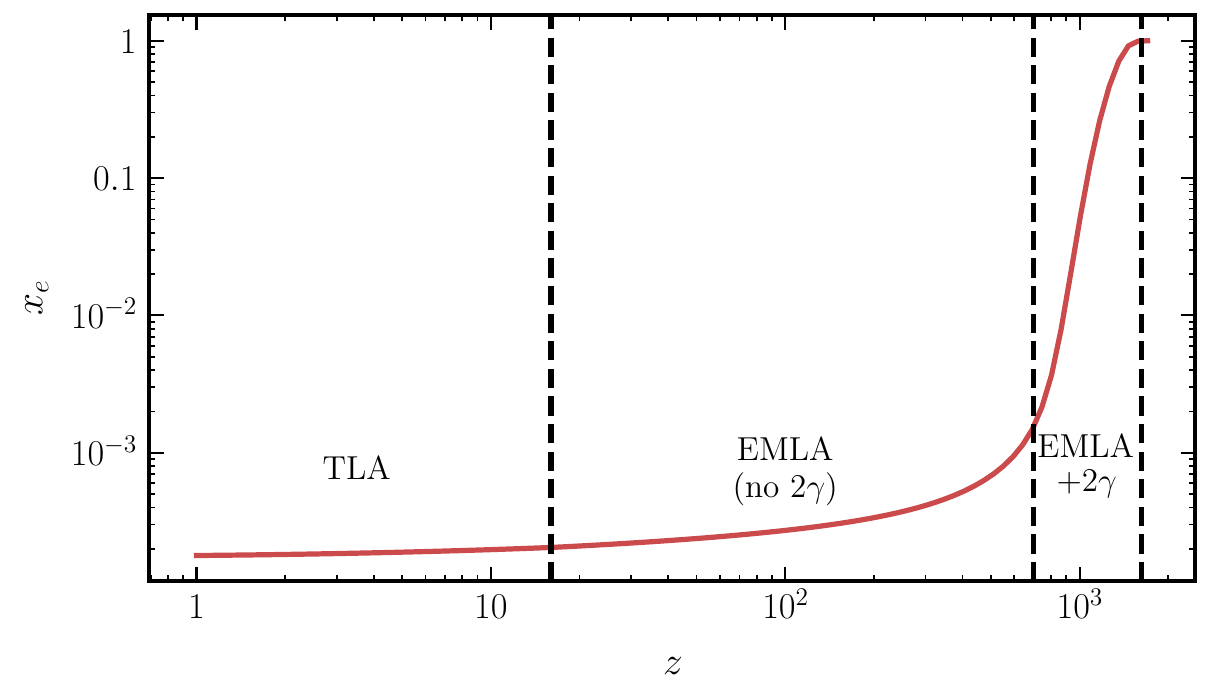}
    \caption{The free electron fraction $x_e$ as a function of redshift $z$ after helium recombination and before reionization, as computed by HyRex. The relevant cosmological parameters chosen are $\textrm{Y}_\textrm{He}=0.245$, $N_{\rm{eff}}=3.044$, $h=0.7$, $\Omega_bh^2=0.0224$, and $\Omega_{\rm{CDM}}h^2=0.12$.  Dashed lines separate Post-Saha (highest $z$), the EMLA + two-photon phase, the EMLA phase after two-photon processes shut off, and the final TLA phase at low redshifts.}
    \label{fig:hydrogen_recomb}
\end{figure}

The post-Saha phase is short, as shown by the rightmost segment of figure~\ref{fig:hydrogen_recomb}.  However, it is critical for numerical stability; the EMLA phases are stiff, and so a brief post-Saha phase is required to move into a more stable region.

We move to the EMLA phases once $\Delta x_e$ exceeds $10^{-5}$ by default.  EMLA is again described by an ODE provided in ref.~\cite{Ali_Haimoud_2010}, though its form is dramatically simplified in HYREC-2 due to the use of precomputed effective recombination coefficients.  We solve eq.~\eqref{eq:xHIIdot}, the EMLA including two-photon processes down to redshift $z=700$, after which two-photon processes no longer contribute~\cite{HYREC}.   
During this epoch the matter temperature $T_m$ departs from the CMB temperature $T_{\rm{CMB}}$ as the baryons and photons decouple, and the matter temperature is tracked via
\begin{equation}
    T_m = T_{\rm{CMB}} \left(1-\frac{H}{\Gamma_C}\right),
\end{equation}
where $\Gamma_C$ is the Compton scattering rate.  With $\sigma$ the Stefan-Boltzmann constant, $\sigma_{\rm{Thomson}}$ the Thomson scattering cross section, $m_e$ the electron mass, and $f_{\rm{He}}$ the helium number fraction, $\Gamma_C$ is given by $\frac{x_e}{1+f_{\rm{He}}+x_e} \frac{ 32\sigma T_{\rm{CMB}}^4 \sigma_{\rm{Thomson}}}{3m_e}$.

After a redshift of 700 we finally switch a simplified ODE for the EMLA where two-photon rates are set to zero.  The matter temperature in this regime evolves according to 
\begin{equation}
\frac{dT_m}{d\ln a} = \frac{-2 H T_m + \Gamma_C (T_{\rm{CMB}} - T_m)}{H}.
\end{equation}

The effective recombination coefficients in HYREC-2 are only tabulated down to a redshift of about 15.  Since $x_e$ is much less than 1 in this regime and reionization processes dominate, we follow the procedure in HYREC-2 and describe recombination with a hydrogen TLA~\cite{Peebles_1968,Zeldovich_1968}.  

This computation is summarized in figure~\ref{fig:hydrogen_recomb}, with each of the epochs described above separated by dashed lines.

After each of these elements is computed, the resulting $x_e$, $T_m$ and log scale factor $\ln a$ are returned back up to the main HyRex module.  They can be used as standalone outputs for analyses sensitive to recombination, or passed up to the ABCMB Einstein-Boltzmann solver and used to compute the matter and CMB power spectra.  

\subsection{Reionization}\label{app:reion}

Once HyRex computes the recombination history, ABCMB patches a reionization correction on to the resulting $x_e$.  As in ref.~\cite{Lewis_2008}, and as implemented in CAMB and as an option in CLASS, we use a simple $\tanh$ model for hydrogen and HeI reionization,
\begin{equation}
x_e^{\rm{reion}}=\frac{1+f_{\rm{He}}}{2}\left(1+\tanh\left(\frac{y_{\rm{reion}} - y}{\Delta y_{\rm{reion}}}\right)\right)\label{eq:h_reion},
\end{equation}
where
\begin{align}
y &= (1+z)^{\beta_{\rm reion}}\nonumber\\
y_{\rm{reion}} &= (1+z_{\rm{reion}})^{\beta_{\rm reion}}\nonumber\\
\Delta y_{\rm{reion}} &= \beta_{\rm reion} \Delta z_{\rm{reion}}\left(1+z_{\rm{reion}}\right)^{\beta_{\rm reion}-1},\label{eq:y_reio}
\end{align}
to the end of the $x_e$ history.  The redshift of reionization $z_{\rm{reion}}$, reionization width $\Delta z_{\rm{reion}}$ and the exponent $\beta_{\rm reion}$ (usually $3/2$) can be specified by the user.

We include a second reionization step for reionization of HeII to HeIII:
\begin{equation}
    x_e^{\rm{HeII\,reion}} = \frac{f_{\rm{He}}}{2} \left(1 + \tanh\left(\frac{z_{\rm{HeII\,reion}} - z}{\Delta z_{\rm{HeII\,reion}}}\right)\right)\label{eq:he_reion},
\end{equation}
where the redshift $z_{\rm{HeII\,reion}}$ and width $\Delta z_{\rm{HeII\, reion}}$ are user input; they are independent from the hydrogen reionization parameters.

We do not track the matter temperature during reionization.  While we might track the small effect on the matter temperature from Compton scattering of electrons with CMB photons, detailed simulations are required to model the effects of photoheating from high-energy free electrons, which is the dominant effect (see e.g.\ refs.~\cite{Wu_2019,Villasenor_2021}).  Since these effects are not relevant for accurate CMB calculations, we eschew them entirely.

\begin{figure}[t]
    \centering
    \includegraphics[width=0.7\linewidth]{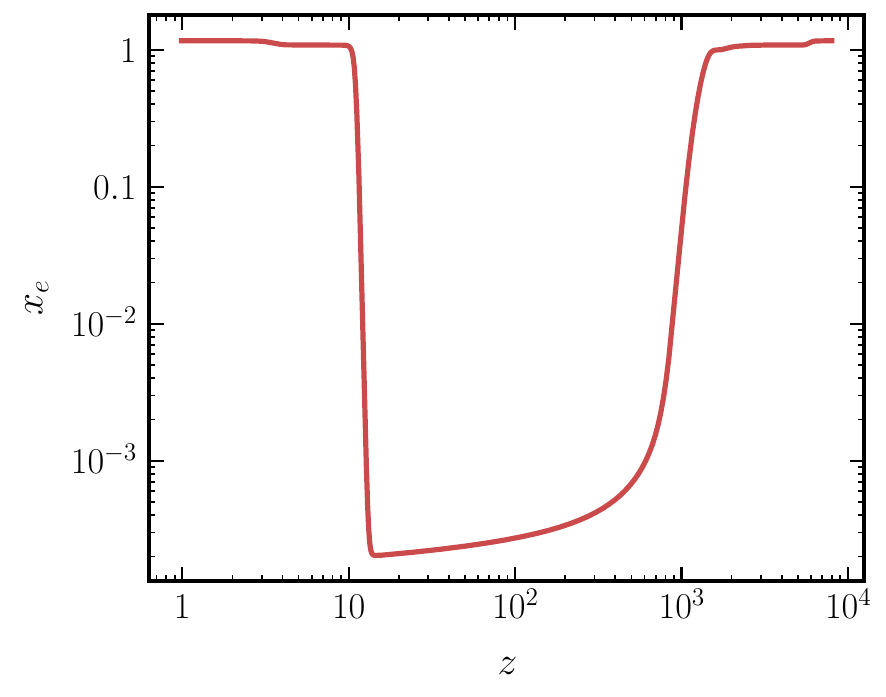}
    \caption{The full $x_e$ history as computed by HyRex, including helium recombination, hydrogen recombination, and reionization.  The relevant input parameters chosen to produce this result are $\textrm{Y}_\textrm{He}=0.245$, $N_{\rm{eff}}=3.044$, $h=0.7$, $\Omega_bh^2=0.0224$, $\Omega_{\rm{CDM}}h^2=0.12$, $z_{\rm{reio}}=11$, and $\Delta z_{\rm{reio}}=0.5$.}
    \label{fig:full_xe}
\end{figure}

The full $x_e$ output of HyRex, including the reionization model appended to the solution by ABCMB, is shown in figure~\ref{fig:full_xe}.

\subsection{Perturbations}\label{sec:perturbations}

After obtaining the electron fraction and matter temperature histories described in the previous section during the initialization of a \texttt{Background} object, ABCMB turns to the \texttt{Perturbations} module. Its purpose is to concurrently evolve the linear perturbations for each fluid, starting from the initial conditions specified by their respective fluid modules. At the end of its computation, the evolution histories of every fluid is packaged into a table over a grid of time and wavenumbers, and forwarded to the spectrum module for computing power spectra.

This section describes the physical and computational approach in general; specific perturbations equations and other fluid-specific quantities are provided in more detail in appendix~\ref{app:species}.  Units used by ABCMB to implement what follows are \si{eV} for energy, \si{eV}/\si{cm^3} for energy density, \si{Mpc} for time and length, and \si{Mpc^{-1}}for conformal Hubble.

\subsubsection{Linear perturbations}\label{sec:linear_perturbations}
ABCMB works entirely in the synchronous gauge and begins with adiabatic initial conditions for all fluids. The modularity of fluid species makes it simple to replace the perturbation equations with the Newtonian or other gauges, as well as isocurvature perturbations, though these are not currently implemented in ABCMB\@.

For the perturbation equations currently implemented, we follow the formalism of ref.~\cite{Ma:1995ey}.  In the synchronous gauge and Fourier space, all collisionless fluid species obey
\begin{align}
    \delta' &= -(1+w)\left(\frac{\theta}{\mathcal{{H}}} + \frac{h_m'}{2} \right) - 3\left(c_s^2-w\right)\delta \\
    \theta'&= -(1+3w)\theta - \frac{w'} {1+w}\theta + \frac{c_s^2}{1+w}\frac{k^2}{\HH}\delta - \frac{k^2}{\HH}\sigma\, ,
\end{align}
where $\delta$, $\theta$ $\sigma$ are the density, velocity and shear perturbations respectively, $w$ is the equation of state, $c_s^2$ is the sound speed squared, $\HH=aH$ is the conformal Hubble parameter, and $h_m$ is the first of two metric perturbations in the synchronous gauge (to be disambiguated from the dimensionless Hubble constant $h$). The derivatives are taken with respect to $x\equiv \ln{a}$, thus the reader may notice a factor of $\HH$ difference from definitions encountered elsewhere in the literature where derivatives are written with respect to conformal time. Accompanying these are the scalar metric perturbation equations
\begin{align}
    h_m' &= \frac{2k^2 \eta}{\HH^2} + \frac{8\pi Ga^2}{\HH^2}\sum_i \bar{\rho}_i\delta_i \label{eq:hprime}\\
    \eta' &= \frac{4\pi Ga^2}{\HH k^2}\sum_i (\bar{\rho}_i+\bar{p}_i)\theta_i \label{eq:etaprime}\, ,
\end{align}
where $\bar{\rho}$ and $\bar{p}$ denote the background energy density and pressure, and the sum is over all fluid species. 

These equations are promoted to a Boltzmann hierarchy (so that shear and higher moments are also included) when the fluid is relativistic and free streaming. In $\lcdm$ this is relevant for photons, massless neutrinos, and, at early times, massive neutrinos. In addition, a collision term must be included in the velocity perturbation for inter-fluid scattering, such as that of the baryon-photon scattering. We include a full list of perturbation equations for all currently implemented species in appendix~\ref{app:species}.

\subsubsection{Massive neutrinos}\label{app:massive_neutrinos}
\label{sec:mnu_integrals}
Unlike other $\lcdm$ species, massive neutrinos transition from relativistic at early times to non-relativistic at late times. Consequently, its mass enters the Fermi-Dirac distribution in a non-trivial manner, and any integrated quantities involving the distribution have no analytical solutions. Numerical solutions to the integrals over momenta require discretized momentum bins, each of which requires a separate Boltzmann hierarchy, rendering the calculations expensive. 

To this end, Gaussian quadrature has proven to be an efficient approximation to the integral, requiring as few as three momentum bins to accurately compute these integrals \cite{Howlett_2012}. This is the method we adopt in ABCMB\@. We work with the temperature normalized variables
\begin{align}
    x &\equiv \frac{a m}{T_0},\ \  q\equiv \frac{a p}{T_0},\ \ \epsilon\equiv \sqrt{x^2+q^2}
\end{align}
in place of the mass $m$, physical momentum $p$, and energy respectively at scale factor $a$ and present neutrino temperature $T_0$. The massive neutrino contribution to the perturbed stress energy can be written as
\begin{align}
    \delta \rho &= 4 \pi \frac{T_0^4}{a^4}\int_0^\infty dq\ q^2 \epsilon f_0(q) \Psi_0(q)\nonumber \\
    (\rho + p)\theta &= 4 \pi \frac{T_0^4}{a^4}\int_0^\infty dq\ q^3 f_0(q) k\Psi_1(q)\nonumber \\
    (\rho + p)\sigma &=  \frac{8\pi T_0^4}{3a^4}\int_0^\infty dq\ \frac{q^4}{\epsilon} f_0(q) \Psi_2(q)\, ,
    \label{eq:mnu_integral}
\end{align}
where $\Psi_l$'s form the Boltzmann hierarchy for momentum bin $q$ and wavenumber $k$.  $f_0$ is the Fermi-Dirac distribution
\begin{align}
    f_0(q) &= \frac{g_s}{(2\pi)^3}\frac{1}{e^{\epsilon}+1}\, ,
    \label{eq:mnu_f0}
\end{align}
with $g_s=2$.  Ref.~\cite{Howlett_2012} found that integrals of the following form can be approximated with the discrete sum
\begin{align}
    \int_0^\infty dq\frac{q^4 e^q}{(1+e^q)^2}\left(\frac{q}{\epsilon}\right)^w \Psi_l &\approx 4 \sum_i K_i \left(\frac{q_i} {\epsilon_i}\right)^w \Psi_l\, ,
\end{align}
which achieves accuracy to better than $2\times 10^{-4}$ using a three point quadrature sampling strategy with weights
\begin{align}
    K_i &= \{0.0687359,\ 3.31435,\ 2.29911\}\nonumber \\
    q_i &= \{0.913201,\ 3.37517,\ 7.79184\}\nonumber\, .
\end{align}
With these weights and roots we can rewrite eq.~\eqref{eq:mnu_integral} as
\begin{align}
    \delta \rho &= \frac{4}{\pi^2}\frac{T_0^4}{a^4}\sum_i K_i \frac{(1+e^{-q_i}) \epsilon_i}{q_i^2} \Psi_0(q_i)\nonumber \\
    (\rho + p)\theta &= \frac{4}{\pi^2}\frac{T_0^4}{a^4}\sum_i K_i \frac{(1+e^{-q_i})}{q_i} k\Psi_1(q_i)\nonumber \\
    (\rho + p)\sigma &= \frac{8}{3\pi^2}\frac{T_0^4}{a^4}\sum_i K_i \frac{(1+e^{-q_i})}{\epsilon_i} \Psi_2(q_i)\, .
    \label{eq:mnu_sum}
\end{align}
These expressions allow us to limit the massive neutrino Boltzmann hierarchies to only 3 systems of $\ell_{\rm max}$ equations, minimizing their impact on performance. In fact, we may go a bit further and express the background quantities with Gaussian quadrature, so that a numerical integral may be replaced by a discrete sum. We write the energy density and pressure as
\begin{align}
    \rho &= 4 \pi \frac{T_0^4}{a^4}\int_0^\infty dq\ q^2 \epsilon f_0(q) = \frac{4}{\pi^2} \frac{T_0^4}{a^4} \sum_i K_i \frac{(1+e^{-q_i}) \epsilon_i}{q_i^2} \nonumber \\
    p &= \frac{4\pi}{3}\frac{T_0^4}{a^4}\int_0^\infty dq\ \frac{q^4}{\epsilon} f_0(q) = \frac{4}{3\pi^2}\frac{T_0^4}{a^4}\sum_i K_i \frac{(1+e^{-q_i})}{\epsilon_i}\, .
\end{align}
Since these background quantities do not depend on the Boltzmann hierarchies, we can use the more accurate 5-point Gauss-Laguerre sampling with the weights and roots
\begin{align}
    K_i &= \{0.0081201,\ 0.689407,\ 2.8063,\ 2.05156,\ 0.12681\}\nonumber \\
    q_i &= \{0.583165,\ 2.0,\ 4.0,\ 7.26582,\ 13.0\}\nonumber\, ,
\end{align}
while minimally impacting performance. As with other species, we include our handling of the Boltzmann hierarchies themselves in appendix~\ref{app:species}.

\subsection{Transfer functions and power spectra}\label{app:spectra}
Currently ABCMB implements the standard scalar primordial spectrum, characterized by the amplitude $A_s$ and scalar tilt $n_s$ as
\begin{align}
    \mathcal{P}(k) = A_s \left(\frac{k}{k_{\rm pivot}}\right)^{n_s-1}\, \label{eq:PK_pivot},
\end{align}
where we take $k_{\rm pivot} = 0.05\ {\rm Mpc}^{-1}$ by default. The matter power spectrum is computed as
\begin{align}
    P(k, z) = \frac{2 \pi^2}{k^3} \mathcal{P}(k)\left[\frac{\sum_{i} \rho_i(z)\delta_i(k, z) }{\sum_i \rho_i(z)} \right]^2 \, ,
    \label{eq:PK}
\end{align}
where the sums are over all species that are non-relativistic.

The calculation of the CMB power spectra consists of a time integral over the photon streaming history, as well as a line of sight integral to project the spectrum onto the spherical sky. To perform these steps, we follow the formalism of ref.~\cite{Lesgourgues:2013bra} for computing the CMB power spectra in non-flat geometries, where for now we take the zero curvature limit to recover the flat space solutions. 

This calculation begins with the CMB source functions, separated into the temperature and polarization contributions. In synchronous gauge they read
\begin{align}
    S_{T0} &= g\left(\frac{\delta_\gamma}{4} + \HH \alpha'\right) + g(\eta - \HH \alpha' -2\HH \alpha) + 2\HH e^{-\kappa}(\eta' - \HH'\alpha - \HH \alpha')\nonumber \\
    &\phantom{S_{T0}S_{T0}S_{T0}}+ \HH \left[g\left(\frac{\theta_b'}{k^2}+\alpha'\right) + g'\left(\frac{\theta_b}{k^2}+\alpha\right)\right]\nonumber \\
    S_{T1} &= e^{-\kappa}k(\HH \alpha' + 2\HH \alpha - \eta)\nonumber \\
    S_{T2} &= \frac{g}{8}(2\sigma_\gamma + G_{\gamma0}+G_{\gamma2})\nonumber \\
    S_{E} &= \sqrt{6}\ S_{T2}\, ,\label{eq:source_functions}
\end{align}
where
\begin{align}
    \kappa(\ln{a}) &= \int_{\ln{a}}^0 \frac{d\ln{a'}}{\HH \tau_c} =\int_{\ln{a}}^0 d\ln{a'}\frac{an_e\sigma_T}{\HH}\nonumber \\
    g &= \frac{1}{\tau_c}e^{-\kappa}= an_e\sigma_T e^{-\kappa}\nonumber \, ,
\end{align}
$G_{\gamma l}$'s are the CMB polarization Boltzmann hierarchy, and $\alpha \equiv \HH\frac{h_m'+6\eta'}{2k^2}$. Here $\tau_c = (a n_e \sigma_T)^{-1}$ is the Compton scattering time, with $\sigma_T$ the Thomson cross section.  These source functions are the same as those used in CLASS, and the separated temperature source functions differ from the single source function implemented in CAMB by integrating the transfer function by parts. This is both to disentangle the physical contributions to the CMB power spectrum (SW, ISW, Doppler and polarization), and to avoid computing second or third time derivatives of various quantities. With these, transfer functions $\Delta$ can be obtained after integrating over the CMB streaming history:
\begin{align}
    \Delta_{T0}(k, \ell) &=\int d\tau\ S_{T0}(k, \tau) j_{\ell}(x) \nonumber \\
    \Delta_{T1}(k, \ell) &= \int d\tau\ S_{T1}(k, \tau) \frac{d}{dx}j_\ell(x)\nonumber \\
    \Delta_{T2}(k, \ell) &= \int d\tau\ S_{T2}(k, \tau) \frac{1}{2}\left[3\frac{d^2}{dx^2}j_\ell(x) + j_\ell(x)\right]\nonumber \\
    \Delta_{E}(k, \ell) &= \int d\tau\ S_{E}(k, \tau) \sqrt{\frac{3(\ell+2)!}{8(\ell-2)!}}\frac{1}{x^2}j_\ell(x)\, ,\nonumber 
\end{align}
where $x \equiv k(\tau_0-\tau)$, $\tau_0$ the conformal time today, and $j_\ell$ is the spherical Bessel function of the first kind.

The spherical Bessel functions of arbitrary $\ell$ are not currently implemented in JAX\@. Furthermore, existing numerical algorithms for computing Bessel functions are costly, especially for the CMB which requires repeated evaluations. We find the following strategy for computing $j_\ell(x)$ to be efficient, which involves splitting the function over three intervals:
\begin{itemize}
    \item The Bessel function is set to $0$ in the interval $0<x\leq x_{\rm start}$. Here $x_{\rm start}$ is defined such that $j_\ell(x_{\rm start})$ exceeds a threshold value of $10^{-10}$ for the first time. Below this value the bessel function is exponentially suppressed. 
    \item We use linear interpolation in the interval $x_{\rm start}<x\leq x_5$, where $x_5$ is the location of the fifth local maximum, i.e. ``peak" of the Bessel function. For each $\ell$ we numerically find the two boundaries of this interval, and store $j_\ell$ computed with \texttt{scipy.special.spherical\_jn} over 5000 points, corresponding to a fine resolution of $\sim1000$ points per oscillation. 
    \item For $x>x_5$ we use an asymptotic expansion given by \cite{watson1944bessel}
    \begin{align}
        j_\ell(x>x_5)&\approx \sqrt{\frac{1}{x\sqrt{x^2-(\ell+\frac{1}{2})^2}}} \cos{\left(Q_{\ell+\frac{1}{2}}(x)-\frac{\pi}{4}\right)} \nonumber \\
        Q_\ell(x)&\equiv \sqrt{x^2-\ell^2}-\frac{\ell\pi}{2} + \ell\arcsin{\left(\frac{\ell}{x}\right)}\nonumber \, ,
    \end{align}
    which we find to be accurate and inexpensive to compute for sufficiently large $x$. 
\end{itemize}

The total temperature transfer function is the sum of the components $\Delta_T \equiv \Delta_{T0} +\Delta_{T1}+\Delta_{T2}$, which is obtained after their respective source functions are integrated. This then finally enters the angular power spectrum via
\begin{align}
    C_\ell^{XY} = 4\pi \int\frac{dk}{k} \mathcal{P}(k) \Delta_X(k, \ell)\Delta_{Y}(k, \ell)\, \label{eq:c_ell_def},
\end{align}
where $X, Y\in \{T, E\}$. With these equations ABCMB is able to compute the TT, TE, and EE spectra using the results from perturbations module. 

\subsection{Lensing}\label{app:lensing}
The matter field between the last scattering surface and the observer gravitationally lenses the CMB signal. The effect is most significant at small angular scales, modifying the power spectrum by as much as $\OO{(0.1)}$ at $\ell \gtrsim 2000$. As was done in CAMB and CLASS, we implement the effect of lensing using the spherical-sky correlation function formalism of ref.~\cite{Challinor_2005}, and we outline the important steps in this section. 

Lensing is sourced by the gravitational lensing potential $\psi$, such that an unlensed signal of the CMB $\Theta(\hat{n})$ is deflected to $\tilde{\Theta}(\hat{n}')$, where $\hat{n}'= \hat{n}+\vec{\nabla}\psi$. The lensing potential obeys the Poisson equation, which in Fourier space reads
\begin{align}
    k^2 \psi &= 4 \pi  G a^2 \rho \delta\, .
\end{align}
Lensing is most efficient after radiation domination ends. For this reason, we have omitted the small contribution from anisotropic stress, and contributions to $\rho \delta$ include only non-relativistic species to good approximation. We can thus write the power spectrum of the lensing potential as
\begin{align}
    P_{\psi}(k, a) &= \frac{9 \Omega_m(a)^2 \HH^4}{8\pi^2 k} P(k, a)\, ,
\end{align}
where $P(k, a)$ is the matter power spectrum in eq.~\eqref{eq:PK}, and $\Omega_m(a)=\rho_m(a)/\rho_{\rm tot}(a)$ (we use this definition since its evolution with $a$ is more apparent, as $\rho_{\rm{crit}}$ is only defined today). 

For the CMB on the spherical sky, we need the angular lensing power spectrum. This can be obtained with the power spectrum defined above and the transfer function formalism of the previous section. However, the lensing power spectrum varies slowly as a function of $k$ compared to the highly oscillatory Bessel functions at high $\ell$. Thus the lensing spectrum can be efficiently computed under the Limber approximation as
\begin{align}
    C_{\ell}^{\psi\psi} &= \frac{8\pi^2}{\left(\ell+\frac{1}{2}\right)^3} \int_0^{\chi_*} \frac{d\chi}{\chi} P_{\psi}\left(k=\frac{\ell+\frac{1}{2}}{\chi}, \tau_0-\chi\right)\left(\frac{\chi_*-\chi}{\chi_*}\right)^2\, ,
\end{align}
where $\chi \equiv \tau_0-\tau$ is the look back time, and $\chi_*$ is defined at recombination. The Limber approximation is highly accurate for $\ell >10$, and leads to an order $10\%$ error below $\ell=10$.  However, the lensing signal is extremely weak at very large angular scales, and the error virtually unobservable. In figures~\ref{figure:tt_residual} and \ref{figure:ee_residual} we show the residual error in the lensed TT and EE spectra compared to CLASS\@. Here the Limber approximation is used by ABCMB and not used by CLASS for $\ell\leq10$, yet the errors remain comparable at all $\ell$'s to the unlensed comparisons in the same figures. This indicates that the use of Limber approximation at low-$\ell$ does not dominate the error. 

With the lensing power spectrum and angular spectrum defined this way, the lensed CMB power spectra are obtained by implementing Eqs.(35-66) in ref.~\cite{Challinor_2005}. In the very last step, the lensed power spectrum $\tilde{C}$ is obtained from the lensed correlation function $\tilde{\xi}$ via an angular Fourier transform, involving an integral of the form
\begin{align}
    \tilde{C}_\ell &= \int_{-1}^{1}d\mu\ \tilde{\xi}(\mu) d_\ell(\mu)\nonumber\, ,
\end{align}
for each TT, TE, and EE\@. Here tilde denotes lensed quantities, and $d_\ell$ is a Wigner matrix element depending on the source (temperature or polarization). Ref.~\cite{Challinor_2005} points out that instead of performing the full integration over $[-1, 1]$, the lensed power spectrum can be more easily obtained by solving the difference $\tilde{C}_\ell - C_\ell$. The analogous integrand for this difference rapidly vanishes away from $\mu=1$, allowing one to speed up the integration by restricting the domain down to a smaller interval, typically over $\sim[0.98, 1]$. In CLASS, this approximation is implemented under the ``fast lensing" setting, as opposed to the ``accurate lensing" setting where the full integral is solved with a high order Gauss-Legendre quadrature. In practice, we find that this approximation introduces error of several percent for $\ell > 3000$. As modern precision measurements easily surpass this angular scale, we conclude that the fast lensing approximation should never be used in ABCMB and thus not implemented. 

\subsection{Approximation schemes}\label{app:approx}

By default, CLASS uses several approximations that considerably accelerate the power spectra computations. Most notably, the costly Boltzmann hierarchies in the perturbation evolution are aided by the tight-coupling approximation (TCA), the ultra-relativistic fluid approximation (UFA), and the radiation streaming approximation (RSA). While they stem from different physical arguments, all three approximations effective decouple the first three moments in the hierarchy from the remaining moments, applicable to photons and both massive and massless neutrinos. After this decoupling, the higher moments are either replaced by some analytical approximation, or trivially set to zero. In either case, the reduction in the number of equations leads to a notable speedup. In addition, these approximations can help in stabilizing the differential equation solver. TCA in particular has been vital in reducing the stiffness of the baryon-photon coupling equations at early times, introduced by the largeness of the Thomson scattering rate and the smallness of the baryon-photon velocity difference. 

At the moment, ABCMB does not utilize any of these three approximations in its perturbation module. Most notably, the diffrax adaptive step integrator used in our differential equation solutions is able to solve the baryon-photon equations before and during recombination without transforming the equations with TCA\@. The resulting baryon and photon perturbations are highly accurate compared to the CLASS output where TCA is used during recombination. In fact, foregoing TCA in ABCMB is computationally advantageous. This is because in using TCA, we must break the solution into two eras, where the derivative equations and the number of moments being evolved are distinct, and a patching is needed at the break point. In practice, this separate-then-combine approach introduces considerable overhead for the diffrax solver, lengthening both the compile and the computation times, while providing no improvements to accuracy. 

RSA and UFA, though not currently included, may still be beneficial to ABCMB's performance. These approximations turn on at late times (after recombination), where the higher $l$ terms in the Boltzmann hierarchies of the radiation species are negligible deep in matter domination. In fact, terminating the solver for the higher $l$ moments here help reduce the error incurred by the Boltzmann hierarchy cutoff, which is known to cause a periodic error for the perturbation modes at a frequency of $(k \tau) \sim l_{\rm max}$. In addition, a combination of RSA and UFA can accelerate CLASS computations by as much as 70-80\%.  We leave implementation of RSA and UFA in ABCMB to future work. 

Lastly, CLASS uses the time cut and multipole cut approximations to reduce the limits of integration for the CMB transfer functions. In ABCMB, we use simple trapezoid rules over the full range of $k$ and $\ln{a}$. Even still, we find that the transfer integral run time is vastly subdominant to the perturbations module on the GPU\@. We thus conclude that these approximations are not necessary and exclude them in the current version. 


\section{Currently implemented species}
\label{app:species}

As discussed in the main text, fluid-specific physics are packaged within individual objects responsible for all features of a given fluid.  Here we present a list of all background and perturbation equations for all species currently implemented in ABCMB\@.  

\subsection{Dark energy}
The \texttt{DarkEnergy} fluid is responsible only for energy density and pressure, given that standard cosmology does not introduce any spatial fluctuations to the dark energy field. The equations are then simply
\begin{align}
    \rho_{\Lambda} &= \rho_{\Lambda, 0} =\Omega_\Lambda \rhoc\nonumber\\
    P_\Lambda&= -\rho_\Lambda\, ,
\end{align}
where we used the standard $w=-1$ equation of state for the dark energy, and $\rhoc = \frac{3 H_0^2}{8 \pi G}$.

\subsection{Cold dark matter}

CDM behaves like a non-relativistic fluid through all of cosmology, with vanishing sound speed and equation of state and energy density that redshifts simply. For the background we have
\begin{align}
    \rho_{\rm cdm} &= \Omega_{\rm cdm}\rhoc a^{-3}\nonumber \\
    P_{\rm cdm}&=0 \, .
\end{align}
The CDM density perturbation follows only the gravitational metric perturbation. Its velocity perturbation vanishes in the synchronous gauge as the rest frame of CDM is used to define the coordinate system of the gauge. The only perturbation equation is therefore
\begin{align}
    \delta_{\rm cdm}' &= -\frac{1}{2}h_m'\, ,
\end{align}
where we emphasize again that $'$ in this work denotes derivative with respect to $\ln{a}$ and $h_m$ is the usual metric perturbation, disambiguated from the dimensionless Hubble constant $h$. 

\subsection{Baryons}
Like CDM, baryons are also non-relativistic throughout the period relevant for the CMB\@. However, interactions with photons via Thomson scattering induces a non-trivial velocity perturbation. Although pressure can still be ignored, we must account for the baryon sound speed contribution to the velocity perturbation despite its small value. The relevant background quantities are
\begin{align}
    \rho_{b} &= \Omega_{b}\rhoc a^{-3}\nonumber \\
    P_{b}&=0 \nonumber \\
    c_s^2 &= \frac{T_b}{\mu_b}\left(1-\frac{1}{3}\frac{d\ln{T_b}}{d\ln{a}}\right)\nonumber \\
    \mu_b &\equiv \frac{\rho_b}{n_b} =\frac{m_\textrm{H}}{(1+x_e)(1-\textrm{Y}_\textrm{He}) + \frac{m_{\textrm{H}}}{m_\textrm{He}}\textrm{Y}_\textrm{He}}\, ,
\end{align}
where $\mu_b$ is the mean baryon mass, receiving contributions from hydrogen and helium atoms/ions as well as free electrons at a given time. The density and velocity perturbations are
\begin{align}
    \delta_b' &= -\frac{\theta_b}{\HH} - \frac{1}{2}h_m'\nonumber \\
    \theta_b' &= -\theta_b +\frac{c_s^2k^2\delta_b}{\HH} + \frac{4\rho_\gamma}{3\rho_b}\frac{1}{\HH \tau_c}(\theta_\gamma - \theta_b)\, ,
\end{align}
where the subscript $\gamma$ denotes photon related quantities that we define next. 

\subsection{Photons}
The photon fluid is not parameterized by a density parameter, but instead by its mean temperature today. The energy density and pressure are expressed in terms of the temperature assuming a Bose-Einstein distribution
\begin{align}
    \rho_{\gamma} &= \frac{\pi^2}{15} T_{\gamma, 0}^4 a^{-4}\nonumber \\
    P_{\gamma}&=\frac{1}{3} \rho_\gamma\, .
\end{align}

Photons require perturbations beyond the density and velocity moments due to their free-streaming nature at late times. The photon Boltzmann hierarchy is divided into the temperature moments $F_{\gamma l}$ and polarization moments $G_{\gamma l}$. The moments $l$ label the expansion coefficients in the Legendre polynomial basis, and the resulting Boltzmann equations for each coefficient are
\begin{align}
   \delta_\gamma' &= -\frac{4}{3\HH}\theta_\gamma - \frac{2}{3}h_m'\nonumber \\
   \theta_\gamma'&= \frac{k^2}{\HH}\left(\frac{1}{4}\delta_\gamma - \sigma_\gamma\right) + \frac{1}{\HH \tau_c}(\theta_b-\theta_\gamma)\nonumber \\
   \sigma_\gamma'&= \frac{4}{15 \HH}\theta_\gamma - \frac{3}{10}\frac{k}{\HH}F_{\gamma3} + \frac{2}{15}h_m' +\frac{4}{5}\eta' - \frac{9}{10}\frac{1}{\HH \tau_c} + \frac{1}{20}\frac{1}{\HH \tau_c}(G_{\gamma 0}+G_{\gamma 2})\nonumber \\
   F_{\gamma l}'&=\frac{1}{2l+1}\frac{k}{\HH} (l F_{\gamma l-1} - (l+1)F_{\gamma l+1}) - \frac{1}{\HH \tau_c}F_{\gamma l} && 3\leq l < l_{\rm max} \nonumber \\
    F_{\gamma l_{\rm max}}' &= \frac{k}{\HH}F_{\gamma l_{\rm max}-1} -\frac{(l_{\rm max}+1)}{\HH \tau}F_{\gamma l_{\rm max}} - \frac{1}{\HH \tau_c}F_{\gamma l_{\rm max}}\nonumber \\
   G_{\gamma l}' &= \frac{1}{2l+1}\frac{k}{\HH} (l G_{\gamma l-1} - (l+1)G_{\gamma l+1}) - \frac{1}{\HH \tau_c}G_{\gamma l} \nonumber\\
   &\phantom{G_{\gamma l}' = \frac{1}{2l+1}\frac{k}{\HH}}+ \frac{1}{2}\frac{1}{\HH \tau_c}(2\sigma_\gamma + G_{\gamma 0}+G_{\gamma 2}) ( \delta_{l, 0}+\frac{1}{2}\delta_{l, 2})  &&  l < l_{\rm max} \nonumber\\
   G_{\gamma l_{\rm max}}' &= \frac{k}{\HH}G_{\gamma l_{\rm max}-1} -\frac{(l_{\rm max}+1)}{\HH \tau}G_{\gamma l_{\rm max}} - \frac{1}{\HH \tau_c}G_{\gamma l_{\rm max}}\, ,
\end{align}
where for consistency with ref.~\cite{Ma:1995ey} we used $\delta_\gamma \equiv F_{\gamma0}$, $\theta_\gamma \equiv \frac{3}{4}k F_{\gamma 1}$, and $\sigma_\gamma \equiv F_{\gamma 2}/2$. 

\subsection{Massless neutrinos}
Massless neutrinos require two parameters, $N_{\rm massless}$ and $T_{\rm massless}$. These define the background quantities
\begin{align}
    \rho_{\nu, {\rm massless}} &= N_{\rm massless} \frac{7\pi^2}{120} T_{\rm massless}^4 a^{-4} \nonumber \\
    P_{\nu, {\rm massless}} &= \frac{1}{3} \rho_{\nu, {\rm massless}}\, .
\end{align}
In the absence of massive neutrinos, $N_{\rm massless}$ is the literal number of massless neutrino species. This means $N_{\rm massless} = 3$ in standard cosmology, and the additional correction in $N_{\rm eff}>3$ is encoded in the late-time temperature $T_{\rm massless}$.  These parameters are discussed in more detail in section~\ref{sec:neutrinos}.

Neutrinos are free-streaming and relativistic at all times, thus generating shear and higher moments in the Boltzmann hierarchy. They are given by
\begin{align}
    \delta_\nu' &= -\frac{4}{3\HH}\theta_\nu - \frac{2}{3}h_m'\nonumber \\
   \theta_\nu'&= \frac{k^2}{\HH}\left(\frac{1}{4}\delta_\nu - \sigma_\nu\right)\nonumber \\
   \sigma_\nu'&= \frac{4}{15 \HH}\theta_\nu - \frac{3}{10}\frac{k}{\HH}F_{\nu3} + \frac{2}{15}h_m' +\frac{4}{5}\eta'\nonumber \\
   F_{\nu l}'&=\frac{1}{2l+1}\frac{k}{\HH} (l F_{\nu l-1} - (l+1)F_{\nu l+1}) && 3\leq l < l_{\rm max} \nonumber \\
    F_{\nu l_{\rm max}}' &= \frac{k}{\HH}F_{\nu l_{\rm max}-1} -\frac{(l_{\rm max}+1)}{\HH \tau}F_{\nu l_{\rm max}}\, .
\end{align}

\subsection{Massive neutrinos}
We provide an overview of integrated massive neutrino quantities in Sec.~\ref{sec:mnu_integrals}, including the energy density and pressure. These integrals rely on the Boltzmann hierarchy moments, which for massive neutrinos depend on both the wavenumber $k$ and the momentum $q$. The momentum dependence is separable, and each value $q$ labels a separate hierarchy which obeys
\begin{align}
    \Psi_{0}' &= -\frac{qk}{\epsilon \HH}\Psi_1 + \frac{1}{6}h_m' \frac{d\ln{f_0}}{d\ln{q}}\nonumber \\
    \Psi_1' &= \frac{1}{3}\frac{qk}{\epsilon \HH}(\Psi_0 - 2\Psi_2) \nonumber \\
    \Psi_2' &= \frac{1}{5}\frac{qk}{\epsilon \HH} (2\Psi_1 - 3\Psi_3) - \left(\frac{1}{15}h_m'+ \frac{2}{5}\eta'\right)\frac{d\ln{f_0}}{d\ln{q}}\nonumber \\
    \Psi_l' &= \frac{1}{2l+1}\frac{qk}{\epsilon \HH} (l \Psi_{l-1} - (l+1)\Psi_{l+1})&& 3\leq l < l_{\rm max}\nonumber \\
    \Psi_{l_{\rm max}}'&=\frac{qk}{\epsilon \HH} \Psi_{l_{\rm max}-1} - \frac{l_{\rm max}+1}{\HH \tau} \Psi_{l_{\rm max}}\, ,
\end{align}
where $\frac{d\ln{f_0}}{d\ln{q}}=-\frac{q}{1+e^{-q}}$, with $f_0$ the Fermi-Dirac distribution given by eq.~\eqref{eq:mnu_f0}. As a reminder $q=a p/T_0$ is the temperature normalized momentum, $T_0$ is the massive neutrino temperature today, and $\epsilon\equiv \sqrt{q^2+(a m/T_0)^2}$.

\section{Gradients with respect to cosmological parameters}\label{app:gradients}
\begin{figure}[h]
\centering
\includegraphics[width=\textwidth]{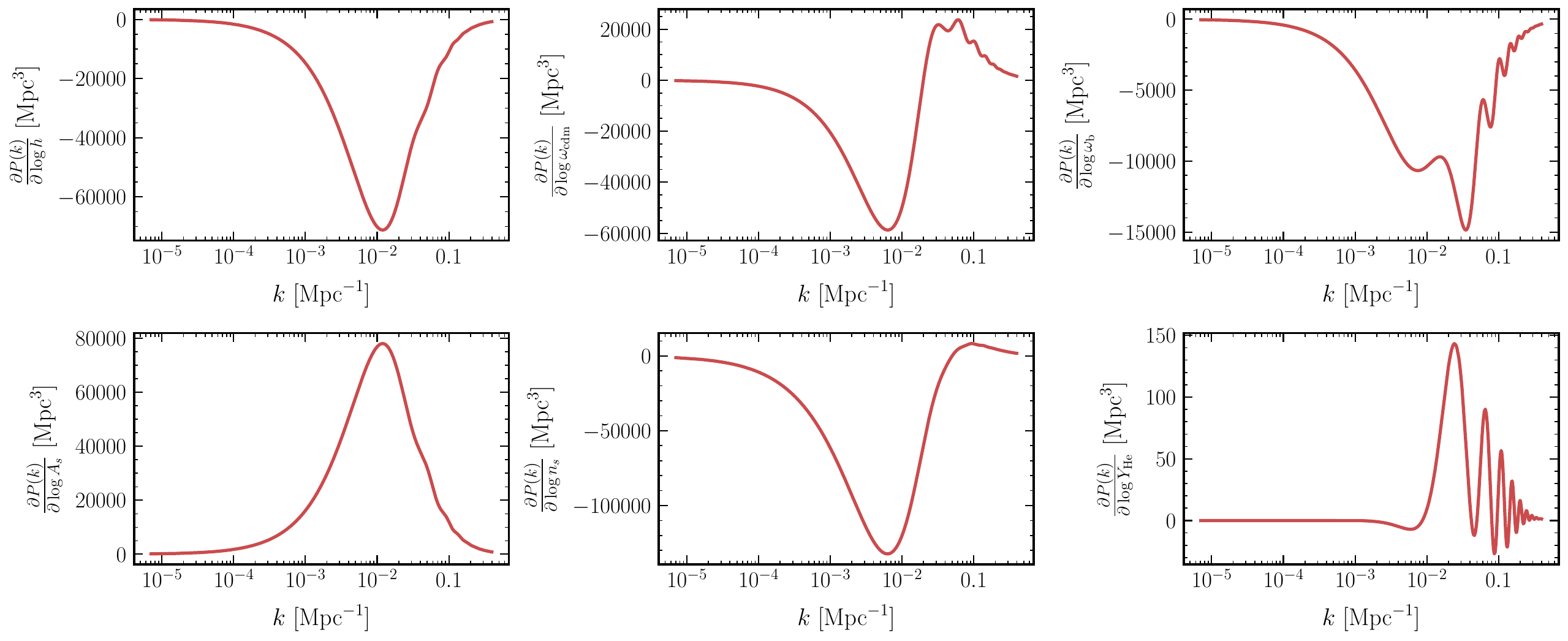}
\caption{The $\lcdm$ matter power spectrum gradients with respect to six cosmological parameters.} 
\label{figure:dPk}
\end{figure}

\begin{figure}[h]
\centering
\includegraphics[width=\textwidth]{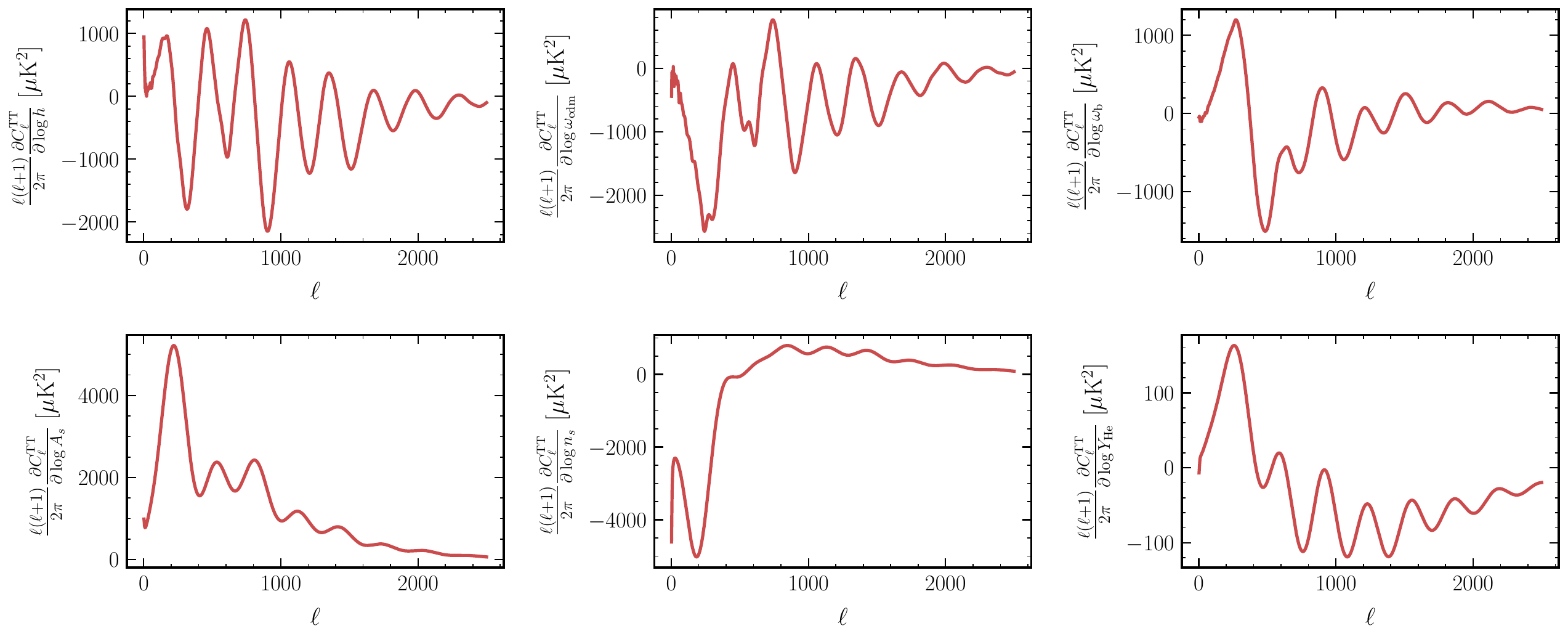}
\caption{The $\lcdm$ TT spectrum gradients with respect to six cosmological parameters.} 
\label{figure:dTT}
\end{figure}

\begin{figure}[h]
\centering
\includegraphics[width=\textwidth]{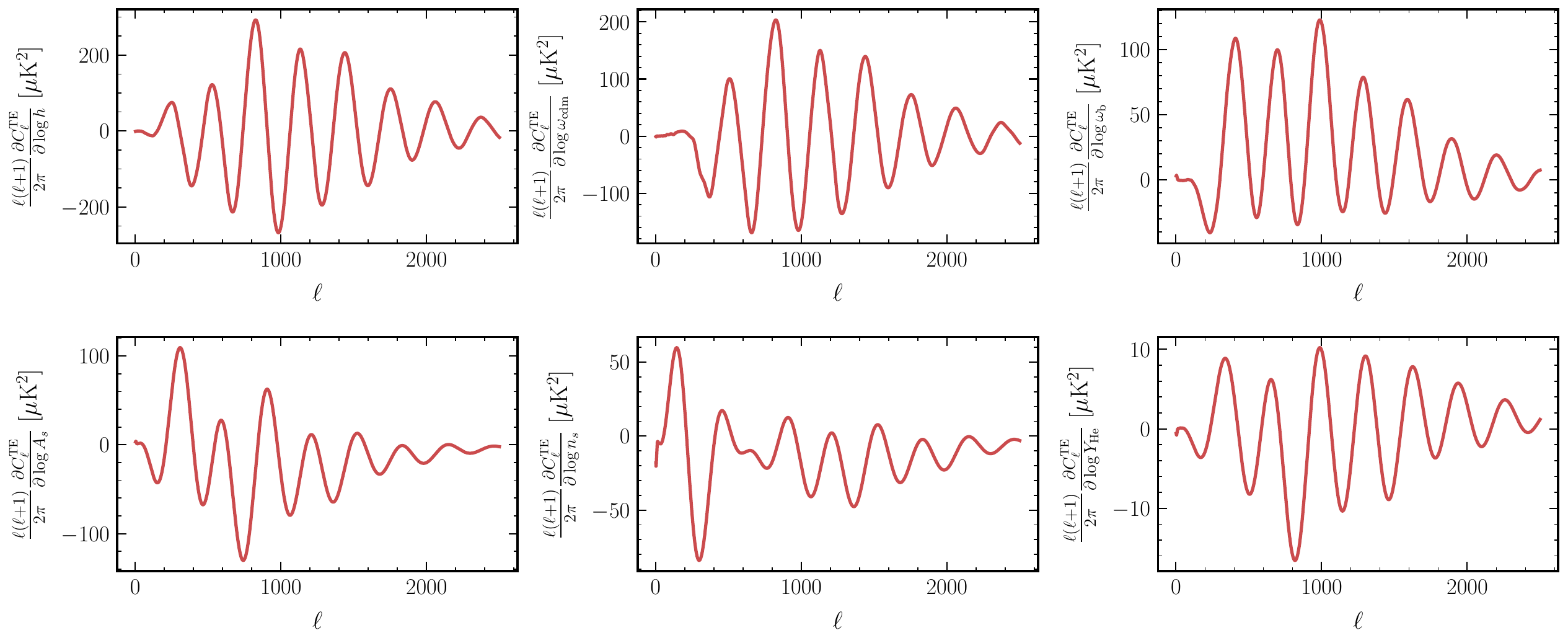}
\caption{The $\lcdm$ TE spectrum gradients with respect to six cosmological parameters.} 
\label{figure:dTE}
\end{figure}

\begin{figure}[h]
\centering
\includegraphics[width=\textwidth]{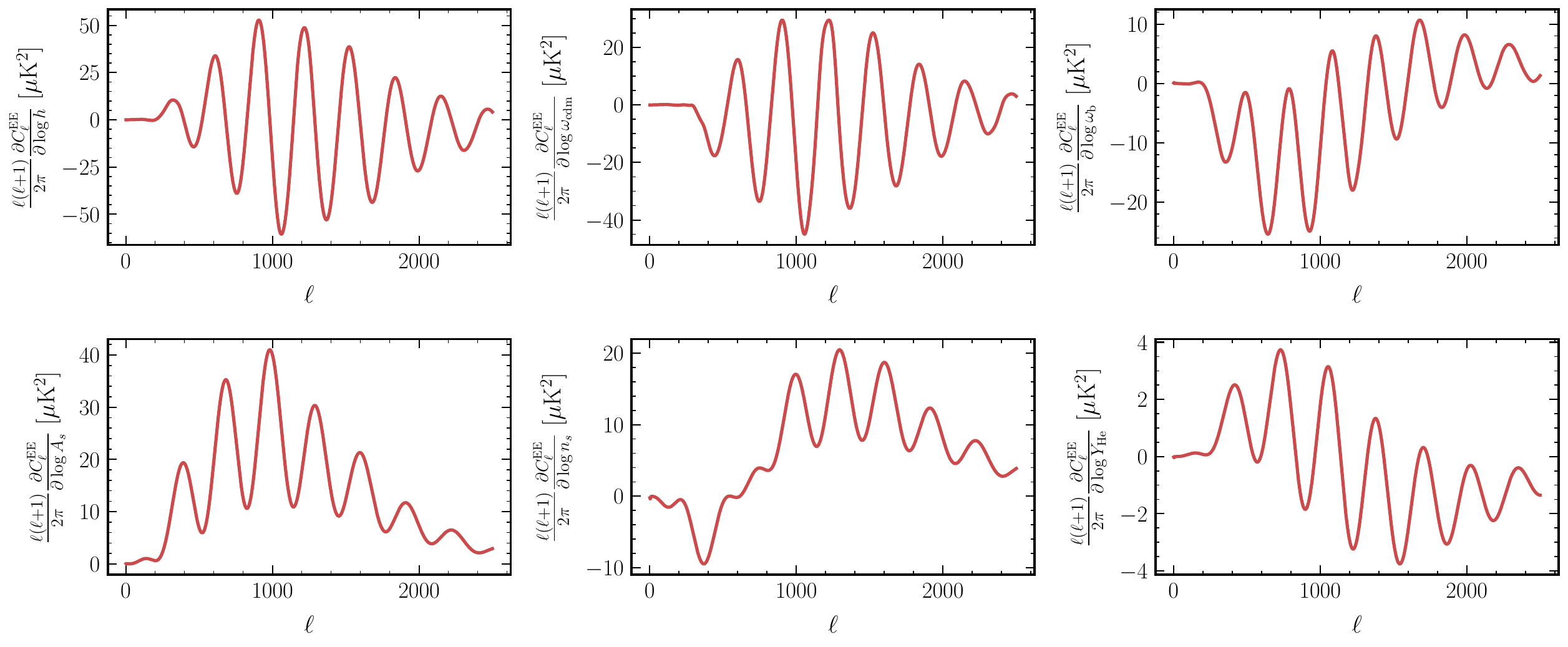}
\caption{The $\lcdm$ EE spectrum gradients with respect to six cosmological parameters.}
\label{figure:dEE}
\end{figure}

For reference, we include additional plots of gradients with respect to cosmological parameters in this appendix.  Each of these can be computed by the user with ABCMB using the built-in \texttt{jax.jacfwd}.  In particular, we compute the gradients of the output $P(k)$ (figure~\ref{figure:dPk}), $C_\ell^{\rm{TT}}$ (figure~\ref{figure:dTT}), $C_\ell^{\rm{TE}}$ (figure~\ref{figure:dTE}), and $C_\ell^{\rm{EE}}$ (figure~\ref{figure:dEE}) with respect to all input cosmological parameters using forward AD\@.

\bibliographystyle{jcap}
\bibliography{references}

@article{HYREC,
   title={HyRec: A fast and highly accurate primordial hydrogen and helium recombination code},
   volume={83},
   ISSN={1550-2368},
   url={http://dx.doi.org/10.1103/PhysRevD.83.043513},
   DOI={10.1103/physrevd.83.043513},
   number={4},
   journal={Physical Review D},
   publisher={American Physical Society (APS)},
   author={Ali-Haïmoud, Yacine and Hirata, Christopher M.},
   year={2011},
   month=feb }

@article{Akita:2020szl,
    author = "Akita, Kensuke and Yamaguchi, Masahide",
    title = "{A precision calculation of relic neutrino decoupling}",
    eprint = "2005.07047",
    archivePrefix = "arXiv",
    primaryClass = "hep-ph",
    doi = "10.1088/1475-7516/2020/08/012",
    journal = "JCAP",
    volume = "08",
    pages = "012",
    year = "2020"
}

@article{Froustey:2020mcq,
    author = "Froustey, Julien and Pitrou, Cyril and Volpe, Maria Cristina",
    title = "{Neutrino decoupling including flavour oscillations and primordial nucleosynthesis}",
    eprint = "2008.01074",
    archivePrefix = "arXiv",
    primaryClass = "hep-ph",
    doi = "10.1088/1475-7516/2020/12/015",
    journal = "JCAP",
    volume = "12",
    pages = "015",
    year = "2020"
}

@article{Bennett:2020zkv,
    author = "Bennett, Jack J. and Buldgen, Gilles and De Salas, Pablo F. and Drewes, Marco and Gariazzo, Stefano and Pastor, Sergio and Wong, Yvonne Y. Y.",
    title = "{Towards a precision calculation of $N_{\rm eff}$ in the Standard Model II: Neutrino decoupling in the presence of flavour oscillations and finite-temperature QED}",
    eprint = "2012.02726",
    archivePrefix = "arXiv",
    primaryClass = "hep-ph",
    reportNumber = "CPPC-2020-10",
    doi = "10.1088/1475-7516/2021/04/073",
    journal = "JCAP",
    volume = "04",
    pages = "073",
    year = "2021"
}

@article{Barron:2025dys,
    author = "Barron, Jared and Curtin, David and Liu, Hongwan and Munoz, Julian and Roy, Sandip",
    title = "{Constraining Dark Acoustic Oscillations with the High-Redshift UV Luminosity Function}",
    eprint = "2512.01998",
    archivePrefix = "arXiv",
    primaryClass = "astro-ph.CO",
    reportNumber = "CERN-TH-2025-200",
    month = "12",
    year = "2025"
}

@article{HYREC2,
   title={hyrec-2: A highly accurate sub-millisecond recombination code},
   volume={102},
   ISSN={2470-0029},
   url={http://dx.doi.org/10.1103/PhysRevD.102.083517},
   DOI={10.1103/physrevd.102.083517},
   number={8},
   journal={Physical Review D},
   publisher={American Physical Society (APS)},
   author={Lee, Nanoom and Ali-Haïmoud, Yacine},
   year={2020},
   month=oct }

@article{LINX_long,
  title = {Fast, differentiable, and extensible big bang nucleosynthesis package},
  author = {Giovanetti, Cara and Lisanti, Mariangela and Liu, Hongwan and Mishra-Sharma, Siddharth and Ruderman, Joshua T.},
  journal = {Phys. Rev. D},
  volume = {112},
  issue = {6},
  pages = {063531},
  numpages = {33},
  year = {2025},
  month = {Sep},
  publisher = {American Physical Society},
  doi = {10.1103/f3tj-r882},
  url = {https://link.aps.org/doi/10.1103/f3tj-r882}
}

@article{LINX_short,
  title = {Cosmological parameter estimation with a joint-likelihood analysis of the cosmic microwave background and big bang nucleosynthesis},
  author = {Giovanetti, Cara and Lisanti, Mariangela and Liu, Hongwan and Mishra-Sharma, Siddharth and Ruderman, Joshua T.},
  journal = {Phys. Rev. D},
  volume = {112},
  issue = {6},
  pages = {063530},
  numpages = {10},
  year = {2025},
  month = {Sep},
  publisher = {American Physical Society},
  doi = {10.1103/wspy-s948},
  url = {https://link.aps.org/doi/10.1103/wspy-s948}
}

@misc{ole,
      title={{OL\'E -- Online Learning Emulation in Cosmology}}, 
      author={Sven Günther and Lennart Balkenhol and Christian Fidler and Ali Rida Khalife and Julien Lesgourgues and Markus R. Mosbech and Ravi Kumar Sharma},
      year={2025},
      eprint={2503.13183},
      archivePrefix={arXiv},
      primaryClass={astro-ph.CO},
      url={https://arxiv.org/abs/2503.13183}, 
}

@article{Hahn_2024,
   title={DISCO-DJ I: a differentiable Einstein-Boltzmann solver for cosmology},
   volume={2024},
   ISSN={1475-7516},
   url={http://dx.doi.org/10.1088/1475-7516/2024/06/063},
   DOI={10.1088/1475-7516/2024/06/063},
   number={06},
   journal={Journal of Cosmology and Astroparticle Physics},
   publisher={IOP Publishing},
   author={Hahn, Oliver and List, Florian and Porqueres, Natalia},
   year={2024},
   month=jun, pages={063} }

@software{li_2023,
  author       = {Zack Li and
                  Jamie Sullivan and
                  Marius Millea},
  title        = {Bolt.jl},
  month        = nov,
  year         = 2023,
  publisher    = {Zenodo},
  version      = {v0.0.1-alpha2},
  doi          = {10.5281/zenodo.10065126},
  url          = {https://doi.org/10.5281/zenodo.10065126},
}

@misc{sletmoen_2025,
      title={{SymBoltz.jl: a symbolic-numeric, approximation-free and differentiable linear Einstein-Boltzmann solver}}, 
      author={Herman Sletmoen},
      year={2025},
      eprint={2509.24740},
      archivePrefix={arXiv},
      primaryClass={astro-ph.CO},
      url={https://arxiv.org/abs/2509.24740}, 
}

@article{Bonici_2024,
   title={{Capse.jl: efficient and auto-differentiable CMB power spectra emulation}},
   volume={7},
   ISSN={2565-6120},
   url={http://dx.doi.org/10.21105/astro.2307.14339},
   DOI={10.21105/astro.2307.14339},
   journal={The Open Journal of Astrophysics},
   publisher={Maynooth University},
   author={Bonici, Marco and Baxter, Eric and Bianchini, Federico and Ruiz-Zapatero, Jaime},
   year={2024},
   month=jan }

@article{Spurio_Mancini_2022,
   title={<scp>CosmoPower</scp>: emulating cosmological power spectra for accelerated Bayesian inference from next-generation surveys},
   volume={511},
   ISSN={1365-2966},
   url={http://dx.doi.org/10.1093/mnras/stac064},
   DOI={10.1093/mnras/stac064},
   number={2},
   journal={Monthly Notices of the Royal Astronomical Society},
   publisher={Oxford University Press (OUP)},
   author={Spurio Mancini, Alessio and Piras, Davide and Alsing, Justin and Joachimi, Benjamin and Hobson, Michael P},
   year={2022},
   month=jan, pages={1771–1788} }

@article{Piras_2023,
   title={{CosmoPower-JAX: high-dimensional Bayesian inference with differentiable cosmological emulators}},
   volume={6},
   ISSN={2565-6120},
   url={http://dx.doi.org/10.21105/astro.2305.06347},
   DOI={10.21105/astro.2305.06347},
   journal={The Open Journal of Astrophysics},
   publisher={Maynooth University},
   author={Piras, Davide and Spurio Mancini, Alessio},
   year={2023},
   month=jun }

@misc{lesgourgues_2011a,
      title={{The Cosmic Linear Anisotropy Solving System} ({CLASS}) {I}: Overview}, 
      author={Julien Lesgourgues},
      year={2011},
      eprint={1104.2932},
      archivePrefix={arXiv},
      primaryClass={astro-ph.IM}
}

@article{lesgourgues_2011b,
   title={{The Cosmic Linear Anisotropy Solving System} ({CLASS}).
 Part {II}: Approximation schemes},
   volume={2011},
   ISSN={1475-7516},
   url={http://dx.doi.org/10.1088/1475-7516/2011/07/034},
   DOI={10.1088/1475-7516/2011/07/034},
   number={07},
   journal={Journal of Cosmology and Astroparticle Physics},
   publisher={IOP Publishing},
   author={Diego Blas and Julien Lesgourgues and Thomas Tram},
   year={2011},
   month=jul, pages={034–034} }

@misc{lesgourgues_2011c,
    author = "Lesgourgues, Julien",
    title = "{The Cosmic Linear Anisotropy Solving System ({CLASS}) {III}: Comparision with {CAMB} for Lambda{CDM}}",
    eprint = "1104.2934",
    archivePrefix = "arXiv",
    primaryClass = "astro-ph.CO",
    reportNumber = "CERN-PH-TH-2011-083, LAPTH-011-11",
    month = "4",
    year = "2011"
}

@article{lesgourgues_2011d,
   title={{The Cosmic Linear Anisotropy Solving System} ({CLASS}) {IV}: efficient implementation of non-cold relics},
   volume={2011},
   ISSN={1475-7516},
   url={http://dx.doi.org/10.1088/1475-7516/2011/09/032},
   DOI={10.1088/1475-7516/2011/09/032},
   number={09},
   journal={Journal of Cosmology and Astroparticle Physics},
   publisher={IOP Publishing},
   author={Lesgourgues, Julien and Tram, Thomas},
   year={2011},
   month=sep, pages={032–032} }

@article{Lewis_1999,
      author         = "Lewis, Antony and Challinor, Anthony and Lasenby,
                        Anthony",
      title          = "{Efficient computation of CMB anisotropies in closed FRW
                        models}",
      journal        = "\apj",
      volume         = "538",
      year           = "2000",
      pages          = "473-476",
      doi            = "10.1086/309179",
      eprint         = "astro-ph/9911177",
      archivePrefix  = "arXiv",
      primaryClass   = "astro-ph",
      SLACcitation   = "%%CITATION = ASTRO-PH/9911177;%%"
}

@article{Howlett_2012,
      author         = "Howlett, Cullan and Lewis, Antony and Hall, Alex and
                        Challinor, Anthony",
      title          = "{CMB power spectrum parameter degeneracies in the era of
                        precision cosmology}",
      journal        = "\jcap",
      volume         = "1204",
      year           = "2012",
      pages          = "027",
      doi            = "10.1088/1475-7516/2012/04/027",
      eprint         = "1201.3654",
      archivePrefix  = "arXiv",
      primaryClass   = "astro-ph.CO",
      SLACcitation   = "%%CITATION = ARXIV:1201.3654;%%"
}

@software{jax2018github,
  author = {James Bradbury and Roy Frostig and Peter Hawkins and Matthew James Johnson and Chris Leary and Dougal Maclaurin and George Necula and Adam Paszke and Jake Vander{P}las and Skye Wanderman-{M}ilne and Qiao Zhang},
  title = {{JAX}: composable transformations of {P}ython+{N}um{P}y programs},
  url = {http://github.com/google/jax},
  version = {0.3.13},
  year = {2018},
}

@software{deepmind2020jax,
  title = {The {D}eep{M}ind {JAX} {E}cosystem},
  author = {DeepMind},
  url = {http://github.com/google-deepmind},
  year = {2020},
}

@article{kidger2021equinox,
    author={Patrick Kidger and Cristian Garcia},
    title={{E}quinox: neural networks in {JAX} via callable {P}y{T}rees and filtered transformations},
    year={2021},
    journal={Differentiable Programming workshop at Neural Information Processing Systems 2021}
}

@phdthesis{kidger2021on,
    title={{O}n {N}eural {D}ifferential {E}quations},
    author={Patrick Kidger},
    year={2021},
    school={University of Oxford},
}

@software{RECFAST,
       author = {{Seager}, Sara and {Sasselov}, Dimitar D. and {Scott}, Douglas},
        title = "{{RECFAST: Calculate the Recombination History of the Universe}}",
 howpublished = {Astrophysics Source Code Library, record ascl:1106.026},
         year = 2011,
        month = jun,
          eid = {ascl:1106.026},
archivePrefix = {ascl},
       eprint = {1106.026},
       adsurl = {https://ui.adsabs.harvard.edu/abs/2011ascl.soft06026S}
}

@article{Ali_Haimoud_2010,
   title={Ultrafast effective multilevel atom method for primordial hydrogen recombination},
   volume={82},
   ISSN={1550-2368},
   url={http://dx.doi.org/10.1103/PhysRevD.82.063521},
   DOI={10.1103/physrevd.82.063521},
   number={6},
   journal={Physical Review D},
   publisher={American Physical Society (APS)},
   author={Ali-Haïmoud, Yacine and Hirata, Christopher M.},
   year={2010},
   month=sep }

@article{Hu_1995,
   title={Effect of physical assumptions on the calculation of microwave background anisotropies},
   volume={52},
   ISSN={0556-2821},
   url={http://dx.doi.org/10.1103/PhysRevD.52.5498},
   DOI={10.1103/physrevd.52.5498},
   number={10},
   journal={Physical Review D},
   publisher={American Physical Society (APS)},
   author={Hu, Wayne and Scott, Douglas and Sugiyama, Naoshi and White, Martin},
   year={1995},
   month=nov, pages={5498–5515} }

@article{simons,
   title={{The Simons Observatory: science goals and forecasts}},
   volume={2019},
   ISSN={1475-7516},
   url={http://dx.doi.org/10.1088/1475-7516/2019/02/056},
   DOI={10.1088/1475-7516/2019/02/056},
   number={02},
   journal={Journal of Cosmology and Astroparticle Physics},
   publisher={IOP Publishing},
   author={Ade, Peter and others},
   collaboration={Simons Observatory},
   year={2019},
   month=feb, pages={056–056} }

@misc{benabed_clipy,
  title        = {{clipy: Pure Python \texttt{clik} implementation with JAX support}},
  author       = {Benabed, Karim},
  howpublished = {\url{https://github.com/benabed/clipy}},
  year         = {2025},
  note         = {Version 0.15 (latest release as of July 2025); MIT License},
}

@article{Balkenhol_2024,
   title={{candl: cosmic microwave background analysis with a differentiable likelihood}},
   volume={686},
   ISSN={1432-0746},
   url={http://dx.doi.org/10.1051/0004-6361/202449432},
   DOI={10.1051/0004-6361/202449432},
   journal={Astronomy \& Astrophysics},
   publisher={EDP Sciences},
   author={Balkenhol, L. and Trendafilova, C. and Benabed, K. and Galli, S.},
   year={2024},
   month=may, pages={A10} }

@software{halofit,
       author = {{Peacock}, J.~A. and {Smith}, R.~E.},
        title = {{HALOFIT}: Nonlinear distribution of cosmological mass and galaxies},
 howpublished = {Astrophysics Source Code Library, record ascl:1402.032},
         year = 2014,
        month = feb,
          eid = {ascl:1402.032},
archivePrefix = {ascl},
       eprint = {1402.032},
       adsurl = {https://ui.adsabs.harvard.edu/abs/2014ascl.soft02032P}
}

@misc{zhou2024,
      title={{Searching for Dark Matter Interactions with ACT, SPT and DES}}, 
      author={Zilu Zhou and Neal Weiner},
      year={2024},
      eprint={2409.06771},
      archivePrefix={arXiv},
      primaryClass={hep-ph},
      url={https://arxiv.org/abs/2409.06771}, 
}

@Article{         harris2020array,
 title         = {Array programming with {NumPy}},
 author        = {Charles R. Harris and others},
 year          = {2020},
 month         = sep,
 journal       = {Nature},
 volume        = {585},
 number        = {7825},
 pages         = {357--362},
 doi           = {10.1038/s41586-020-2649-2},
 publisher     = {Springer Science and Business Media {LLC}},
 url           = {https://doi.org/10.1038/s41586-020-2649-2}
}

@ARTICLE{2020SciPy-NMeth,
  author  = {{SciPy 1.0 Contributors}},
  title   = {{{SciPy} 1.0: Fundamental Algorithms for Scientific
            Computing in Python}},
  journal = {Nature Methods},
  year    = {2020},
  volume  = {17},
  pages   = {261--272},
  adsurl  = {https://rdcu.be/b08Wh},
  doi     = {10.1038/s41592-019-0686-2},
}

@Article{Hunter:2007,
  Author    = {Hunter, J. D.},
  Title     = {Matplotlib: A {2D} graphics environment},
  Journal   = {Computing in Science \& Engineering},
  Volume    = {9},
  Number    = {3},
  Pages     = {90--95},
  publisher = {IEEE COMPUTER SOC},
  doi       = {10.1109/MCSE.2007.55},
  year      = 2007
}

@article{Ma:1995ey,
    author = "Ma, Chung-Pei and Bertschinger, Edmund",
    title = "{Cosmological perturbation theory in the synchronous and conformal Newtonian gauges}",
    eprint = "astro-ph/9506072",
    archivePrefix = "arXiv",
    doi = "10.1086/176550",
    journal = "Astrophys. J.",
    volume = "455",
    pages = "7--25",
    year = "1995"
}

@manual{sphinx,
  title  = {Sphinx Documentation Generator},
  author = {Georg Brandl and others},
  year   = {2023},
  url    = {https://www.sphinx-doc.org/},
  note   = {Accessed: 2024-08-21}
}

@article{Lewis_2008,
   title={Cosmological parameters from {WMAP} 5-year temperature maps},
   volume={78},
   ISSN={1550-2368},
   url={http://dx.doi.org/10.1103/PhysRevD.78.023002},
   DOI={10.1103/physrevd.78.023002},
   number={2},
   journal={Physical Review D},
   publisher={American Physical Society (APS)},
   author={Lewis, Antony},
   year={2008},
   month=jul }

@article{Ali-Haimoud:2010puu,
    author = "Ali-Haimoud, Yacine and Grin, Daniel and Hirata, Christopher M.",
    title = "{Radiative transfer effects in primordial hydrogen recombination}",
    eprint = "1009.4697",
    archivePrefix = "arXiv",
    primaryClass = "astro-ph.CO",
    doi = "10.1103/PhysRevD.82.123502",
    journal = "Phys. Rev. D",
    volume = "82",
    pages = "123502",
    year = "2010"
}

@ARTICLE{Peebles_1968,
       author = {{Peebles}, P.~J.~E.},
        title = "{Recombination of the Primeval Plasma}",
      journal = {\apj},
         year = 1968,
        month = jul,
       volume = {153},
        pages = {1},
          doi = {10.1086/149628},
       adsurl = {https://ui.adsabs.harvard.edu/abs/1968ApJ...153....1P}
}

@ARTICLE{Zeldovich_1968,
       author = {{Zeldovich}, Y.~B. and {Kurt}, V.~G. and {Syunyaev}, R.~A.},
        title = "{Recombination of Hydrogen in the Hot Model of the Universe}",
      journal = {Zhurnal Eksperimentalnoi i Teoreticheskoi Fiziki},
         year = 1968,
        month = jul,
       volume = {55},
        pages = {278-286},
       adsurl = {https://ui.adsabs.harvard.edu/abs/1968ZhETF..55..278Z}
}

@article{Grin_2010,
   title={Cosmological hydrogen recombination: The effect of extremely high-n states},
   volume={81},
   ISSN={1550-2368},
   url={http://dx.doi.org/10.1103/PhysRevD.81.083005},
   DOI={10.1103/physrevd.81.083005},
   number={8},
   journal={Physical Review D},
   publisher={American Physical Society (APS)},
   author={Grin, Daniel and Hirata, Christopher M.},
   year={2010},
   month=apr }

@article{Hirata_2009,
   title={Lyman-$\alpha$ transfer in primordial hydrogen recombination},
   volume={80},
   ISSN={1550-2368},
   url={http://dx.doi.org/10.1103/PhysRevD.80.023001},
   DOI={10.1103/physrevd.80.023001},
   number={2},
   journal={Physical Review D},
   publisher={American Physical Society (APS)},
   author={Hirata, Christopher M. and Forbes, John},
   year={2009},
   month=jul }

@misc{list_2025,
      title={{DISCO-DJ II}: a differentiable particle-mesh code for cosmology}, 
      author={Florian List and Oliver Hahn and Thomas Flöss and Lukas Winkler},
      year={2025},
      eprint={2510.05206},
      archivePrefix={arXiv},
      primaryClass={astro-ph.CO},
      url={https://arxiv.org/abs/2510.05206}, 
}

@article{Villasenor_2021,
   title={{Effects of Photoionization and Photoheating on Ly-alpha Forest Properties from Cholla Cosmological Simulations}},
   volume={912},
   ISSN={1538-4357},
   url={http://dx.doi.org/10.3847/1538-4357/abed5a},
   DOI={10.3847/1538-4357/abed5a},
   number={2},
   journal={The Astrophysical Journal},
   publisher={American Astronomical Society},
   author={Villasenor, Bruno and Robertson, Brant and Madau, Piero and Schneider, Evan},
   year={2021},
   month=may, pages={138} }

@article{Wu_2019,
   title={Simulating the effect of photoheating feedback during reionization},
   volume={488},
   ISSN={1365-2966},
   url={http://dx.doi.org/10.1093/mnras/stz1726},
   DOI={10.1093/mnras/stz1726},
   number={1},
   journal={Monthly Notices of the Royal Astronomical Society},
   publisher={Oxford University Press (OUP)},
   author={Wu, Xiaohan and Kannan, Rahul and Marinacci, Federico and Vogelsberger, Mark and Hernquist, Lars},
   year={2019},
   month=jun, pages={419–437} }

@software{conlin_2025,
  author       = {Conlin, Rory},
  title        = {interpax},
  month        = oct,
  year         = 2025,
  publisher    = {Zenodo},
  version      = {v0.3.11},
  doi          = {10.5281/zenodo.17378822},
  url          = {https://doi.org/10.5281/zenodo.17378822},
  swhid        = {swh:1:dir:54aedc47fdfc0088311b2be7997dcc0c8cd36398
                   ;origin=https://doi.org/10.5281/zenodo.10028967;vi
                   sit=swh:1:snp:c885a77053c09625ee7b51e70d3c8df9997c
                   58ba;anchor=swh:1:rel:a9515db646cc0e6ccdfaaadc75f4
                   f87e9599f2e7;path=f0uriest-interpax-e64569b
                  },
}

@article{Gariazzo:2021iiu,
    author = "Gariazzo, S. and F. de Salas, P. and Pisanti, O. and Consiglio, R.",
    title = "{PArthENoPE revolutions}",
    eprint = "2103.05027",
    archivePrefix = "arXiv",
    primaryClass = "astro-ph.IM",
    doi = "10.1016/j.cpc.2021.108205",
    journal = "Comput. Phys. Commun.",
    volume = "271",
    pages = "108205",
    year = "2022"
}

@article{Consiglio_2018,
   title={{PArthENoPE} reloaded},
   volume={233},
   ISSN={0010-4655},
   url={http://dx.doi.org/10.1016/j.cpc.2018.06.022},
   DOI={10.1016/j.cpc.2018.06.022},
   journal={Computer Physics Communications},
   publisher={Elsevier BV},
   author={Consiglio, R. and de Salas, P.F. and Mangano, G. and Miele, G. and Pastor, S. and Pisanti, O.},
   year={2018},
   month=dec, pages={237–242} }

@article{Pisanti:2007hk,
    author = "Pisanti, O. and Cirillo, A. and Esposito, S. and Iocco, F. and Mangano, G. and Miele, G. and Serpico, P. D.",
    title = "{PArthENoPE: Public Algorithm Evaluating the Nucleosynthesis of Primordial Elements}",
    eprint = "0705.0290",
    archivePrefix = "arXiv",
    primaryClass = "astro-ph",
    reportNumber = "DSF-13-07, FERMILAB-PUB-07-079-A, SLAC-PUB-12488",
    doi = "10.1016/j.cpc.2008.02.015",
    journal = "Comput. Phys. Commun.",
    volume = "178",
    pages = "956--971",
    year = "2008"
}

@article{Lesgourgues:2013bra,
    author = "Lesgourgues, Julien and Tram, Thomas",
    title = "{Fast and accurate CMB computations in non-flat FLRW universes}",
    eprint = "1312.2697",
    archivePrefix = "arXiv",
    primaryClass = "astro-ph.CO",
    reportNumber = "CERN-PH-TH-2013-298, LAPTH-071-13",
    doi = "10.1088/1475-7516/2014/09/032",
    journal = "JCAP",
    volume = "09",
    pages = "032",
    year = "2014"
}

@article{Aghanim_2018,
   title={{Planck2018 results: VI. Cosmological parameters}},
   volume={641},
   ISSN={1432-0746},
   url={http://dx.doi.org/10.1051/0004-6361/201833910},
   DOI={10.1051/0004-6361/201833910},
   journal={Astronomy \& Astrophysics},
   publisher={EDP Sciences},
   author={Aghanim, N. and others},
   collaboration={Planck},
   year={2020},
   month=sep, pages={A6} }

@article{Doux_2018,
    author = {Doux, Cyrille and Penna-Lima, Mariana and Vitenti, Sandro D P and Tréguer, Julien and Aubourg, Eric and Ganga, Ken},
    title = {Cosmological constraints from a joint analysis of cosmic microwave background and spectroscopic tracers of the large-scale structure},
    journal = {Monthly Notices of the Royal Astronomical Society},
    volume = {480},
    number = {4},
    pages = {5386-5411},
    year = {2018},
    month = {08},
    issn = {0035-8711},
    doi = {10.1093/mnras/sty2160},
    url = {https://doi.org/10.1093/mnras/sty2160},
    eprint = {https://academic.oup.com/mnras/article-pdf/480/4/5386/25985754/sty2160.pdf},
}

@article{Challinor_2005,
   title={Lensed {CMB} power spectra from all-sky correlation functions},
   volume={71},
   ISSN={1550-2368},
   url={http://dx.doi.org/10.1103/PhysRevD.71.103010},
   DOI={10.1103/physrevd.71.103010},
   number={10},
   journal={Physical Review D},
   publisher={American Physical Society (APS)},
   author={Challinor, Anthony and Lewis, Antony},
   year={2005},
   month=may }

@misc{camphuis2025spt3gd1cmbtemperature,
      title={{SPT-3G D1: CMB temperature and polarization power spectra and cosmology from 2019 and 2020 observations of the SPT-3G Main field}}, 
      author={E. Camphuis and others},
      collaboration={SPT-3G},
      year={2025},
      eprint={2506.20707},
      archivePrefix={arXiv},
      primaryClass={astro-ph.CO},
      url={https://arxiv.org/abs/2506.20707}, 
}

@article{Mangano_2005,
   title={Relic neutrino decoupling including flavour oscillations},
   volume={729},
   ISSN={0550-3213},
   url={http://dx.doi.org/10.1016/j.nuclphysb.2005.09.041},
   DOI={10.1016/j.nuclphysb.2005.09.041},
   number={1–2},
   journal={Nuclear Physics B},
   publisher={Elsevier BV},
   author={Mangano, Gianpiero and Miele, Gennaro and Pastor, Sergio and Pinto, Teguayco and Pisanti, Ofelia and Serpico, Pasquale D.},
   year={2005},
   month=nov, pages={221–234} }

@INCOLLECTION{2011hmcm.book..113N,
       author = {{Neal}, Radford},
        title = "{MCMC Using Hamiltonian Dynamics}",
    booktitle = {Handbook of Markov Chain Monte Carlo},
         year = 2011,
        pages = {113-162},
          doi = {10.1201/b10905},
       adsurl = {https://ui.adsabs.harvard.edu/abs/2011hmcm.book..113N}
}

@article{betancourt2015hamiltonian,
  title={{Hamiltonian Monte Carlo} for hierarchical models},
  author={Betancourt, Michael and Girolami, Mark},
  journal={Current trends in Bayesian methodology with applications},
  volume={79},
  number={30},
  pages={2--4},
  year={2015},
  publisher={CRC Press Boca Raton, FL}
}

@misc{janken2025clientnewtoolemulating,
      title={{CLiENT}: A new tool for emulating cosmological likelihoods using deep neural networks}, 
      author={Luca Janken and Steen Hannestad and Thomas Tram and Andreas Nygaard},
      year={2025},
      eprint={2512.17509},
      archivePrefix={arXiv},
      primaryClass={astro-ph.CO},
      url={https://arxiv.org/abs/2512.17509}, 
}

@article{Louis_2025,
doi = {10.1088/1475-7516/2025/11/062},
url = {https://doi.org/10.1088/1475-7516/2025/11/062},
year = {2025},
month = {nov},
publisher = {IOP Publishing},
volume = {2025},
number = {11},
pages = {062},
author = {Louis, Thibaut and others},
collaboration={Atacama Cosmology Telescope},
title = {{The Atacama Cosmology Telescope: DR6 power spectra, likelihoods and $\Lambda$CDM parameters}},
journal = {Journal of Cosmology and Astroparticle Physics},
}

@article{Bianchini_2020,
   title={{Constraints on Cosmological Parameters from the 500 deg$^2$ SPTPOL Lensing Power Spectrum}},
   volume={888},
   ISSN={1538-4357},
   url={http://dx.doi.org/10.3847/1538-4357/ab6082},
   DOI={10.3847/1538-4357/ab6082},
   number={2},
   journal={The Astrophysical Journal},
   publisher={American Astronomical Society},
   author={Bianchini, F. and others},
   year={2020},
   collaboration={SPT},
   month=jan, pages={119} }

@misc{aver2026_IV,
      title={{The LBT Y$_\mathrm{p}$ Project IV: A New Value of the Primordial Helium Abundance}}, 
      author={Erik Aver and Evan D. Skillman and Richard W. Pogge and Noah S. J. Rogers and Miqaela K. Weller and Keith A. Olive and Danielle A. Berg and John J. Salzer and John H. Miller Jr. and José Eduardo Méndez-Delgado},
      year={2026},
      eprint={2601.22238},
      archivePrefix={arXiv},
      primaryClass={astro-ph.CO},
      url={https://arxiv.org/abs/2601.22238}, 
}

@misc{yeh2026_V,
      title={{The LBT $Y_{\rm p}$ Project V: Cosmological Implications of a New Determination of Primordial $^4$He}}, 
      author={Tsung-Han Yeh and Keith A. Olive and Brian D. Fields and Erik Aver and Richard W. Pogge and Noah S. J. Rogers and Evan D. Skillman and Miqaela K. Weller},
      year={2026},
      eprint={2601.22239},
      archivePrefix={arXiv},
      primaryClass={astro-ph.CO},
      url={https://arxiv.org/abs/2601.22239}, 
}

@article{Pitrou_2018,
   title={Precision big bang nucleosynthesis with improved Helium-4 predictions},
   volume={754},
   ISSN={0370-1573},
   url={http://dx.doi.org/10.1016/j.physrep.2018.04.005},
   DOI={10.1016/j.physrep.2018.04.005},
   journal={Physics Reports},
   publisher={Elsevier BV},
   author={Pitrou, Cyril and Coc, Alain and Uzan, Jean-Philippe and Vangioni, Elisabeth},
   year={2018},
   month=sep, pages={1–66} }

@article{Poulin:2018cxd,
    author = "Poulin, Vivian and Smith, Tristan L. and Karwal, Tanvi and Kamionkowski, Marc",
    title = "{Early Dark Energy Can Resolve The Hubble Tension}",
    eprint = "1811.04083",
    archivePrefix = "arXiv",
    primaryClass = "astro-ph.CO",
    doi = "10.1103/PhysRevLett.122.221301",
    journal = "Phys. Rev. Lett.",
    volume = "122",
    number = "22",
    pages = "221301",
    year = "2019"
}

@article{Boddy:2018wzy,
    author = "Boddy, Kimberly K. and Gluscevic, Vera and Poulin, Vivian and Kovetz, Ely D. and Kamionkowski, Marc and Barkana, Rennan",
    title = "{{Critical assessment of CMB limits on dark matter-baryon scattering: New treatment of the relative bulk velocity}}",
    eprint = "1808.00001",
    archivePrefix = "arXiv",
    primaryClass = "astro-ph.CO",
    doi = "10.1103/PhysRevD.98.123506",
    journal = "Phys. Rev. D",
    volume = "98",
    number = "12",
    pages = "123506",
    year = "2018"
}

@article{Poulin:2018dzj,
    author = "Poulin, Vivian and Smith, Tristan L. and Grin, Daniel and Karwal, Tanvi and Kamionkowski, Marc",
    title = "{Cosmological implications of ultralight axionlike fields}",
    eprint = "1806.10608",
    archivePrefix = "arXiv",
    primaryClass = "astro-ph.CO",
    doi = "10.1103/PhysRevD.98.083525",
    journal = "Phys. Rev. D",
    volume = "98",
    number = "8",
    pages = "083525",
    year = "2018"
}

@article{DiValentino:2017oaw,
    author = "Di Valentino, Eleonora and B{\o}ehm, C{\'e}line and Hivon, Eric and Bouchet, Fran{\c{c}}ois R.",
    title = "{Reducing the $H_0$ and $\sigma_8$ tensions with Dark Matter-neutrino interactions}",
    eprint = "1710.02559",
    archivePrefix = "arXiv",
    primaryClass = "astro-ph.CO",
    doi = "10.1103/PhysRevD.97.043513",
    journal = "Phys. Rev. D",
    volume = "97",
    number = "4",
    pages = "043513",
    year = "2018"
}

@article{Lesgourgues:2015wza,
    author = "Lesgourgues, Julien and Marques-Tavares, Gustavo and Schmaltz, Martin",
    title = "{Evidence for dark matter interactions in cosmological precision data?}",
    eprint = "1507.04351",
    archivePrefix = "arXiv",
    primaryClass = "astro-ph.CO",
    doi = "10.1088/1475-7516/2016/02/037",
    journal = "JCAP",
    volume = "02",
    pages = "037",
    year = "2016"
}

@article{Buen-Abad:2017gxg,
    author = "Buen-Abad, Manuel A. and Schmaltz, Martin and Lesgourgues, Julien and Brinckmann, Thejs",
    title = "{Interacting Dark Sector and Precision Cosmology}",
    eprint = "1708.09406",
    archivePrefix = "arXiv",
    primaryClass = "astro-ph.CO",
    doi = "10.1088/1475-7516/2018/01/008",
    journal = "JCAP",
    volume = "01",
    pages = "008",
    year = "2018"
}

@article{Herold_2025,
   title={{Revisiting the impact of neutrino mass hierarchies on neutrino mass constraints in light of recent DESI data}},
   volume={111},
   ISSN={2470-0029},
   url={http://dx.doi.org/10.1103/PhysRevD.111.083518},
   DOI={10.1103/physrevd.111.083518},
   number={8},
   journal={Physical Review D},
   publisher={American Physical Society (APS)},
   author={Herold, Laura and Kamionkowski, Marc},
   year={2025},
   month=apr }

@article{Aloni_2022,
   title={A Step in understanding the Hubble tension},
   volume={105},
   ISSN={2470-0029},
   url={http://dx.doi.org/10.1103/PhysRevD.105.123516},
   DOI={10.1103/physrevd.105.123516},
   number={12},
   journal={Physical Review D},
   publisher={American Physical Society (APS)},
   author={Aloni, Daniel and Berlin, Asher and Joseph, Melissa and Schmaltz, Martin and Weiner, Neal},
   year={2022},
   month=jun }

@misc{hough2026kosmulatorpythonframeworkcosmological,
      title={{Kosmulator: A Python framework for cosmological inference with MCMC}}, 
      author={Renier T. Hough and Robert Rugg and Shambel Sahlu and Amare Abebe},
      year={2026},
      eprint={2602.08424},
      archivePrefix={arXiv},
      primaryClass={astro-ph.CO},
      url={https://arxiv.org/abs/2602.08424}, 
}

@book{watson1944bessel,
  author    = {Watson, G. N.},
  title     = {A Treatise on the Theory of Bessel Functions},
  edition   = {2},
  year      = {1944},
  publisher = {Cambridge University Press},
  address   = {Cambridge, England}
}

@article{El_Gammal_2023,
   title={{Fast and robust Bayesian inference using Gaussian processes with GPry}},
   volume={2023},
   ISSN={1475-7516},
   url={http://dx.doi.org/10.1088/1475-7516/2023/10/021},
   DOI={10.1088/1475-7516/2023/10/021},
   number={10},
   journal={Journal of Cosmology and Astroparticle Physics},
   publisher={IOP Publishing},
   author={El Gammal, Jonas and Schöneberg, Nils and Torrado, Jesús and Fidler, Christian},
   year={2023},
   month=Oct, pages={021} }

@article{Nygaard_2023,
   title={{CONNECT: a neural network based framework for emulating cosmological observables and cosmological parameter inference}},
   volume={2023},
   ISSN={1475-7516},
   url={http://dx.doi.org/10.1088/1475-7516/2023/05/025},
   DOI={10.1088/1475-7516/2023/05/025},
   number={05},
   journal={Journal of Cosmology and Astroparticle Physics},
   publisher={IOP Publishing},
   author={Nygaard, Andreas and Holm, Emil Brinch and Hannestad, Steen and Tram, Thomas},
   year={2023},
   month=May, pages={025} }

\end{document}